\documentclass[preprint]{aastex}

\usepackage[onecolumn]{emulateapj5}

\def\HII{{\ion{H}{2}}}
\def\OII{[{\ion{O}{2}}]}
\def\OIIHa{[{\ion{O}{2}}]/H$\alpha$}
\def\OIINIIHa{[{\ion{O}{2}}]/(H$\alpha$+[{\ion{N}{2}}])}
\def\OIIHb{[{\ion{O}{2}}]/H$\beta$}
\def\HaHb{H$\alpha$/H$\beta$}
\def\OIIOIII{[{\ion{O}{2}}]/[{\ion{O}{3}}]}
\def\OIIIHb{[{\ion{O}{3}}]/H$\beta$}
\def\OIII5007Hb{[{\ion{O}{3}}]~$\lambda5007$/H$\beta$}
\def\4959_5007{[\ion{O}{3}]~$\lambda \lambda$4959,5007}
\def\OIII49595007{[\ion{O}{3}]~$\lambda \lambda 4959,5007$}
\def\ratioR23{([\ion{O}{2}]~$\lambda$3727 +
[\ion{O}{3}]~$\lambda\lambda$4959,5007)/H$\beta$}
\def\R23{${\rm R}_{23}$}
\def\dS23{${\rm S}_{23}$}
\def\OIIIl{[\ion{O}{3}]~$\lambda$5007}

\def\Msun{${\rm M}_{\odot}$}
\def\Lsun{${\rm L}_{\odot}$}
\def\NII{[{\ion{N}{2}}]}
\def\OIIIOII{[\ion{O}{3}]/[\ion{O}{2}]}

\def\NIIOII{[\ion{N}{2}]/[\ion{O}{2}]}

\def\OH{$\log({\rm O/H})+12$}
\def\NIISII{[\ion{N}{2}]/[\ion{S}{2}]}

\def\ratioS23{([\ion{S}{2}]~$\lambda \lambda$6717,31 +
[\ion{S}{3}]~$\lambda\lambda$9069,9532)/H$\beta$}
\def\NIIHa{[\ion{N}{2}]/H$\alpha$}

\def\Hb{{H$\beta$}}
\def\O4363{[{\ion{O}{3}}]~$\lambda$4363}
\def\OIII{[{\ion{O}{3}}]}
\def\OI{[{\ion{O}{1}}]~$\lambda$6300}
\def\Ha{{H$\alpha$}}
\def\L60{L$_{60}$}

\def\Lsun{L$_{\odot}$}

\def\EBV{E($B-V$)}

\shorttitle{}
\shortauthors{}

\slugcomment{To appear in the April 2004 issue of the Astronomical Journal}

\begin{document}

\title{[OII] as a Star Formation Rate Indicator}

\author{Lisa J. Kewley\altaffilmark{1}}
\affil {Harvard-Smithsonian Center for Astrophysics}
\authoraddr{ 60 Garden Street MS-20, Cambridge, MA 02138}
\email {lkewley@cfa.harvard.edu}
\altaffiltext{1}{CfA Fellow}

\author{Margaret J. Geller}
\affil{Smithsonian Astrophysical Observatory}
\authoraddr{ 60 Garden Street MS-20, Cambridge, MA 02138}

\author{Rolf A. Jansen}
\affil{Dept. of Physics \& Astronomy, Arizona State University}
\authoraddr{P.O. Box 871504, Tempe, AZ 85287-1504}

\begin{abstract}
We investigate the \OII\ emission-line as a 
star formation rate (SFR) indicator 
using integrated spectra of 97 galaxies from the Nearby Field 
Galaxies Survey (NFGS).  The sample includes all Hubble types and contains
SFRs ranging from 0.01 to 100 \Msun~yr$^{-1}$.
We compare the Kennicutt [\ion{O}{2}] and H$\alpha$ SFR calibrations
and show that there are two significant effects which produce 
disagreement between SFR(\OII) and SFR(\Ha): reddening and metallicity.
 Differences in the ionization state of the ISM do not
contribute significantly to the observed difference between
SFR(\OII) and SFR(H$\alpha$) for the NFGS galaxies with 
metallicities \OH$\gtrsim8.5$.
The Kennicutt \OII--SFR relation assumes a typical 
reddening for nearby galaxies;  in practice, the reddening 
differs significantly from sample to sample.  We derive a new SFR(\OII) 
calibration which does not contain a reddening assumption.  
Our new SFR(\OII) calibration also provides an optional correction for 
metallicity. Our SFRs derived from \OII\ agree with those
derived from \Ha\ to within
0.03-0.05~dex.
We show that the reddening, E($B-V$), increases with intrinsic (i.e. reddening
corrected) \OII\ luminosity for the NFGS sample.  
We apply our SFR(\OII) calibration with metallicity correction to two samples:
high-redshift $0.8<z<1.6$ galaxies from the NICMOS \Ha\ survey,
and $0.5<z<1.1$ galaxies from the Canada-France Redshift Survey. 
The SFR(\OII) and SFR(\Ha) for these samples agree to within 
the scatter observed for the NFGS sample, indicating that 
our SFR(\OII) relation can be 
applied to both local and high-$z$ galaxies.
Finally, we apply our SFR(\OII) to estimates of 
the cosmic star formation history.
After reddening and metallicity corrections, the star formation rate
densities derived from \OII\ and \Ha\ agree to within $\sim30$\%.
\end{abstract}


\section{Introduction}

Observing the star formation rate since 
the earliest times in the Universe is crucial to our understanding 
of the formation 
and evolution of galaxies.  The star formation rate 
indicators developed four decades ago provided a first quantitative 
measure of the global star formation in galaxies 
\citep{Tinsley68,Tinsley72,Searle73}.
These indicators were based on stellar population
synthesis models of galaxy colors.  More recent and 
precise star formation rate
indicators rely on optical emission-lines, 
UV continuum, radio, and infrared fluxes 
\citep[e.g.,][]{Kennicutt83,Donas84,Rieke78}.  These indicators, 
applied to nearby samples, provide insight into the 
properties of galaxies along the Hubble sequence 
\citep[see][for a review]{Kennicutt98}.   

The advent of large spectroscopic surveys enabled significant progress 
in our understanding of global galaxy evolution as a function of redshift.
\citet{Lilly95} studied the cosmic evolution of the field galaxy 
population to a redshift of $z\sim 1$ using the Canada-France Redshift 
Survey (CFRS).  They showed that the field galaxy population 
evolves and that this evolution is strongly related to galaxy color.
\citet{Ellis96} confirmed this observation using Autofib redshift 
survey data over a similar redshift range.  Ellis et al. concluded that 
the steepening of the luminosity function with look-back time is a direct 
consequence of the increasing space density of blue star forming galaxies 
at moderate redshift.  

Deep surveys like the Hubble Deep Fields allowed the study of
star formation history of galaxies over an even wider redshift range.  
\citet{Madau96} estimated the star formation history of galaxies 
between $z=0$ and $z=4$.  Using the Hubble Deep Fields and UV surveys from 
\citet{Lilly96}, Madau et al. argued that the peak star formation rate
occurs at redshifts from $z=1.3-2.7$.  Many large, deep spectroscopic 
surveys carried out recently have sparked an explosion of research into 
the star formation history of the universe 
\citep[for example,][]{Hammer97,Rowan-Robinson01,Cole01,Baldry02,Lanzetta02,
Rosa-Gonzalez02,Tresse02,Hippelein03}.  

Cosmic star formation history
studies over a large redshift range require the use
of different star formation rate indicators.
Unfortunately, there are significant discrepancies among SFR estimates 
made using different indicators.  
To obtain a more reliable
view of the cosmic star formation history, 
it is essential to gain a detailed  understanding of and to reach 
agreement among the star formation indicators at multiple wavelengths .

The hydrogen Balmer 
line \Ha~6563\AA\ is currently the most reliable tracer of star formation, 
provided \Ha\ can be corrected for reddening. 
In the ionization-bounded nebulae of \HII\ regions and
star-forming galaxies, the Balmer emission line luminosity scales directly 
with the total ionizing flux of the embedded stars.   For 
many years there was an apparent disagreement between the star formation 
rate derived from \Ha\ and those derived at other wavelengths, including the 
far-infrared (FIR).  Correction of \Ha\ for stellar 
absorption and reddening brings the \Ha\ and FIR SFRs into agreement for 
active star-forming galaxies \citep[e.g.,][]{Rosa-Gonzalez02,Charlot02,
Dopita02} and for the normal star-forming galaxies of all Hubble types 
in the Nearby Field Galaxy Survey \citep{Kewley02a,Jansen00a,Jansen00b}.

Although \Ha\ provides a useful SFR indicator for nearby galaxies, it is
not easily observable for more distant galaxies.  \Ha\ redshifts out of 
the visible band for $z\gtrsim0.4$.  An alternative diagnostic for the
$z \sim 0.4 - 1.5$ range is the \OII~$\lambda \lambda 3726,3729$ 
doublet.   Several authors have calibrated the \OII\ star formation
rate \citep[e.g.,][]{Gallagher89,Kennicutt98,Rosa-Gonzalez02}.
Unfortunately, the \OII\ emission-line is plagued by problems including 
reddening and abundance dependence \citep{Jansen01, Charlot02}.   
Most previous comparisons of \OII\ with other SFR indicators
were based on spectra which lack sufficient spatial coverage, 
signal-to-noise, and/or wavelength coverage to make a detailed
correction for reddening 
\citep[eg.,][]{Kennicutt83o,Hopkins01,Charlot02,Buat02}.
For example, \citet{Charlot02} had to assume an `average' attenuation 
${\rm A_{v}}=1$ for the galaxies in the Stromlo-APM survey. 
They found 
significant discrepancies between the \Ha\ and \OII\ SFRs which they 
attributed to variations in the effective gas parameters (ionization, 
metallicity, and dust content) of the galaxies.  
Similarly, \citet{Teplitz03} showed
that there is a large disagreement among the SFR estimates based on 
\Ha\ and those based on the \OII\ emission line.   

At temperatures typical of star-forming regions (10000-20000 K), the 
excitation energy 
between the two upper D levels for \OII\ and the lower S level 
is roughly the thermal electron energy $kT$. The \OII\ 
doublet is therefore closely linked to the electron temperature and
consequently abundance.  In fact, the \OII\ emission-line doublet 
enters most reliable optical abundance diagnostics developed over
the last two decades 
\citep[eg.,][]{Pagel80,Edmunds84,
Dopita86,Torres-Peimbert89,Skillman89,McGaugh91,Zaritsky94,Charlot01,
Kewley02b}.  Most of these abundance diagnostics use the intensity ratio 
\ratioR23, commonly known as \R23.  \citet{Jansen01} showed that the
\OIIHa\ ratio is strongly correlated with the \R23\ ratio for the 
Nearby Field Galaxies Survey (NFGS).  Jansen et al. concluded
that \OII\ is affected by metallicity and that \Hb\ is a significantly
better tracer of star formation when detected at a sufficient signal-to-noise (S/N) and spectral resolution to correct for underlying stellar absorption.   None of the SFR 
calibrations so far take abundance into account, 
largely because a set of high S/N integrated (global) 
spectra for galaxies spanning a large range of SFRs is 
required to derive a reliable calibration.   Such a sample has not been 
available.  

Here, we investigate \OII\ as a star formation rate diagnostic using 
integrated (global) spectra for the NFGS.  Our spectra have the advantage 
that: 
(1) they contain
both \OII\ and \Ha, (2) we can  correct \Ha\ and \Hb\ for underlying stellar
absorption, (3) we can measure the reddening using the 
Balmer Decrement, and (4) we can resolve \Ha\ and \NII.
Furthermore, we can calculate global galaxy abundances from the 
common diagnostic \R23\ to investigate the relationship between SFR(\OII) and
abundance.  

We describe the sample selection and optical spectra in 
Section~\ref{sample}.  The commonly-used SFR indicators are discussed
in Section~\ref{oiisfr}.  In Section~\ref{newoii}, we explore the 
discrepancy between the SFR(\OII) and SFR(\Ha) and derive a new 
SFR(\OII) calibration which takes abundance and reddening into account.
In Section~\ref{models}, we use theoretical population synthesis and 
photoionization models to investigate the theoretical dependence 
of SFR(\OII) on the ionized gas properties and we 
derive a theoretical SFR(\OII) diagnostic.
In Section~\ref{high_z}, we discuss the use of \OII\ as a SFR indicator
in more distant samples and apply our SFR(\OII) calibrations to the 
$0.8<z<1.6$ sample of \citet{Hicks02} and the $0.5<z<1.1$ sample of
\citet{Tresse02}.  Based on these findings, in 
Section~\ref{cosmicsfr}, we investigate the implications for cosmic 
star formation history studies which use \OII\ as a SFR indicator.
Throughout this paper, we adopt the flat $\Lambda$-dominated cosmology
as measured by the WMAP experiment ($h=0.72$, $\Omega_{m}=0.29$; 
Spergel et al. 2003).

\section{Sample Selection and Spectrophotometry \label{sample}}

The NFGS is ideal for investigating star formation rates because it
is an  objectively selected sample for which integrated
spectra are available. \citet{Jansen00a} provide a detailed discussion 
of the NFGS sample selection.  Briefly, Jansen et al. 
selected 198 nearby galaxies in an objective (unbiased) manner by sorting 
the CfA catalog into 1 mag-wide bins 
of $M_{Z}$.  Within each bin, the sample was sorted according to their
CfA1 morphological type.  To avoid a strict diameter limit, which 
might introduce a bias against the inclusion of low surface brightness 
galaxies in the sample, Jansen et al. used a
radial velocity limit, ${\rm V_{LG} (km\,s^{-1}) > 10^{-0.19-0.2M_{z}}}$
(with respect to the Local Group standard of rest).  To avoid a sampling
bias favoring a cluster population, they excluded galaxies in the
direction of the Virgo Cluster.  Finally, Jansen et al. selected 
every $Nth$ galaxy in each bin to approximate the local galaxy 
luminosity function \citep[e.g.,][]{Marzke94}.  The final 198-galaxy sample 
represents the full range in 
Hubble type and absolute magnitude present in the CfA1 galaxy survey 
\citep{Davis83,Huchra83}.   

Both integrated and nuclear spectrophotometry are available for almost all
galaxies in the NFGS sample, including integrated \Ha, 
\Hb, and \OII\ fluxes \citep{Jansen00b}.  
The integrated spectra typically cover 82$\pm$7\% of each galaxy.
We have calibrated the integrated fluxes to absolute fluxes 
by careful comparison with B-band surface photometry \citep[described in ][;
 hereafter Paper I]{Kewley02a}.  
The \Ha\ and \Hb\ emission-line fluxes were carefully corrected
for underlying stellar absorption as described in Paper I.  

A total of 116 galaxies in the NFGS have spectra with measurable \Ha,
\Hb, \NII~$\lambda6584$ fluxes, \OII~$\lambda3727$, and 
\OIII~$\lambda5007$ emission lines. 
Due to low S/N ratios in the \OIII~$\lambda4959$ emission-line, we used
the theoretical ratio \OIII~$\lambda5007$/\OIII~$\lambda4959 \sim 3$
to calculate the \OIII~$\lambda4959$ flux. 

The NFGS emission-line fluxes have been 
corrected for Galactic extinction by two methods: 
(1) using the HI maps of \citet{Burnstein84}, listed in the Third Reference Catalogue of Bright Galaxies \citep{deVaucouleurs91}, 
and (2) using the COBE and IRAS maps (plus the Leiden-Dwingeloo maps 
of HI emission) of \citet{Schlegel98}.  The average Galactic extinction is
E($B-V$)=$0.014 \pm 0.003$ (method 1) or E($B-V$)=$0.016 \pm 0.003$ (method 2).

We corrected the emission line 
fluxes for reddening using the Balmer decrement and the \citet{Cardelli89} 
(CCM) reddening curve.  We assumed an ${\rm R_{V}=Av/{\rm E}(B-V)} = 3.1$ and an 
intrinsic H$\alpha$/H$\beta$ ratio of 2.85 (the Balmer decrement for case B 
recombination at T$=10^4$K and $n_{e} \sim 10^2 - 10^4 {\rm cm}^{-3}$;
Osterbrock 1989).  After underlying Balmer absorption was removed,
ten galaxies have Balmer decrements less than 2.85.  A Balmer decrement
less than 2.85 results 
from a combination of: (1) intrinsically low reddening, (2) errors in the
stellar absorption correction, and (3) errors in the line flux calibration and 
 measurement.  Errors
in the stellar absorption correction and flux calibration are discussed in
detail in Paper I, and are $\sim$12-17\% on average, with a maximum
error of $\sim30$\%.
For the S/N of our data, the lowest E(B-V) measurable is 0.02.  We therefore
assign these ten galaxies an upper limit of E(B-V)$<0.02$.  The difference
between applying a reddening correction with an E(B-V) of 0.02 and 0.00 
is minimal: an E(B-V) of 0.02 corresponds to an attenuation factor of 1.04
at \Ha\ and 1.09 at \OII\ using the CCM curve.

To rule out the presence of AGN in the NFGS sample, we used the
theoretical optical classification scheme developed by \citet{Kewley01a}. 
The optical diagnostic diagrams indicate that the global spectra of
97/116 NFGS galaxies are dominated by star formation.  These 97 galaxies
(Table~\ref{sample_table}) constitute the sample we analyse here. 
The spectra of the remaining 19 galaxies are
either dominated by AGN (5/19) or are ``ambiguous'' galaxies (14/19).
Ambiguous galaxies have line ratios that indicate the presence 
of an AGN in one or two out of the three optical diagnostic diagrams.  
Because these galaxies are likely to contain both starburst and AGN activity
\citep[see e.g.,][]{Kewley01a,Hill99}, we do not include them
in the following analysis.  

\section{\boldmath The K98 [OII] and \Ha\ SFR indicators \label{oiisfr}}
 
The development of SFR calibrations has been an intense topic of
research for more than three decades \citep[see][for a review]{Kennicutt98}
(hereafter K98).  We start with the K98 SFR relations for \OII\ and 
\Ha\ because these relations are applied in many current SFR studies.

The K98 \Ha\ SFR calibration is derived from evolutionary
synthesis models that assume solar metallicity and no dust.  K98 assumed 
that the total integrated stellar luminosity shortward of the Lyman limit is
re-emitted in the nebular emission lines. The K98 relation between \Ha\ 
luminosity and SFR is:

\begin{equation}
{\rm SFR (M_{\odot} yr^{-1}) = 7.9 \times 10^{-42}\, L(H\alpha)\, (ergs\,s^{-1})}
\label{eq_SFR_Ha}
\end{equation}

\noindent
where L(\Ha) denotes the intrinsic \Ha~6563\AA\ luminosity.  
Paper~I shows that, once the SFR(\Ha) is
corrected for underlying Balmer absorption and reddening, the mean SFR
derived from \Ha\ agrees with the mean SFR derived from the far-infrared
luminosity to within 10\% \citep{Kewley02a}.   

The \OII\ SFR calibration is much less straightforward.  The
K98 \OII\ calibration is:
\begin{equation}
{\rm SFR (M_{\odot} yr^{-1}) = (1.4\pm0.4) \times 10^{-41}\, 
L([OII])\, (ergs\,s^{-1})}  \label{eq_SFR_OII}
\end{equation}

\noindent
K98 derived this calibration from the 
K98 SFR(\Ha) relation (equation~\ref{eq_SFR_Ha}) and 
two previous \OII\ calibrations by \citet[][; hereafter G89]{Gallagher89} 
 and \citet[][; hereafter K92]{Kennicutt92}.   
The error estimate in equation~(\ref{eq_SFR_OII}) reflects the difference  
between the G89 and K92 samples.  
There are a number of important points to consider regarding the K92 and
G89 calibrations:

1. The K92 calibration is based on the \OIINIIHa\ ratio.
K92 assumes an 
average value for \NIIHa\ of 0.5 because \Ha\ and \NII\ are blended for 
many galaxies in the K92 sample.  Recent higher resolution spectroscopy and
theoretical photoionization models show that the average \NIIHa\ ratio 
is around $0.5\pm 0.2$ for most optical and infrared selected samples with
metallicities exceeding 0.5$\times$solar.  For metallicities below 0.5$\times$solar, 
the \NIIHa\ ratio may be as low as 0.01 \citep{Kewley01b}.  
The mean \NIIHa\ ratio for any particular sample depends on the sample 
selection criteria and on the diagnostic used to remove galaxies containing 
AGN from the sample.  For the NFGS, the \NIIHa\ ratio 
ranges between 0.03 and 0.50 (Jansen et al.\ 2000b) with a mean value of
$0.27 \pm 0.01$, significantly different from the K92 value of 0.5.

2. The K92 calibration is derived by starting with an \Ha\ SFR 
calibration from population synthesis models assuming no dust.  K92 uses the 
average \OIIHa\ ratio for their sample to convert the \Ha\ SFR calibration
into an SFR calibration based on \OII.
The average \OIIHa\ ratio used by K92 is uncorrected for reddening.  
Such reddening corrections were difficult 
to make at the time: lower resolution and lower S/N spectra limited the ability 
to obtain a reliable estimate for the Balmer decrement.
The K92 \OII\ SFR calibration therefore assumes an average reddening for
the sample.  The G89 \OII\ SFR calibration is derived in a similar manner,
but with an uncorrected average \OIIHb.  The effect of reddening is less 
severe for the \OIIHb\ ratio than for the \OIIHa\ ratio.  (Note that the 
K98 \OII\ SFR indicator requires correction for reddening at \Ha\ 
rather than at \OII\ because the \Ha\ SFR calibrations are 
calculated from stellar population synthesis models assuming no dust.)

The problem with applying the K92 or G89 \OII\ indicators to individual
galaxies or to samples of galaxies is that the reddening between 
\OII\ and \Ha\ may 
not be the same as the average reddening for either of the K92 or G89 samples.  
Indeed, \citet{Aragon03} showed that prior to reddening 
correction, there is a significant difference between the average 
\OIIHa\ ratio for the NFGS sample and for galaxies in the 
\Ha\ selected Universidad Complutense de Madrid (UCM) Survey.  

\begin{figure}
\epsscale{0.7}
\plotone{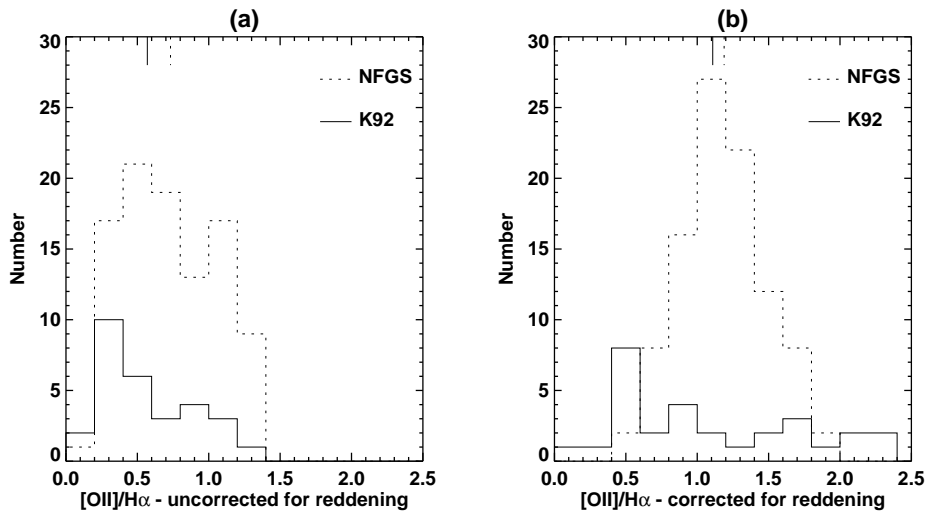}
\caption{Histograms showing the \OIIHa\ distribution for the NFGS and
K92 samples (a) prior to reddening correction, and (b) after reddening 
correction.  The vertical lines at the
top indicate the mean \OIIHa\ for each sample.  Prior to reddening
correction the means are: $0.57 \pm 0.06$ (K92)
and $0.73 \pm 0.03$ (NFGS).  After reddening correction, the means
are:  $1.1 \pm 0.1$ (K92)
and $1.2 \pm 0.03$ (NFGS).
\label{OIIHa_hist}}
\end{figure}

In Figure~(\ref{OIIHa_hist}a), we show the difference between
the \OIIHa\ ratio for the NFGS and the K92 high resolution sample, prior to reddening correction.
The mean \OIIHa\ for the K92 sample is $0.57 \pm 0.06$ compared to 
$0.73 \pm 0.03$ for the NFGS.  However, after correction for reddening, 
the \OIIHa\ ratio difference disappears (Figure~\ref{OIIHa_hist}b).  
The mean \OIIHa\ for the K92 sample
after reddening correction is $1.1 \pm 0.1$, compared to $1.2 \pm 0.3$ 
for the NFGS.  \citet{Aragon03} found a similar agreement between the 
mean \OIIHa\ for the NFGS and UCM samples after reddening correction.
These results suggest that the \OII\ SFR indicator can be 
recalibrated in a reddening-independent manner.  The SFR \OII\ indicator
would then be applicable in an unbiased way to a wider range of samples.

3. Hidden in the \OII\ SFR indicator may be errors resulting from
stellar absorption underlying the \Ha\ and \Hb\ emission-lines.
Balmer absorption is difficult to measure reliably 
with lower S/N, lower resolution data.  
This problem does not affect the K98 \Ha\ SFR indicator; it assumes 
that the \Ha\ 
emission-line is corrected for underlying stellar absorption.  This problem 
does however affect the \OIIHa\ ratio derived for the K92 sample and the 
\OIIHb\ ratio for the G89 sample.  Underlying 
stellar absorption reduces the flux of the \Ha\ or \Hb\ emission-line;
without correction, the \OIIHa\ ratio is overestimated.

4. Differences in metallicity also hinder the calibration of \OII.  
K98 states that metal abundances have a relatively small effect on the 
\OII\ calibration over most of the abundance range of interest for the 
K92 galaxies: ($0.05\, Z_{\odot} \le Z \le 1\, Z_{\odot}$).  
However, \citet{Jansen01} find that there is a correlation between \OIIHa\
and the oxygen-abundance sensitive line ratio \R23.  \citet{Charlot02}
 observe a similar dependence.

In summary, the SFR(\OII) estimated using K98 for any individual galaxy may 
not provide the true SFR because of differences in the reddening, Balmer
absorption, the \NIIHa\ ratio, ionization properties and metallicity of
the galaxy compared to the average of the K92 sample.  
For an entire sample, the combination of these effects could result 
in an increase in the error (scatter) in the SFR(\OII) relation, and
possibly systematic shifts, depending on the sample selection criteria.

\begin{figure}
\epsscale{0.7}
\plotone{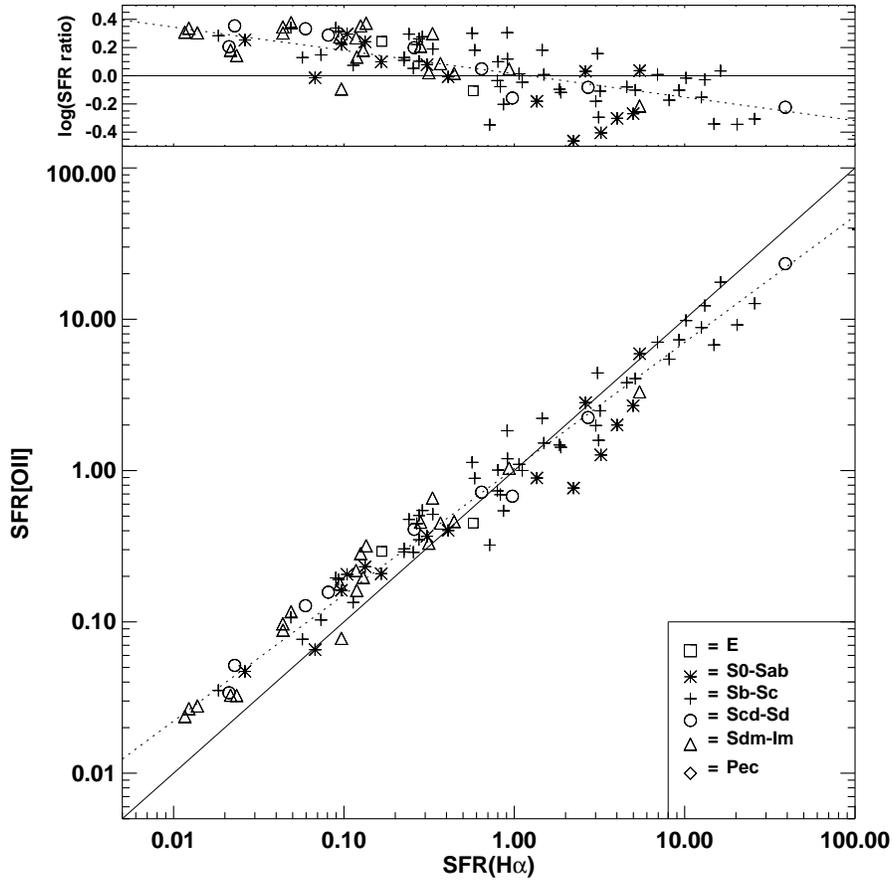}
\caption{\emph{Bottom panel}: A comparison between the \Ha\ and \OII\ SFRs 
based on the K98 calibrations 
(equations~\ref{eq_SFR_Ha} and \ref{eq_SFR_OII})
\Ha\ has been corrected for reddening using the Balmer
decrement.  We corrected the 
\OII\ flux for reddening at \Ha\  
as required by K98. 
  The legend indicates the Hubble type.  
\emph{Top panel}: The \Ha\ SFR versus
the ratio of the two SFRs from Figure (a) : SFR(\OII)$/$SFR(\Ha).
In both panels, the solid line shows
where the data would lie if both SFR indicators agreed (y=x) and
the dotted line shows the least squares fit to the data.
We assume errors of $\sim30\%$ for SFR(\Ha) and SFR(\OII), as
 in \citet{Kewley02a}.  
\label{SFR_Ha_vs_SFR_OII_K98}}
\end{figure}

In Figure~(\ref{SFR_Ha_vs_SFR_OII_K98}), we compare the SFR(\Ha) with
the SFR(\OII) derived using the K98 calibrations.  We corrected the 
\OII\ flux for reddening at \Ha\ as required by K98.  We fit a 
straight line 
to the logarithm of the SFRs using linear least-squares
minimization that includes error estimates for both variables.  We assumed
errors of $\sim$30\% as in \citet{Kewley02a}.  The resulting fit (dotted 
line in Figure~\ref{SFR_Ha_vs_SFR_OII_K98}) has the form:

\begin{equation}
\log[{\rm SFR([OII])}] = (0.83\pm0.02) \log[{\rm SFR(H\alpha)}] + (0.01\pm0.02)
\end{equation}

Figure~\ref{SFR_Ha_vs_SFR_OII_K98} shows that the K98 SFR(\Ha) calibration
predicts a lower SFR than the K98 SFR(\OII) calibration for SFRs below
1~\Msun/yr, but a larger SFR estimate for SFRs above 1~\Msun/yr. 
The rms dispersion around the line of best-fit in 
Figure~\ref{SFR_Ha_vs_SFR_OII_K98} is 0.11 in the log.  
We will investigate the difference in slope and the relatively large 
scatter in Section~\ref{newoii}.
  
Other SFR(\OII) calibrations have been derived in a manner similar to K98.
\citet{Hippelein03} provided an \OII\ SFR based 
on extinction-corrected \OIIHa\ measurements but did not correct for 
Balmer absorption.  \citet{Rosa-Gonzalez02}, however, did correct for 
reddening and underlying Balmer absorption. \citet{Gallagher89}
and \citet{Cowie97} used the \OIIHa\ ratio to obtain the SFR(\OII) for 
different samples.  None of these methods, however, take abundance into 
account.

\section{Derivation of the New SFR(\OII) Indicator \label{newoii}}

As mentioned earlier, in the NFGS spectra, (1) the \NII\ 
and \Ha\ lines are cleanly separated, (2) reddening can be estimated 
from the  Balmer decrement, 
and (3) the stellar absorption under \Ha\ can be measured from the \Hb\ 
emission-line profile.  Furthermore,  theoretical 
strong-line abundance diagnostics now enable the reliable determination 
of abundances from a wide variety of available emission-lines 
\citep[eg.,][]{McGaugh91,Zaritsky94,Charlot02,Kewley02b}.  These diagnostics 
allow us to derive a new SFR(\OII) calibration which  includes an explicit 
correction for abundance.

\subsection{Reddening Correction \label{redden}}

\begin{figure}
\plotone{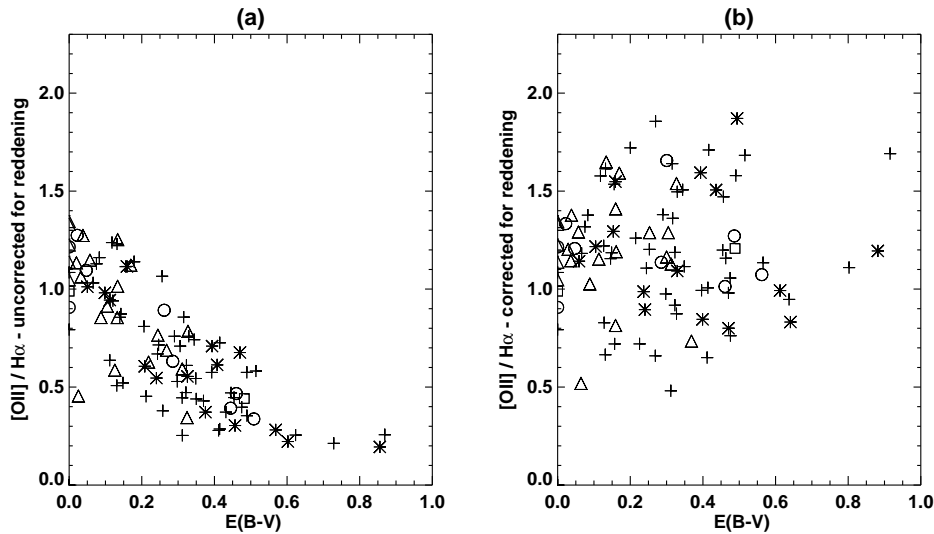}
\caption{The reddening E($B-V$) derived from the Balmer decrement 
compared with the \OIIHa\ ratio; (a) prior to reddening correction, and 
(b) after reddening correction.  All fluxes have been corrected for 
Galactic extinction.  
The estimated errors are $\sim 17$\% and $\sim23\%$ for the uncorrected and 
corrected \OIIHa\ ratios respectively. The estimated error 
in E($B-V$) is $\pm 0.04$.
Symbols are as in Figure~\ref{SFR_Ha_vs_SFR_OII_K98}.
\label{EB_V_vs_OIIHa}}
\end{figure}

In Figure~(\ref{EB_V_vs_OIIHa}a) we show the relationship between
\OIIHa\ and the reddening E($B-V$) derived from the Balmer decrement.
The \OIIHa\ ratio is uncorrected for reddening.
The Spearman Rank correlation coefficient is
-0.80. The two-sided probability of obtaining a value of -0.80 by
chance is almost zero ($\sim 1.1 \times 10^{-22}$), indicating a
strong correlation between \OIIHa\ and E($B-V$).

As described in Section~\ref{sample}, we correct the \OIIHa\ ratio for 
reddening using the CCM reddening curve.   
Figure~(\ref{EB_V_vs_OIIHa}b) 
shows the relationship 
between \OIIHa\ and E($B-V$) after reddening correction.
The Spearman Rank correlation correlation coefficient is -0.02.  The 
probability of obtaining a value of -0.02 by chance is 83\%,
indicating that reddening correction by the CCM method removes the
dependence of \OIIHa\ on E($B-V$). 

The mean \OIIHa\ for the NFGS sample after reddening correction is $1.2 \pm 0.3$.
If we apply this factor to equation~(\ref{eq_SFR_Ha}), we obtain:

\begin{equation}
{\rm SFR([OII]) (M_{\odot} yr^{-1}) = (6.58 \pm 1.65) \times 10^{-42}\, L([OII])\, (ergs\,s^{-1})}
\label{eq_SFR_OII_ratiocorr}
\end{equation}

\noindent
where L(\OII) should be corrected for reddening at \OII.  Note that
in contrast to the K98 SFR(\OII) equation~(\ref{eq_SFR_OII}), our 
equation~(\ref{eq_SFR_OII_ratiocorr}) makes no assumption about the
typical reddening.  

\begin{figure}
\plotone{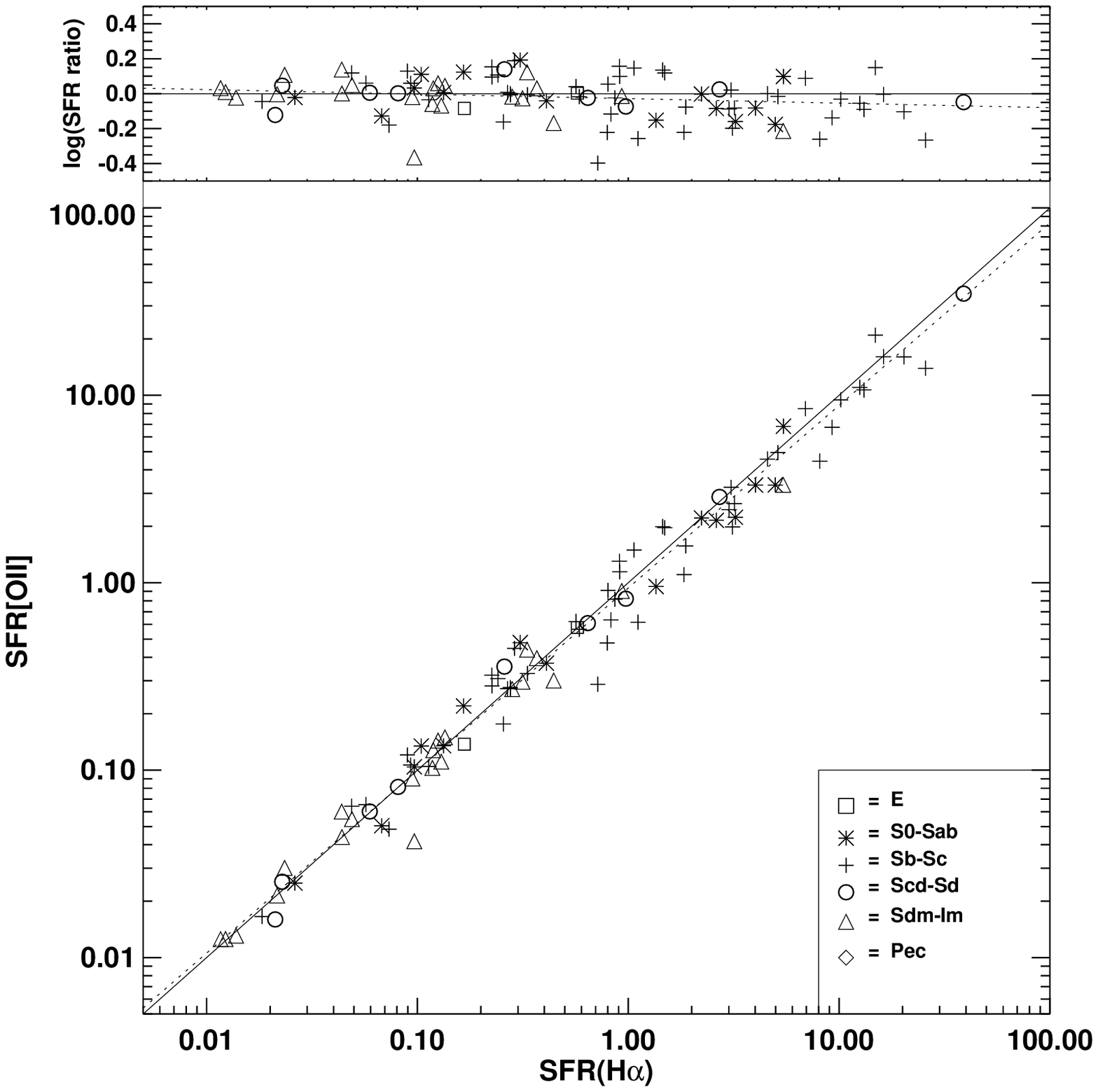}
\caption{{\it Bottom panel:} A comparison between the SFR(\Ha) and our 
SFR(\OII).  SFR(\OII) has been derived using our new \OIIHa\ 
ratio as given in equation~(\ref{eq_SFR_OII_ratiocorr}).
Both \OII\ and \Ha\ have been corrected for reddening using
the Balmer decrement.  The \OII\ luminosity has been corrected for
reddening at \OII, as required by equation~(\ref{eq_SFR_OII_ratiocorr}).  
The solid line is y=x.  The dotted line shows the least squares  fit
to the data.  The legend
indicates the Hubble type.  
{\it Top Panel:} The \Ha\ SFR versus
the ratio of the two SFRs from the bottom panel: SFR(\OII)$/$SFR(\Ha).
The solid line shows
where the data would lie if both SFR indicators agreed.  
Estimated errors are $\sim30\%$ for SFR(\Ha) and $\sim35\%$ for SFR(\OII).
The dotted curve corresponds to the least squares fit.
\label{SFR_Ha_vs_SFR_OII_ratiocorr}}
\end{figure}

Figure~\ref{SFR_Ha_vs_SFR_OII_ratiocorr} 
shows the SFR(\OII) derived with equation~(\ref{eq_SFR_OII_ratiocorr}) 
compared to SFR(\Ha).  The line of best-fit to the data is now:

\begin{equation}
\log[{\rm SFR([OII])}] = (0.97\pm0.02) \log[{\rm SFR(H\alpha)}] + (-0.03\pm0.02)
\end{equation}

\noindent
Clearly the difference in reddening between \OII\ and \Ha\ for the NFGS 
compared to the K92 sample is responsible for the departure of the slope from unity in
Figure~\ref{SFR_Ha_vs_SFR_OII_K98}.  The rms scatter of the data about the 
fit is slightly smaller, 0.08 dex.
In Sections~\ref{ionpar}~and~\ref{abund}, we explore the remaining sources 
of scatter between SFR(\OII) and SFR(\Ha).

\subsection{\OIIHa\ and the Ionization Parameter \label{ionpar}}

There is some concern (eg K98) that the SFR(\OII) calibration is less
precise than the SFR(\Ha) calibration because \OII\ is sensitive
to the excitation state of the gas.  For example, the excitation of \OII\ is 
 particularly high in the diffuse gas in starburst galaxies 
\citep[e.g.,][]{Martin97}.  The ionization parameter is a 
measure of the excitation of the gas, and is defined as

\begin{equation}
q=\frac{S_{{\rm H}^{0}}}{n}  \label{1}
\end{equation}
where $S_{{\rm H}^{0}}$ is the ionizing photon flux per unit 
area, and $n$
is the local number density of hydrogen atoms. The ionization parameter $q$, 
can be physically
interpreted as the maximum velocity of an ionization front 
driven by the local radiation field.  Dividing by the speed of light 
gives the more
commonly used dimensionless ionization parameter; ${\cal U}\equiv q/c.$ 

\begin{figure}
\epsscale{0.5}
\plotone{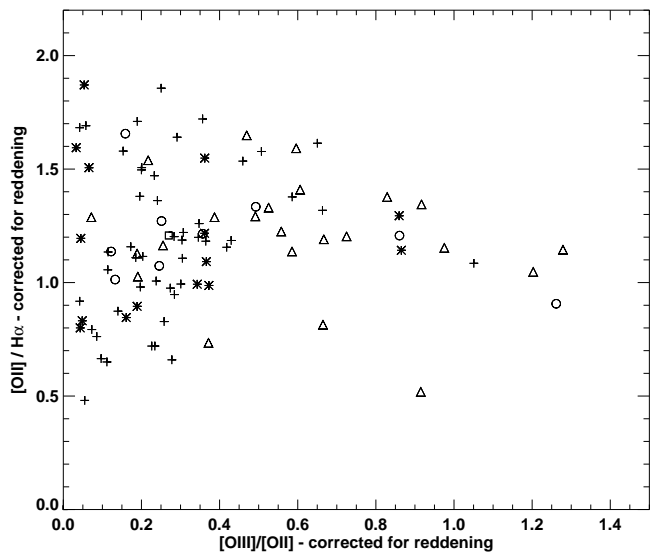}
\caption{The ionization-parameter sensitive ratio 
[\ion{O}{3}]~$\lambda\lambda\,4959,5007/$[\ion{O}{2}]~$\lambda\lambda\,3726,3729$ versus the [\ion{O}{2}]~$\lambda\lambda\,3726,3729$/H$\alpha$ ratio.  
There is no statistically significant dependence of \OIIHa\ on the
ionization parameter.  Estimated errors in \OIIHa\ and \OIIIOII\ 
are $\sim23$\%.
Symbols are as in Figure~\ref{SFR_Ha_vs_SFR_OII_K98}.
\label{OIIIOII_vs_OIIHa}}
\end{figure}

If the \OII\ SFR calibration depends upon the ionization parameter, 
then we expect to observe this dependence in the \OIIHa\ ratio.  
In Figure~\ref{OIIIOII_vs_OIIHa}, we plot the ionization-parameter sensitive
ratio \OIIIOII\ versus \OIIHa.  
The Spearman Rank correlation coefficient is 0.11.  The 
two-sided probability of obtaining this value by chance is 30\%,
indicating that there is no statistically significant dependence of
\OIIHa\ on the ionization parameter as traced by \OIIIOII.  Our local 
sample covers a small range in ionization parameter
($1\times10^{7}$ - $3\times10^{7}$ cm/s; Dopita et al. 2001).  
The majority of the oxygen
emission in the NFGS is likely to result from the O$^{+}$ 
species: the \OI\ emission is weak or
immeasurable, and the majority of the NFGS (72\%) 
have \OIIIOII\ ratios less than 0.5, with a mean(\OIIIOII)$\sim0.38\pm0.03$.

Note that the NFGS is representative of galaxies in the local 
universe.  Samples which have not been objectively selected, and perhaps
those at high redshifts could exhibit different ionization properties from
those observed in the NFGS.   In particular, active starburst galaxies
and blue compact galaxies may contain radiation fields 
characterized by larger ionization parameters than observed for the NFGS
\citep[e.g.,][]{Martin97}.  For example, the K92 local ``high resolution'' 
sample (excluding the galaxies known to contain AGN) has a mean 
\OIIIOII\ ratio of $0.5 \pm 0.2$, compared to the mean NFGS \OIIIOII\ 
ratio of $0.38 \pm 0.03$.  The larger K92 mean \OIIIOII\ is caused by 
one galaxy in the K92 sample that has an extremely large \OIIIOII\ ratio 
of 4.57, a factor of 8 times larger than any other galaxy in the K92 
sample.  If this outlying galaxy is removed, the average
\OIIIOII\ ratio is much lower: \OIIIOII$=0.28 \pm 0.03$.  
Clearly the 
\OIIIOII\ ratios can vary significantly from galaxy to galaxy.  In addition,
 the range in ionization parameter and metallicity covered by a particular 
sample may be influenced by the sample selection criterion. 
 \citet{Lilly03} observed the \OIIIOII\ ratio
for 66 galaxies with redshifts $0.47<z<0.92$.  They find that the \OIIIOII\
ratio (uncorrected for reddening) in these galaxies is $\sim0.1 - 1.3$.
Lilly et al. note that if an average reddening of \EBV$\sim 0.2$ is applied
to the CFRS sample, then the range in \OIIIOII\ for the CFRS sample is 
similar to the range observed in the NFGS.  However, the \OIIIOII\ ratio
is much higher in the five Lyman break galaxies ($z\sim3$) observed by 
\citep{Pettini01} with \OII\ and \OIII\ line fluxes.  These Lyman 
break galaxies all have $1<$\OIIIOII$<10$. 
The dominant process affecting the \OIIHa\ ratio in such galaxies may be 
ionization parameter rather than abundance because relatively large amounts 
of oxygen may exist in \OIII~$\lambda\lambda$4959,5007 and higher levels 
of excitation.

\subsection{\OIIHa\ and the Oxygen Abundance \label{abund}}

We now investigate the dependence of the \OIIHa\ ratio on the
oxygen abundance.  
The oxygen abundance is ideally measured directly from the ionic 
abundances obtained from a determination of the electron temperature 
of the galaxy.  An appropriate correction factor accounts for 
the unseen stages of ionization.  The electron temperature can be 
determined from the ratio of the auroral line \O4363\ to a lower excitation 
line such as \OIIIl. In practice, however, \O4363\ is very weak 
in metal-poor galaxies, and is not observed in higher
metallicity galaxies.  In addition, \O4363\ may be subject to systematic
errors when using global spectra: \citet{Kobulnicky99}
found that for low metallicity galaxies, the \O4363\ diagnostic
systematically underestimates the global oxygen abundance.

Without a reliable electron temperature diagnostic, global abundance 
determinations are dependent on the measurement of the ratios of 
strong emission-lines.  The most commonly-used ratio is \ratioR23 
(otherwise known as \R23), first proposed by \citet{Pagel79}. 

The logic for the use of this ratio is that it provides an estimate of 
the total cooling due to oxygen.  
Because oxygen is one of 
the principle nebular coolants, the \R23\ ratio should be sensitive to 
the oxygen abundance.  
One of the caveats, however,  with using \R23\ is that it
is double valued: at low abundance, the intensity of the oxygen
lines scales roughly with the chemical abundance; at high
abundance the nebular cooling becomes dominated by the infrared fine
structure lines and the electron temperature 
(and therefore \R23)  decreases.
Detailed theoretical model fits to \HII\ regions 
have been used to develop a number of calibrations of \R23\ with abundance
\citep[see e.g.,][for a review]{Kewley02b}.   
Calibrations of \R23 produce oxygen abundances which are generally
comparable in accuracy to direct methods relying on the measurement of
nebular temperature, at least in the cases where these direct methods are
available for comparison \citep{McGaugh91}.  

Because different abundance diagnostics can have systematic problems,
we applied four independent abundance diagnostics; 
(1) the \citet{Kewley02b} \NIIOII\ diagnostic (hereafter KD02), 
(2) the \citet{McGaugh91} \R23\ diagnostic (hereafter M91), 
(3) the \citet{Zaritsky94} \R23\ diagnostic 
(hereafter Z94), and 
(4) the \citet{Charlot01} ``case F'' diagnostic (hereafter C01).      

The KD02 \NIIOII\ calibration is based on a combination of stellar population
synthesis and detailed photoionization models.  
The \NIIOII\ ratio is sensitive
to abundance for \OH$>8.5$ for two reasons: (1) \NII\ is 
predominantly a secondary element for \OH$>8.5$, and therefore \NII\ is
a stronger function of metallicity than \OII, (2) at high 
metallicity, the lower electron temperature decreases the number of 
collisional excitations of the \OII\ lines.    For \OH$<8.5$, \NIIOII\ is
less sensitive to abundance, and is only useful for providing an initial 
guess to a more sensitive abundance diagnostic.
There are 17/97 galaxies with \OH$<8.5$ (log(\NIIOII)$\lesssim-1.1$).  
Four galaxies have very low \NIIOII\ ratios (log(\NIIOII)$<-1.43$).    
Only galaxies with low abundances ($7.5<$\OH$<8.2$) are likely to have such 
low \NIIOII\ ratios, but the KD02 \NIIOII\ diagnostic can not provide
a more specific estimate.

The M91 calibration of \R23\ makes use of detailed \HII\ region models
based on the photoionization code CLOUDY \citep{Ferland81}.  The M91 diagnostic includes the effects of dust and variations in the ionization parameter.  
We have used the analytic expressions for the M91 models given in 
\citet{Kobulnicky99}.  An initial guess is required to determine
which branch of the M91 \R23\ curve to use.  We use the \NIIOII\ diagnostic
to provide this initial abundance estimate.

The Z94 calibration of \R23\ is an average of 
the three independent calibrations given by 
\citet{Edmunds84,Dopita86,McCall85},
with the uncertainty reflecting the difference among the three
determinations.   A solution for the ionization parameter is not explicitly
included in the Z94 calibration.  The Z94 diagnostic was calibrated
against \HII\ regions spanning the metallicity range \OH$\gtrsim8.4$.
As a result, the Z94 calibration does not reflect the fact that \R23\
is double-valued with abundance: the use of the Z94 diagnostic assumes that 
all objects have \OH$\gtrsim8.4$.  We use the \NIIOII\ ratio to provide
an initial guess of the abundance to ensure that the Z94 calibration is
not applied to the objects with \OH$<8.4$.

C01 gives a number of calibrations depending on the availability of
observations of particular spectral lines. Their calibrations are based on
a combination of stellar population synthesis and photoionization codes with a simple dust prescription, and include ratios to account for the ionization
parameter.  We use the C01 ``case F'' diagnostic which is based on the \OIIIHb\
ratio for abundance sensitivity and the \OIIOIII5007\ ratio for ionization 
parameter correction.  C01 recommends using the ``case F'' diagnostic 
when the only available emission lines are \OII\, \OIII, and \Hb.
For \OIIOIII5007$<0.8$, the C01 diagnostic uses 
both \OIIIHb\ and \OIIOIII5007, but for \OIIOIII5007$\ge0.8$, only the 
\OIIIHb\ ratio is utilized. Only one of our galaxies has 
 \OIIOIII5007$<0.8$, so the C01 ``case F'' diagnostic is based on
the \OIIIHb\ ratio for the majority of our sample.   An
\OIIOIII5007\ ratio $\ge0.8$ is not unusual: the 
majority of 
\HII\ regions in \citet{vanZee98}, \citet{Kennicutt96}, \citet{Walsh97},
and \citet{Roy97} have \OIIOIII5007~$\ge0.8$ \citep{Dopita00}.  
The C01 ``case F'' diagnostic is potentially
problematic for our sample because the \OIIIHb\ ratio is relatively 
insensitive to metallicity \citep[e.g.][]{Dopita00}.  Nevertheless, we 
include the C01 ``case F'' diagnostic because the various C01 diagnostics 
are becoming widely used.

\begin{figure}
\epsscale{0.8}
\plotone{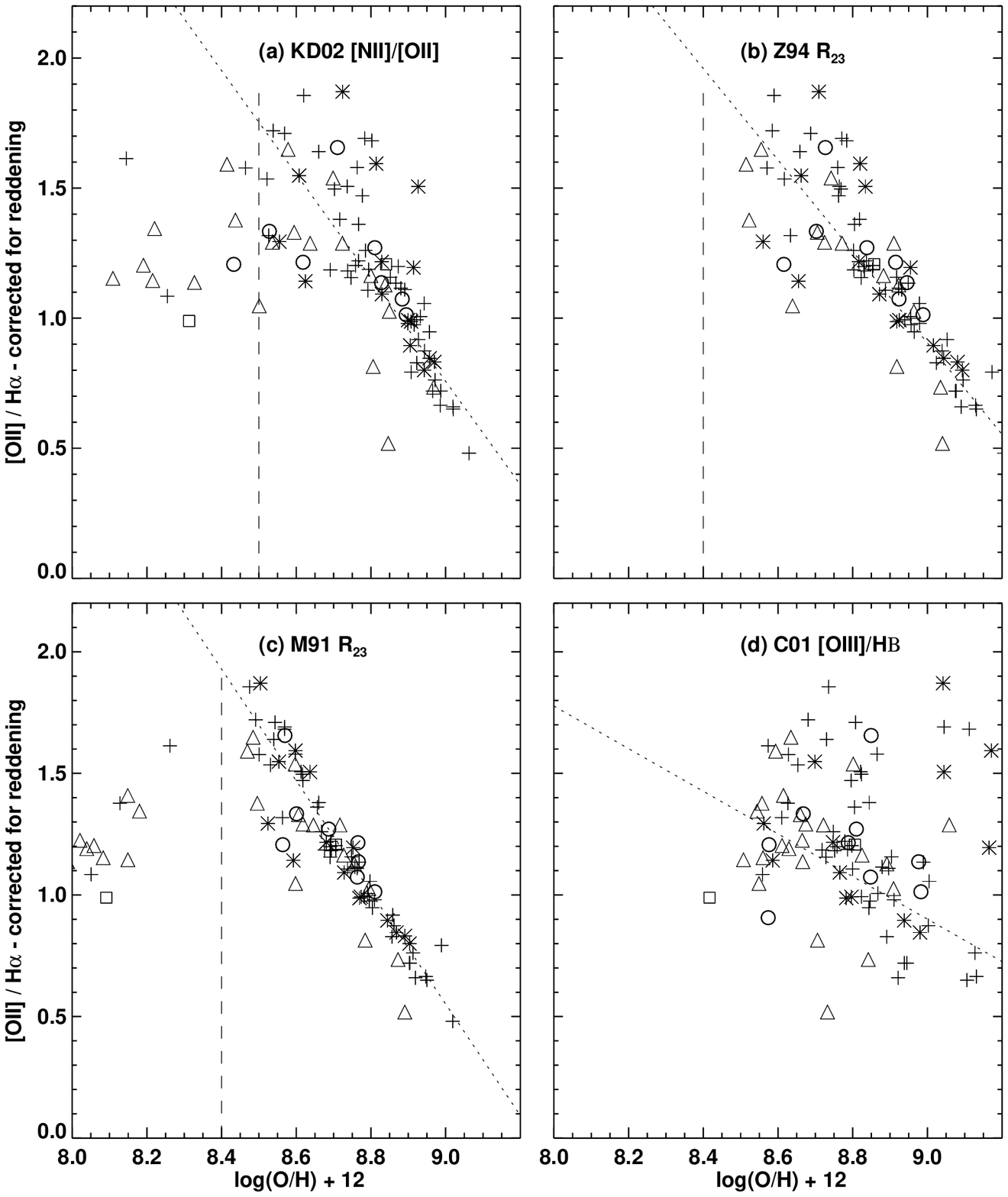}
\caption{Abundance $\log({\rm O/H})+12$ versus the \OIIHa\ ratio.  
Abundances are calculated  according to: (a) the \citet{Kewley02b} (KD02)
\NIIOII\ method, (b) the \citet{McGaugh91} (M91)
\R23\ method, (c) the \citet{Zaritsky94} (Z94)
\R23\ method, and (d) the \citet{Charlot01} (C01)
``case F'' method.  The dotted lines show the best-fit to the data.
For the KD02 diagnostic, we fit only data to the right of the dashed line
because the KD02 diagnostic becomes less sensitive to abundance
below \OH$<8.5$.   For the M91 diagnostic, we fit only data to the right of 
the dashed line because the M91 diagnostic is double-valued
with a local maximum at \OH$=8.4$.  Abundances were not calculated with
the Z94 diagnostic for  \NIIOII\ \OH$<8.4$ because Z94 is unreliable
at these abundances.  Errors in the emission-line fluxes are $\sim12$\%,
as described in \citet{Kewley02a}.  The error in the abundance estimates
is dominated by the inaccuracies of the models used to derive the abundance
diagnostics.  These errors are 
$\lesssim 0.1$~dex for the Z94, C01, and KD02 diagnostics, and $\sim 0.15$~dex
for the M91 method.  The actual error varies depending on the abundance 
range and the method used, as discussed in \citet{Kewley02b}.
\label{OIIHa_vs_abund4}}
\end{figure}

Figures~(\ref{OIIHa_vs_abund4}a-d) show the relationship between the 
metallicity in units of log(O/H)+12 and \OIIHa\ (corrected for reddening) 
for the KD02, Z94, M91 and C01 abundance diagnostics respectively. 
The absolute values of the abundances vary depending on the diagnostic 
\citep{Kewley02b}.  The mean abundances are:
\OH$\sim 8.63$ (M91), $\sim 8.73$ (KD02 \NIIOII), $
\sim 8.60$ (C01), and $\sim8.86$ (Z94).  Note that the Z94 diagnostic is an overestimate 
because Z94 abundances cannot be calculated for galaxies with 
\OH$<8.4$.  The KD02 \NIIOII\ diagnostic is also an upper limit because
of the decreasing sensitivity of \NIIOII\ with smaller abundances.
 
For metallicities \OH$\gtrsim8.4$ (M91, Z94, C01 methods) and \OH$\gtrsim8.5$ (KD02 method), 
we fit a least-squares line of best fit to the relationship between \OIIHa\ and
abundance (dotted line in Figure~\ref{OIIHa_vs_abund4}).  This line has the form:

\begin{equation}
{\rm [O\,II]/H}\alpha = a*[\log({\rm O/H})+12] + b  \label{eq_abund_vs_OIIHa}
\end{equation}

\noindent
where $a$ is the slope and $b$ is the y-intercept.  Table~\ref{coeffs} gives
the slope, y-intercept, and rms for each of the four abundance diagnostics.
Ideally, all diagnostics should produce the same estimate 
for the oxygen abundance for each galaxy.  Unfortunately, abundance
diagnostics are subject to systematic errors resulting 
from either the modeling, or the data used to calibrate the 
diagnostic \citep[see][for a review]{Kewley02b}.  These errors are 
particularly significant for the \R23\
and \OIIIHb\ diagnostics: \R23\ and \OIIIHb\ are double valued with 
abundance and are strongly influenced by the ionization parameter.
Because of these errors, the observed relationship between 
\OIIHa\ and  abundance is influenced by the shape of the 
model curves used to calibrate the diagnostics.
The shape of the model curves demonstrate the 
theoretical temperature sensitivity of \OII\ with increasing abundance.
Because the linear relations are model-dependent,
it is rather remarkable that for \OH$>8.5$, the 
KD02 \NIIOII, M91 \R23, and Z94 \R23 diagnostics 
have the same slope and y-intercept to within the errors. 
Table~\ref{coeffs} lists the Spearman Rank 
coefficients and probabilities.  For the M91, Z94 and KD02 methods,
the correlation coefficient is $\sim 0.79-0.93$, with the probability of 
obtaining 
such a correlation coefficient by chance $\lesssim 10^{-16}$.  
Because the M91, Z94,
and [NII]/[OII] diagnostics are independent, we can be reasonably confident 
that the strong correlation observed between \OIIHa\ and abundance is real.
Indeed, the \R23\ ratio is a valid abundance diagnostic
for precisely this reason.  For the ionization parameter range of our 
sample, the shape of the \R23\ curves derives
from the temperature sensitivity of the \OII\ emission-line compared to
\Hb.  At high metallicities, \R23\ is
strongly sensitive to the metallicity because the \OII\ and (to a lesser
degree) \OIII\ fluxes drop dramatically with the low electron 
temperatures associated with the increasing abundance.
However, at low metallicities (\OH$\lesssim8.4$), the electron temperature 
is high and the \OII\ flux increases slowly with abundance. 
The strong relationship between \OIIHa\ and metallicity should 
also be observed for metallicities derived from non-\R23\ methods.
Figure~(\ref{OIIHa_vs_abund4}a) supports this statement:  
the \OIIHa\ ratio is strongly correlated with the oxygen abundance
derived from \NIIOII.  The Spearman-Rank correlation coefficient is -0.79 and
the probability of obtaining this value by chance is negligible 
(1.75$\times10^{-18}$).    The slope and y-intercept for the \NIIOII-derived
abundances are within the error 
range for the other three diagnostics. 

The C01 \OIIIHb\ diagnostic, however, shows a considerably
larger scatter, with an rms of 0.24 about the best-fit line. 
KD02 showed that the C01 ``case a'' (\NIISII) diagnostic also exhibits a
larger scatter compared to the M91, Z94, or KD02 theoretical methods 
(including, but not limited to \R23).  In addition
to placing most galaxies at abundances \OH$>8.5$, the C01 diagnostic also
predicts that 16 galaxies have very low global abundances (\OH$<8.0$).  
The C01 \OIIIHb\
diagnostic appears to introduce a strong systematic effect in the
abundance estimates.  We will analyze this issue for the C01 diagnostic in 
 Section~\ref{models}.

\subsection{SFR(\OII) and Oxygen Abundance Correction \label{SFR_abundcorr}}

In this section, we apply an abundance
correction to our \OII\ SFR calibration.
SFR(\OII) is normally calibrated using an assumed \OIIHa\ ratio.  
This ratio is not independent of abundance.
We have shown that the actual \OIIHa\ ratio varies considerably for 
the NFGS sample, and that this variation is strongly 
correlated with the oxygen abundance.  For \OH$\gtrsim8.5$, the 
use of any particular \OIIHa\ 
ratio {\it automatically} implies a metallicity which 
may or may not be appropriate for the sample being studied.

Ideally, one should use the 
\OIIHa\ ratio for each galaxy to derive an SFR(\OII) diagnostic.  
However, if \Ha\ were available, it would be used as an SFR 
diagnostic rather than \OII.   
For redshifts $z>0.4$, it is theoretically 
possible to use H$\beta$ as
a SFR diagnostic through the SFR(\Ha) calibration.  In practice,
H$\beta$\ is often contaminated by an unknown amount of underlying stellar
absorption.  In the absence of \Ha\ or a high S/N, high resolution \Hb\, an oxygen abundance estimate can 
be used as a tracer of the \OIIHa\ ratio. 
To obtain the SFR(\OII), we start with the SFR(\Ha) calibration derived by K98:

\begin{equation}
{\rm SFR(H\alpha) (M_{\odot} yr^{-1}) = 7.9 \times 10^{-42}\, L(H\alpha)\, (ergs\,s^{-1})}. \label{SFRHa}
\end{equation}

\noindent
Substituting equation~(\ref{eq_abund_vs_OIIHa}) into equation~(\ref{SFRHa})
with \OIIHa$={\rm L(H\alpha)/L([OII])}$ gives

\begin{equation}
{\rm SFR([OII]) (M_\odot yr^{-1})} = \frac{7.9 \times 10^{-42}\,{\rm L([OII])\,(ergs\,s^{-1})}}{a(\log({\rm O/H})+12) + b}   \label{eq_SFR_abund_corr}
\end{equation}

\noindent
The SFR(\OII) spans four orders of magnitude and is therefore particularly 
sensitive to the values of $a, b$ and $\log({\rm O/H})+12$.  Care should
be taken to use $a$, $b$, and $\log({\rm O/H})+12$ derived from
the same abundance diagnostic (Table~\ref{coeffs}).  This process assumes
(1) that the relationship between \OIIHa\ and metallicity is linear, and
(2) that the abundance diagnostic being applied is reliable. 
Both of these assumptions are only valid for metallicities 
$\log({\rm O/H})+12\gtrsim8.5$ where the \OIIHa\ emission decreases 
with increasing abundance.  The \OII\ flux is not a strong function 
of electron temperature at low metallicities because the nebular cooling
is dominated by hydrogen free-free emission.  

\begin{figure}
\plotone{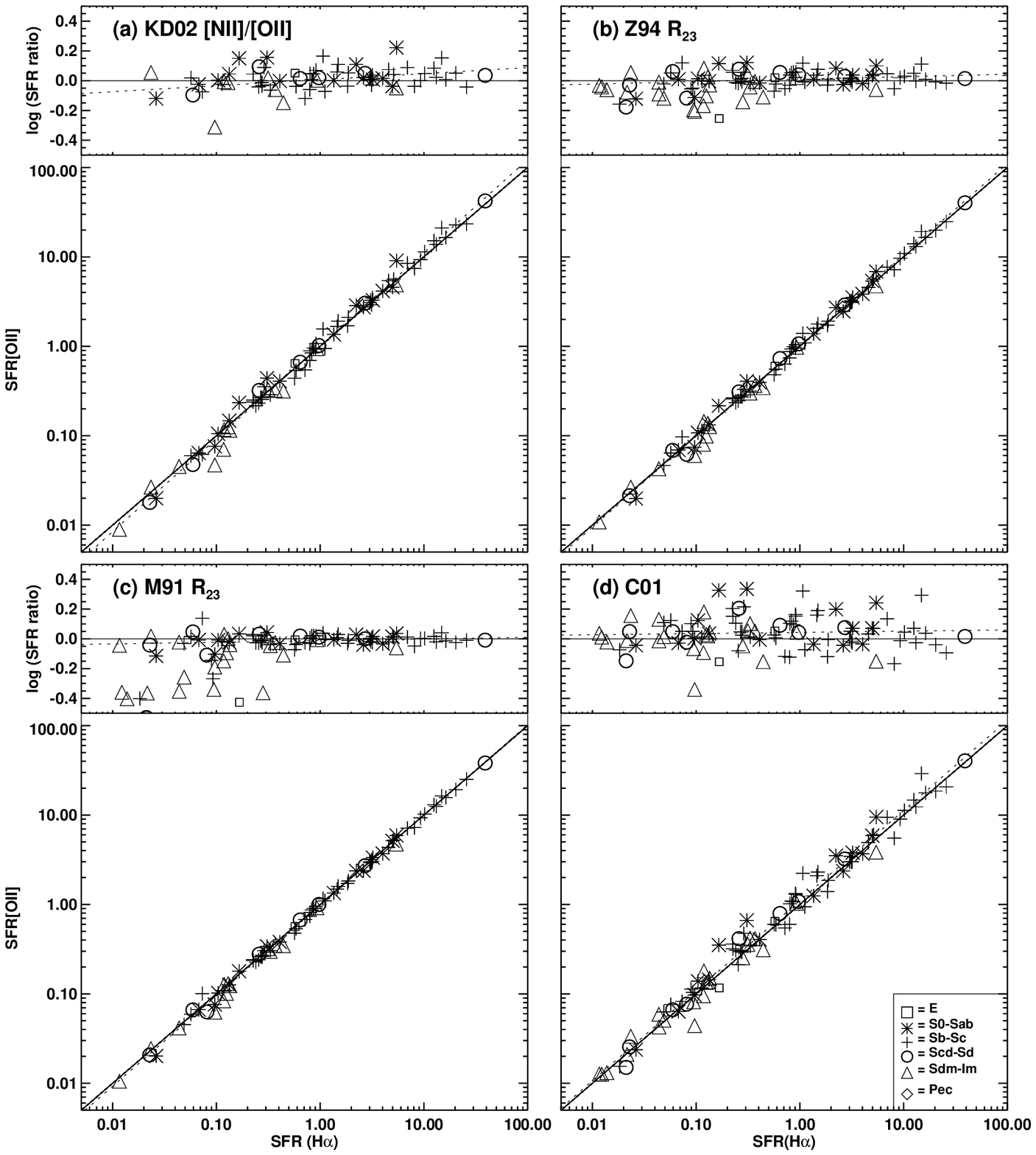}
\caption{\emph{Bottom panels}: SFR(\Ha) versus SFR(\OII) for the four
abundance diagnostic methods.  
The SFR(\OII) ratio is
corrected for reddening and abundance according to 
equation~(\ref{eq_SFR_abund_corr_adopt}).  Abundances are 
calculated using: (a) the \citet[][; KD02]{Kewley02b} \NIIOII\ method, 
(b) \citet[][; Z94]{Zaritsky94}, (c)
\citet[][; M91]{McGaugh91}, and (d) the \citet[][; C01]{Charlot01}
``case F'' method.  
\emph{Top panels}: The  K98 SFR(\Ha) versus the logarithm of the ratio of 
SFR(\OII) and SFR(\Ha) from the bottom panels.  
In each panel, the dotted line shows the 
least squares fit to the data.  Estimated errors are $\sim30\%$ for 
SFR(\Ha) and $\sim35\%$ for SFR(\OII), as in \citet{Kewley02a}.  
\label{SFR_Ha_vs_SFR_OII_4abund}}
\end{figure}

Figure~\ref{SFR_Ha_vs_SFR_OII_4abund} 
shows the
relationship between the K98 \Ha\ SFR and the SFR(\OII) derived
from our new calibration (equation~\ref{eq_SFR_abund_corr}) for each of 
the abundance diagnostics.
In each plot, a dotted line indicates the best fit to the data.  
Table~\ref{SFR_coeffs} gives the slope, y-intercept, and rms for each fit.  
For comparison, Table~\ref{SFR_coeffs} also lists the slope, y-intercept, 
and rms for the K98 SFR(\OII) and SFR(\Ha) plot in Figure~\ref{SFR_Ha_vs_SFR_OII_K98}. 
After correction for oxygen abundance, in all four cases the 
line of best fit to the data has a slope of $\sim 1$ and a y-intercept 
of $\sim 0$ within the errors, 
indicating that the abundance correction does not introduce 
a systematic offset.  For the KD02, M91 and Z94 diagnostics, the 
rms scatter decreases significantly after correction for oxygen abundance 
(0.03-0.05 versus 0.08-0.11).  

\citet{Cardiel03} also observed a decreased scatter
after metallicity correction in a small sample of 7 galaxies with redshifts of
$z\sim 0.4$ and $z\sim0.8$.  Cardiel et al. applied a 
metallicity correction to \OII\ based on \R23\ and found 
excellent agreement between SFR(\OII) and SFR(\Ha).   This result 
gives us confidence that our abundance-corrected SFR(\OII) calibration will 
be applicable to more distant samples than the NFGS.  Indeed, 
we derive our SFR(\OII) calibration only from the strong 
\OIIHa-metallicity correlation.  In theory, this correlation is a result
of the temperature sensitivity of \OII\ relative to \Ha\ and, therefore,
should not be sensitive to redshift.   In practice, however, the 
situation is more complicated.  The abundance diagnostics are based 
on theoretical models calibrated against nearby \HII\ regions
or galaxies.  It is unclear whether the model assumptions apply at
high-$z$.  Model assumptions which may differ at high-$z$ include (but are
not limited to) the gas geometry, dust geometry, density, and the initial
mass function. 

The large scatter (Figure~\ref{SFR_Ha_vs_SFR_OII_4abund}d) for the 
C01 ``case F'' (\OIIIHb) abundance 
diagnostic propagates into the SFR(\OII) calibration based on the C01 
constants and abundance estimate.  We therefore do not recommend the use
of C01 case F to derive an abundance-corrected \OII\ star formation rate if
large scatter is a concern.
As we have seen, the M91, Z94 and KD02 abundance diagnostic methods (and
associated $a$ and $b$ constants) give almost identical relations 
(within the errors) between SFR(\Ha) and SFR(\OII) with a very small scatter 
(0.03-0.05~dex).    
The fact that Z94, M91, and KD02 (\NIIOII) are independent of one another and 
still produce identical relations (within the errors) 
supports the use of  these to correct the NFGS \OII\ SFRs for oxygen 
abundances between \OH$\sim 8.5-9.2$.   

\begin{figure}
\plotone{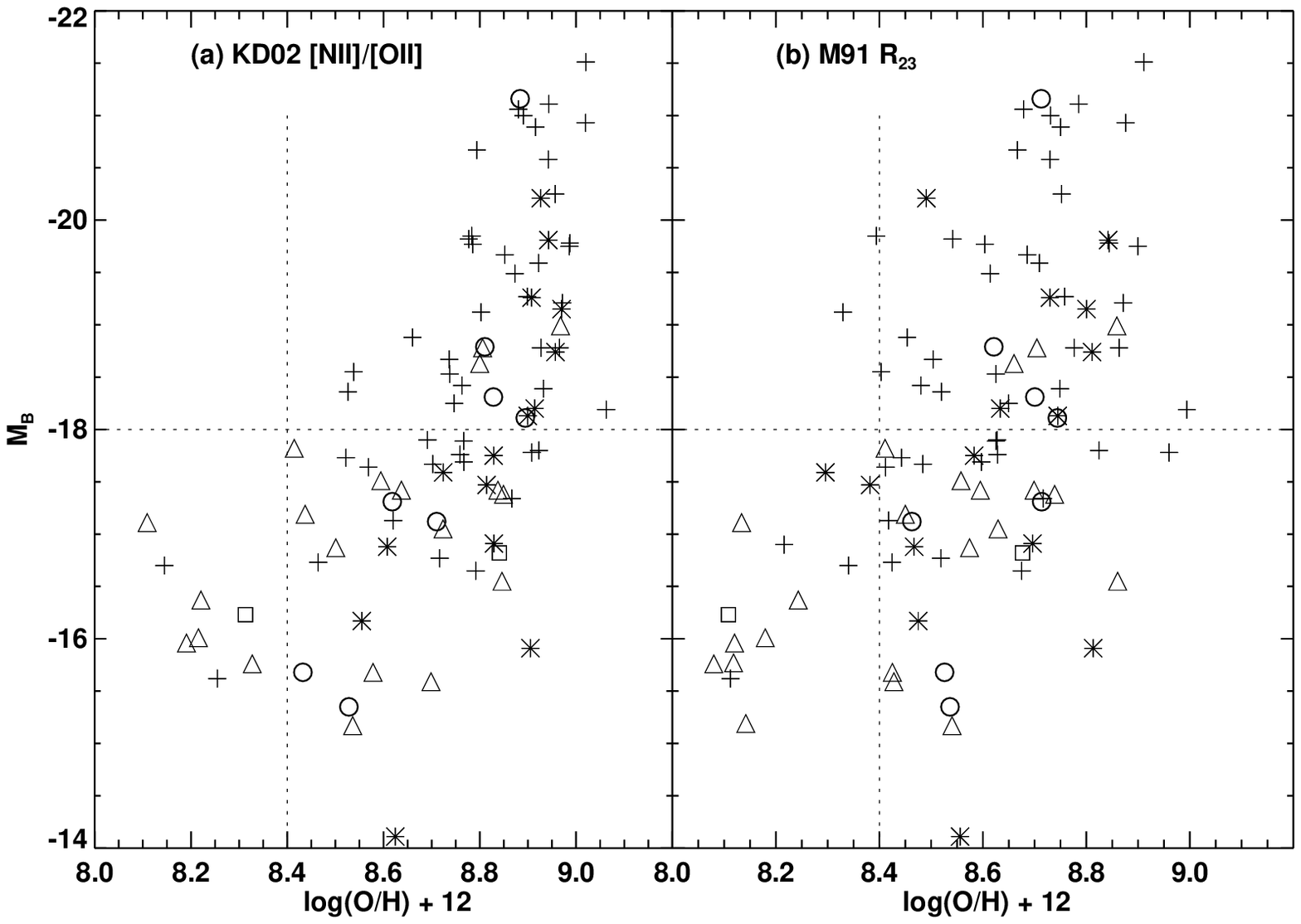}
\caption{comparison between the metallicity in units of \OH\ and the blue
absolute magnitude.  Abundances are 
calculated using: (a) the \citet[][; KD02]{Kewley02b} \NIIOII\ method, 
and (b) the \citet[][; M91]{McGaugh91} \R23\ method.
The error in the blue absolute magnitude is 0.013~mag on average \citep{Jansen00a}.
The error in the abundance estimates is 
$\lesssim 0.1$~dex for the Z94, C01, and KD02 diagnostics, and $\sim 0.15$~dex
for the M91 method.  The actual error varies depending on the abundance 
range and the method used, as discussed in \citet{Kewley02b}.
\label{MB_vs_abund}}
\end{figure}

The drawback to using \R23\ diagnostics is that they are double-valued
with abundance.
The M91 diagnostic requires an initial guess of the oxygen abundance to 
determine which branch of the \R23\ curve to use.  The Z94 calibration is 
only valid for the upper metallicity branch (\OH$>8.4$).  Unfortunately, 
the \OII, \OIII, and \Hb\ lines alone are not sufficient to determine
which branch of the \R23\ curve to use.    For the NFGS sample, we use
the \NIIOII\ line ratio to resolve this problem.   
In local galaxies, 
the luminosity-metallicity (L-Z) correlation may help to break the 
degeneracy.
For example, objects more luminous than ${\rm M_{B}} \simeq -18$ generally
have metallicities greater than \OH$\sim8.4$ (e.g., Z94) and therefore
probably lie on the upper \R23\ branch.  Figure~\ref{MB_vs_abund} supports
this conclusion.  In Figure~\ref{MB_vs_abund}, we compare the absolute magnitude 
${\rm M_{B}}$ for the NFGS galaxies with the abundances
derived with the KD02 \NIIOII\ and the M91 \R23\ methods.  Even
though the KD02 \NIIOII\ method is less sensitive to abundance for
\OH$<8.5$, the KD02 \NIIOII\ method shows a strong correlation between
abundance and ${\rm M_{B}}$.  For abundances estimated using KD02,
all eight galaxies with \OH$<8.4$ have ${\rm M_{B}}< -18$.  For abundances
calculated using M91, 13/15 (87\%) of the galaxies with \OH$<8.4$ have 
${\rm M_{B}}< -18$.  For ${\rm M_{B}}<-18$, therefore, ${\rm M_{B}}$ is
a useful discriminator between the
two \R23\ branches in nearby galaxies and provides a crude estimate of
the abundance in the absence of alternative methods. The error in the
abundance is likely to be $\pm0.2$ in units of log(O/H)+12.  At lower
luminosities (${\rm M_{B}}>-18$), the ${\rm M_{B}}$-metallicity 
relation provides, at most, an upper limit.

It is not clear whether the same 
${\rm M_{B}}$-metallicity relationship applies for galaxies at higher 
redshifts.  The few studies of the luminosity-metallicity (L-Z) relation at larger redshifts 
appear to produce conflicting results.
 \citet{Carollo01} analysed a sample of 13 star forming galaxies between
$0.5<z<1$ and find no significant evolution in the L-Z relation out to 
$z=1$.    \citet{Lin99} examined $>$2000 late-type CNOC2 
(Canadian Network for Observational Cosmology Field Galaxy Redshift Survey) 
galaxies and found no significant luminosity evolution
between $0.12<z<0.55$.  However, results from the DEEP Groth
Strip Survey suggest that the L-Z relation does evolve from the local
relation between $z=0$ to $z=1$ \citep{Kobulnicky03}.   
At larger redshifts $z>2$, 
the L-Z relation appears to be significantly different from the 
local relation \citep{Kobulnicky00,Pettini01}.
Therefore, although potentially useful, the local 
${\rm M_{B}}$-metallicity relationship should not be applied blindly 
to non-local samples.

To conclude, the Z94 abundance estimates agree well with those obtained 
using the KD02 \NIIOII\ diagnostic.  The \NIIOII\ ratio is 
very sensitive to metallicity and is almost independent of ionization 
parameter (KD02).  Therefore, if an initial 
guess of the abundance gives \OH$\gtrsim8.4$, and the ionization parameter 
is likely to cover a small range (similar to the NFGS or local \HII\ 
regions), we recommend using the Z94 \R23\ 
diagnostic.  Using the slope and y-intercept appropriate for Z94 (Table~\ref{coeffs}) gives:

\begin{equation}
{\rm SFR([OII],Z) (M_{\odot} yr^{-1})} = \frac{7.9 \times 10^{-42}\, {\rm L([OII])\, (ergs\,s^{-1})}}{(-1.75 \pm 0.25)*[\log({\rm O/H})+12] + (16.73 \pm 2.23)}. \label{eq_SFR_abund_corr_adopt}
\end{equation}

\noindent
where $\log({\rm O/H})+12$ comes from

\begin{eqnarray}
\log({\rm O/H}) + 12 & = &   9.265-0.33\,{\rm R}_{23}-
                       0.202\,{\rm R}_{23}^{2}-
                       0.207\,{\rm R}_{23}^{3}-
                       0.333\,{\rm R}_{23}^{4}\,\, (Z94)  \label{eq_Z94}
\end{eqnarray}

\noindent
and ${\rm R}_{23}=\log{({\rm [OII]}\,\lambda 3727 + 
{\rm [OIII]}\,\lambda 5007)/{\rm H}\beta}$.  

Given the difficulty in estimating abundances with limited data,
our SFR(\OII,Z) calibration should be useful for deriving an SFR(\OII) 
calibration for samples which have a different mean abundance from the NFGS.  
Equation~(\ref{eq_SFR_OII_ratiocorr}) is based on the mean intrinsic 
\OIIHa\ of the NFGS sample.  Any SFR(\OII) calibration that is derived
from SFR(\Ha) and an \OIIHa\ ratio automatically includes an assumption
about the average abundance.  As we have seen, the remaining cause of
discrepancies in \OIIHa\ from galaxy to galaxy in the NFGS is abundance.  
If the mean abundance of a sample is not the same as for 
the NFGS (using the same abundance diagnostic), then 
equation~(\ref{eq_SFR_abund_corr_adopt}) can be used to 
calculate a new SFR(\OII) calibration based on the mean or assumed abundance
for the new sample.  This approach could be useful in cases where individual
galaxy abundances are not available, but an estimate of the sample mean 
abundance can be made.  Such estimates could be based on known abundances for
similar galaxies, or could be calculated using a subsample of galaxies for 
which abundance measurements are available \citep[e.g.,][]{Lilly03}.  A similar
process can be utilized for deriving a mean sample extinction estimate. 

If no abundance estimate can be made, we recommend using 
equation~(\ref{eq_SFR_OII_ratiocorr}) derived in Section~\ref{redden}.
This SFR indicator is most useful for large samples because, provided
the mean abundance is similar to that observed in the NFGS, the mean 
SFR(\OII) should approximate the mean SFR(\Ha), thus reducing the scatter.
Note that our 
equations~(\ref{eq_SFR_OII_ratiocorr}) and 
(\ref{eq_SFR_abund_corr_adopt}) assume
that the \OIIHa\ ratio does not depend significantly on the ionization parameter.
Investigations using 
SFR(\OII,Z) should also include a calculation of \OIIIOII\ to measure the 
dominant ionization state of oxygen. If \OIIIOII\ covers a wide range, 
and the oxygen abundance covers a relatively small range, then 
equation~(\ref{eq_SFR_OII_ratiocorr}) would be a more appropriate
SFR(\OII) calibration to use.

\section{A Theoretical Calibration of SFR(\OII) and Abundance\label{models}}

In this section, we utilize theoretical models to further investigate 
the relationship between the \OIIHa\ ratio, abundance, and the ionization 
state of the gas.   We use the stellar population synthesis models
Pegase \citep{Fioc97} and Starburst99 \citep{Leitherer99} to provide the
ionizing stellar radiation field for the photoionization code, Mappings III
\citep[eg.,][]{Sutherland93,Groves03}.  Mappings III self-consistently
calculates radiative transfer through gas in the presence of dust.  
Our models, described in 
\citet{Kewley01b,Dopita00}, have been successfully applied to
\HII\ regions \citep{Kewley02b,Dopita00} and nearby starburst galaxies 
\citep{Kewley01b,Calzetti03}.  We use the instantaneous burst models
with an ionization parameter range of $q=5\times10^6 - 8\times10^7$cm/s.  
The models cover 
metallicities of 0.05, 0.1, 0.2, 0.5, 1.0, 1.5, 2.0, and 
3.0$\times$solar, where solar metallicity is defined in \citet{Anders89}.  
The corresponding metallicities in ${\rm log(O/H)+12}$
are 7.6, 7.9, 8.2, 8.6, 8.9, 9.1, 9.2, 9.4.  
Note that for the currently favored value of solar abundance 
\citep[${\rm log(O/H)+12}\sim8.69$; ][]{Allende01}, the model metallicities become 
0.09, 0.2, 0.4, 0.9, 1.7, 2.6, 
3.5, 5.2$\times$solar.  The metallicities correspond to specific 
stellar tracks used in the population synthesis models and to the
nebular abundance of the photoionization models.  Typical metallicities
for \HII\ regions range between  $8.2<{\rm log(O/H)+12}<9.2$ 
\citep[e.g.,][]{Kewley02b}.

\begin{figure}
\epsscale{0.7}
\plotone{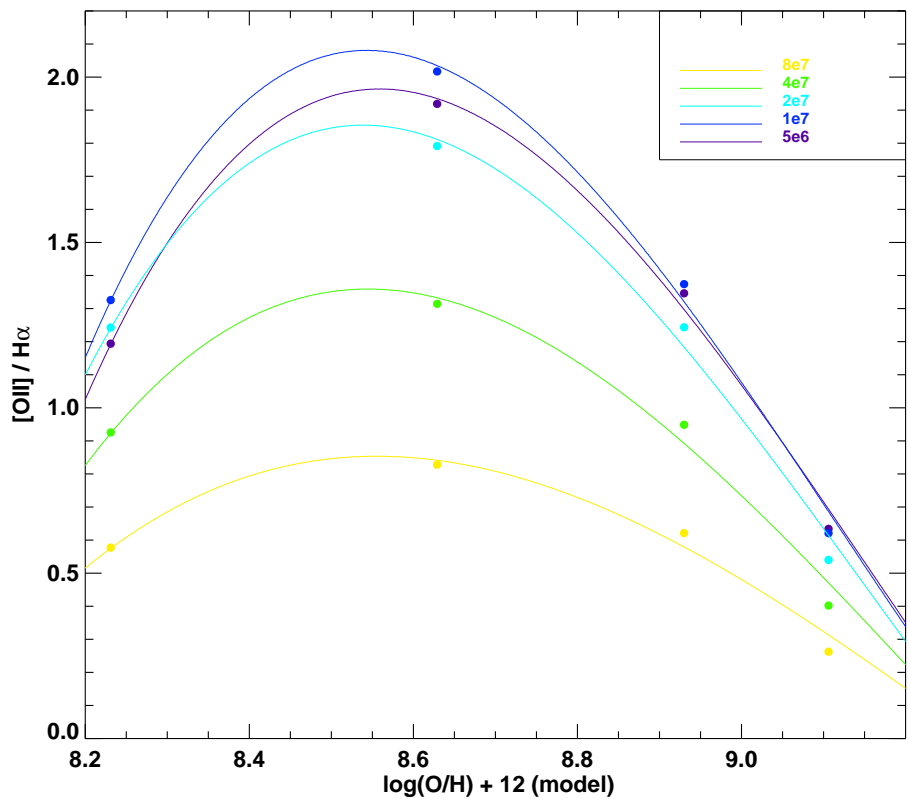}
\caption{Theoretical models of \OIIHa\ as a function of 
the model gas-phase oxygen
abundance compared to the NFGS galaxies.  The colored lines correspond
to different ionization parameters, shown in the legend in cm/s. 
The model grid errors are $\lesssim 0.1$~dex in
$\log({\rm O/H})+12$, and $\lesssim23$\% in \OIIHa.
\label{Model_OIIHa_vs_abund4}}
\end{figure}

Figure~\ref{Model_OIIHa_vs_abund4} shows the theoretical relationship
between \OIIHa\ and metallicity.  We conclude that: (a) both ionization 
parameter and metallicity affect the \OIIHa\ ratio, and (b) for a particular
sample, the relative importance of the ionization parameter compared to 
metallicity is governed by the range in abundances and ionization parameters 
spanned by the sample.  The \OIIHa\ ratio is relatively insensitive to
metallicity for $8.3\lesssim$\OH$\lesssim8.6$.  
However, for \OH$\gtrsim8.6$, \OIIHa\
becomes a strong function of metallicity, particularly for low ionization
parameters.  This behaviour reflects
the temperature sensitivity of the \OII\ emission-line.  As we have 
discussed, for temperatures typical of star-forming regions 
(10000-20000 K), the excitation energy between the two upper D levels 
for \OII\ and the lower S level is of the order of the thermal electron 
energy $kT$. The \OII\ doublet is therefore closely linked to the 
electron temperature.  At low metallicities, the electron temperature is
high, and \OII\ emission increases with metallicity.  In this regime, the
thermal cooling is dominated by hydrogen free-free emission.
 However, when the 
metallicity increases, the number of coolants in the nebula rises, thus
lowering the electron temperature.  The \OII\ emission therefore 
drops rapidly with increasing metallicity.

Figure~\ref{Model_OIIHa_vs_abund4} predicts that samples covering a small
range of metallicities will not show a correlation between \OIIHa\
and abundance because of the range of possible ionization parameters.
Samples spanning metallicities $8.3<$\OH$<8.6$ will also not show a strong
relationship between \OIIHa\ and \OH\  because at these metallicities 
\OIIHa\ is a weak function of \OH\ and \OIIHa\ is more strongly 
affected by the ionization parameter. 
Only samples with a normal range of ionization parameters 
\citep[$q=1\times10^7 - 8\times10^7$cm/s; ][]{Dopita00} and
covering a large range of metallicities exceeding \OH$\sim8.5$ will
exhibit a correlation between \OIIHa\ and metallicity.   

Note that our models do not require a specific method for 
calculating metallicities from the data.  If the metallicities are 
calculated using some reliable abundance diagnostic, our models
predict that galaxies with typical ionization parameters and 
metallicities lie along the curves in 
Figure~\ref{Model_OIIHa_vs_abund4}.  The curve for each ionization parameter
can be characterized by a third order polynomial:

\begin{equation}
\frac{\rm [OII]}{\rm H\alpha}=a+bx+cx^2+dx^3 \label{eq_Model_OIIHa_abund}
\end{equation}

\noindent
where $x=$\OH\ and the coefficients $a,b,c,d$ are displayed in 
Table~\ref{Model_coeffs}.  

\begin{figure}
\epsscale{0.9}
\plotone{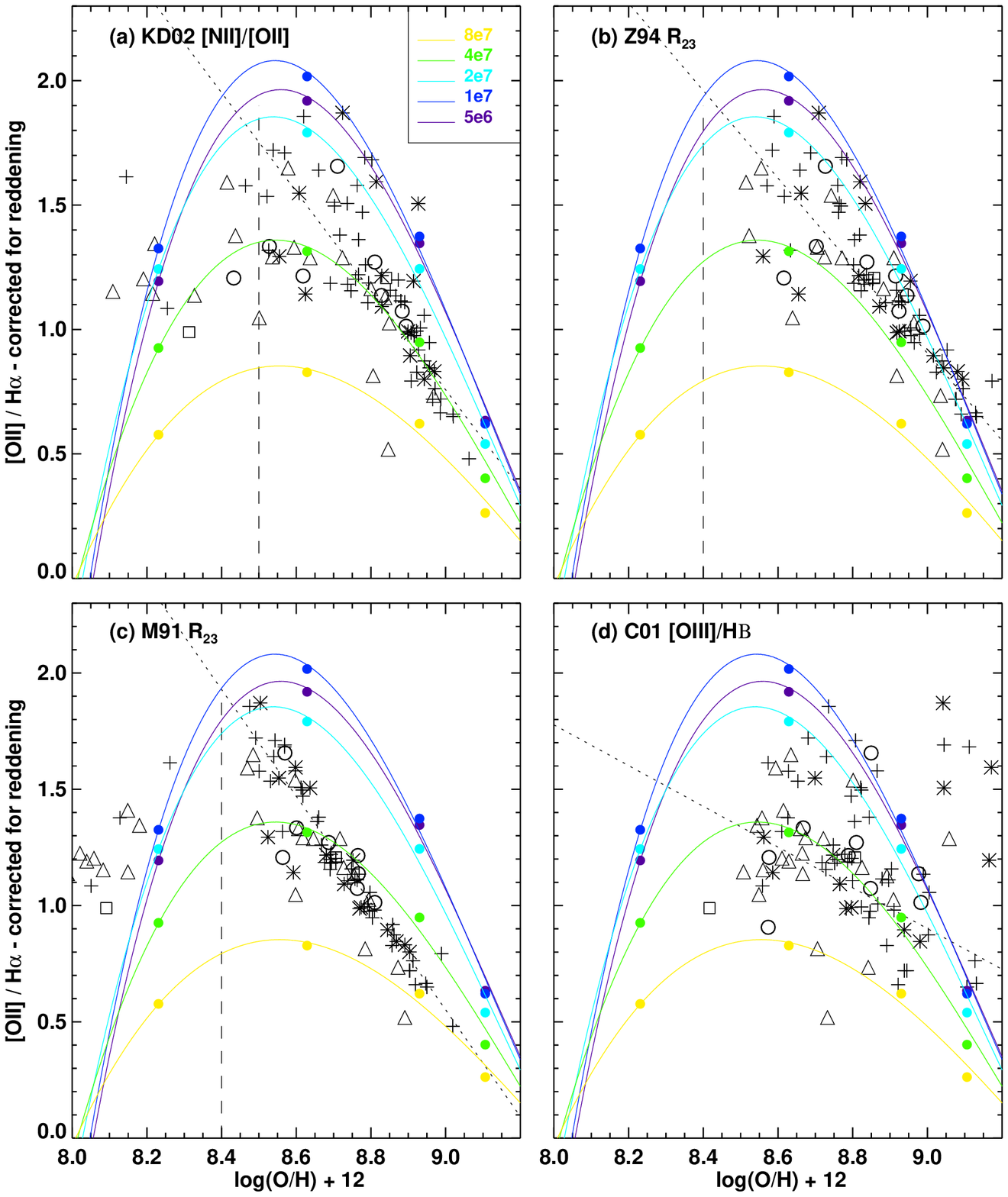}
\caption{Theoretical models of \OIIHa\ as a function of 
the model gas-phase oxygen
abundance compared to the NFGS galaxies.  
The model oxygen abundance is defined by the
metallicity of the stellar tracks in the stellar population
synthesis models, and is the same metallicity used for the surrounding 
nebular gas.  The NFGS abundances were calculated using: (a) the \citet[][; KD02]{Kewley02b} \NIIOII\ method, (b) \citet[][; Z94]{Zaritsky94}, (c)
\citet[][; M91]{McGaugh91}, and (d) the \citet[][; C01]{Charlot01}
``case F'' method.  
\label{Model_OIIHa_vs_abund4_data}}
\end{figure}

In Figure~(\ref{Model_OIIHa_vs_abund4_data}), we compare the theoretical 
models with the abundances derived using the different diagnostics.   
In Figure~(\ref{Model_OIIHa_vs_abund4_data}a) we show
the abundances derived using \NIIOII.
The ionization parameters are typically 
$2 \times 10^{7}-4 \times 10^{7}$~cm/s , 
and the data follow a similar trajectory to the models.  As we
have discussed, the oxygen 
abundances derived using the \NIIOII\ diagnostic show a larger scatter, 
particularly for oxygen abundances \OH$\lesssim8.5$.  
The models show that the \NIIOII\ and \OIIHa\ ratios 
become less sensitive to abundance as metallicity decreases, increasing
the scatter.  
The Z94 method produces abundances similar to the \NIIOII\ method.  The
data follow a similar trajectory to the models and the ionization parameters
are between typically $2 \times 10^{7}-4 \times 10^{7}$~cm/s.

The M91 method (Figure~\ref{Model_OIIHa_vs_abund4_data}c)
shows a systematic offset compared to the Z94 and KD02 methods and to
 the models.  This offset has been observed previously 
\citep{Kewley02b} and is probably a result of the different
stellar atmospheres and stellar models used to derive the ionizing 
radiation fields.  The difference in abundances estimated for the
M91 and KD02 method is $\sim0.1-0.2$ in \OH.  This variation is 
within the errors associated with the diagnostics (0.15 dex for M91 and 
0.1 for KD02).  We note that the
 \R23\ diagnostics may be minimizing the scatter 
by making a hidden assumption about the ionization parameter.  
The \R23\ ratio is sensitive to the ionization parameter and the ionization 
parameter diagnostic \OIIIOII\ is sensitive to abundance.  If the 
ionization parameter correction is not made iteratively, then a 
calibration may, in effect, favor a particular ionization parameter or 
range of ionization parameters.  This effect may contribute to the
difference between the M91 and Z94 diagnostics.  The 
Z94 abundance estimates agree well with 
the ionization-parameter independent diagnostic \NIIOII.

\begin{figure}
\epsscale{0.7}
\plotone{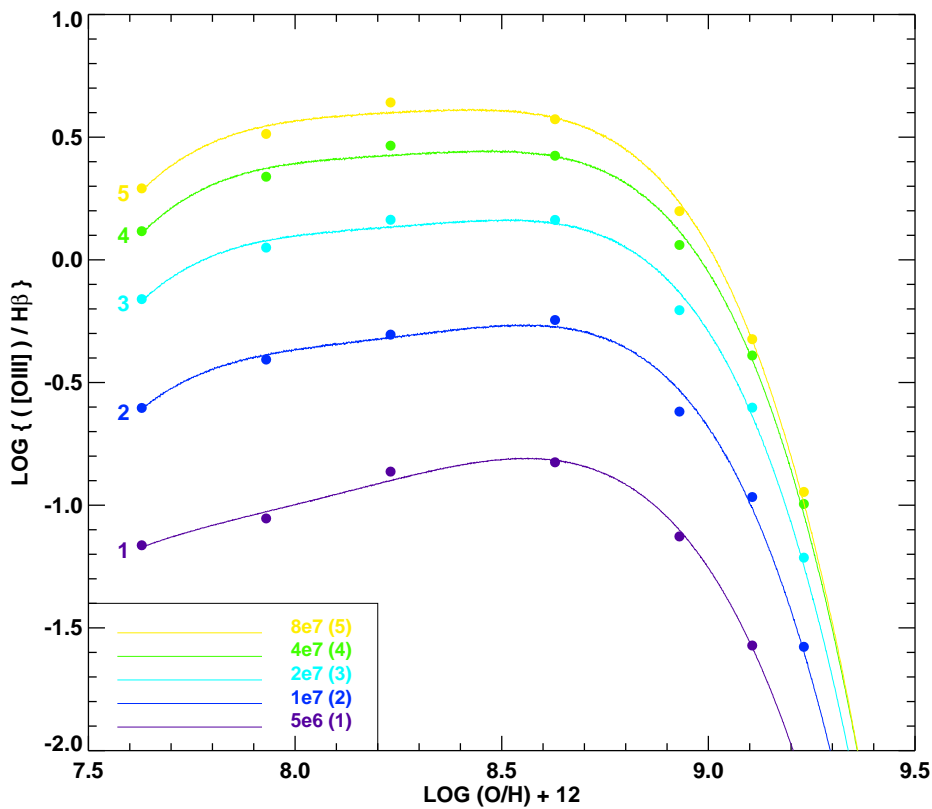}
\caption{Theoretical models of \OIIIHb\ as a function of 
the model gas-phase oxygen
abundance.  The model oxygen abundance is defined by the
metallicity of the stellar tracks in the stellar population
synthesis models, and is the same metallicity used for the surrounding 
nebular gas.  The \OIIIHb\ ratio is almost independent of metallicity
for all but the highest metallicities.  The grid errors are estimated 
to be $\lesssim 0.1$dex in $\log({\rm O/H})+12$, and up to $\sim23$\% 
in \OIIIHb.
\label{Model_OIIIHb_vs_abund}}
\end{figure}

The C01 ``case F'' diagnostic produces some \OIIHa-abundance combinations
which can not be produced using our stellar population 
synthesis+photoionization models, even with the 0.1 dex error estimates.  
As we discussed earlier, the C01 ``Case F'' diagnostic is based on \OIIIHb\ 
for the majority of galaxies in our sample.
Figure~\ref{Model_OIIIHb_vs_abund} shows the theoretical 
relationship between \OIIIHb\ and abundance.  The \OIIIHb\ ratio
is much more sensitive to the ionization parameter than metallicity for
all but the highest metallicities.  In addition, \OIIIHb\ is double valued
with abundance.  Any particular \OIIIHb\ ratio could correspond to a range
of abundance/ionization parameter combinations, and the possibility of 
obtaining an incorrect abundance estimate is high.  

We can conclude from Figures~\ref{Model_OIIHa_vs_abund4_data}a-c  
that the \OIIHa\ ratio depends on
abundance for \OH$\gtrsim8.5$, and that a linear correction for abundance 
is theoretically plausible for samples with metallicities in this range, 
providing the data do not span a larger range in ionization parameters
than is observed in local \HII\ regions.  
For the KD02, Z94, and M91 abundances, the NFGS data follow a trajectory 
with a similar slope to the models for \OH$\gtrsim8.5$.  
The model trajectory for each ionization parameter can be used to 
derive a {\it theoretical} \OII\ SFR calibration.  We begin, once again, 
with the K98 SFR(\Ha) calibration:

\begin{eqnarray}
{\rm SFR(H\alpha)\,\, (M_{\odot} yr^{-1})} &=& 7.9 \times 10^{-42}\, {\rm L(H\alpha)\, (ergs\,s^{-1})} \nonumber \\
   &=&  \frac{7.9 \times 10^{-42}\,{\rm L([OII])\, (ergs\,s^{-1})}}
             {\rm [OII]/H\alpha} \label{eq_SFR_Ha2}
\end{eqnarray}

Substituting equation~(\ref{eq_Model_OIIHa_abund}) into 
equation~(\ref{eq_SFR_Ha2}) for \OIIHa\ then yields:

\begin{equation}
{\rm SFR([OII],Z)}_{t}\,\, {\rm (M_{\odot} yr^{-1})} = \frac{7.9 \times 10^{-42}\,{\rm L([OII])\, (ergs\,s^{-1})}}{a+bx+cx^2+dx^3}
\end{equation}

\noindent
where the constants $a,b,c,d$ are given in Table~\ref{Model_coeffs} 
for each ionization parameter and $x=$\OH.  The subscript $t$ indicates that the calibration is
based on theoretical models. 
The majority of the NFGS have ionization parameters between
$q=2\times10^7 - 4\times10^7$cm/s, according to the 
KD02 \NIIOII\ and Z94 \R23\ diagnostics.   Interpolating between the
$q=2\times10^7$ and 4$\times10^7$cm/s curves gives a curve with
an approximate ionization parameter of $q=3\times10^7$cm/s:

\begin{equation}
{\rm SFR([OII],Z)_{t}\,\, (M_{\odot} yr^{-1})} = \frac{7.9 \times 10^{-42}\,{\rm L([OII])}\, ({\rm ergs\,s^{-1}})}{-1857.24+612.693x-67.0264x^2+2.43209x^3}. 
\label{eq_Grid_SFR_OII}
\end{equation}

\noindent 
This curve provides a useful theoretical description of
the behavior of \OIIHa\ with metallicity for the NFGS sample.
We emphasize that the metallicities of the models 
correspond to the metallicities in the 
stellar tracks and the modeled nebulae, and are independent
of the method used to derive \OH.  Therefore, any method can be used 
to derive \OH, as long as the method is reliable over the expected 
abundance range of the sample.  The Z94 \R23\ diagnostic can easily be 
used if the abundances exceed \OH$>8.4$.  

\begin{figure}
\epsscale{0.9}
\plotone{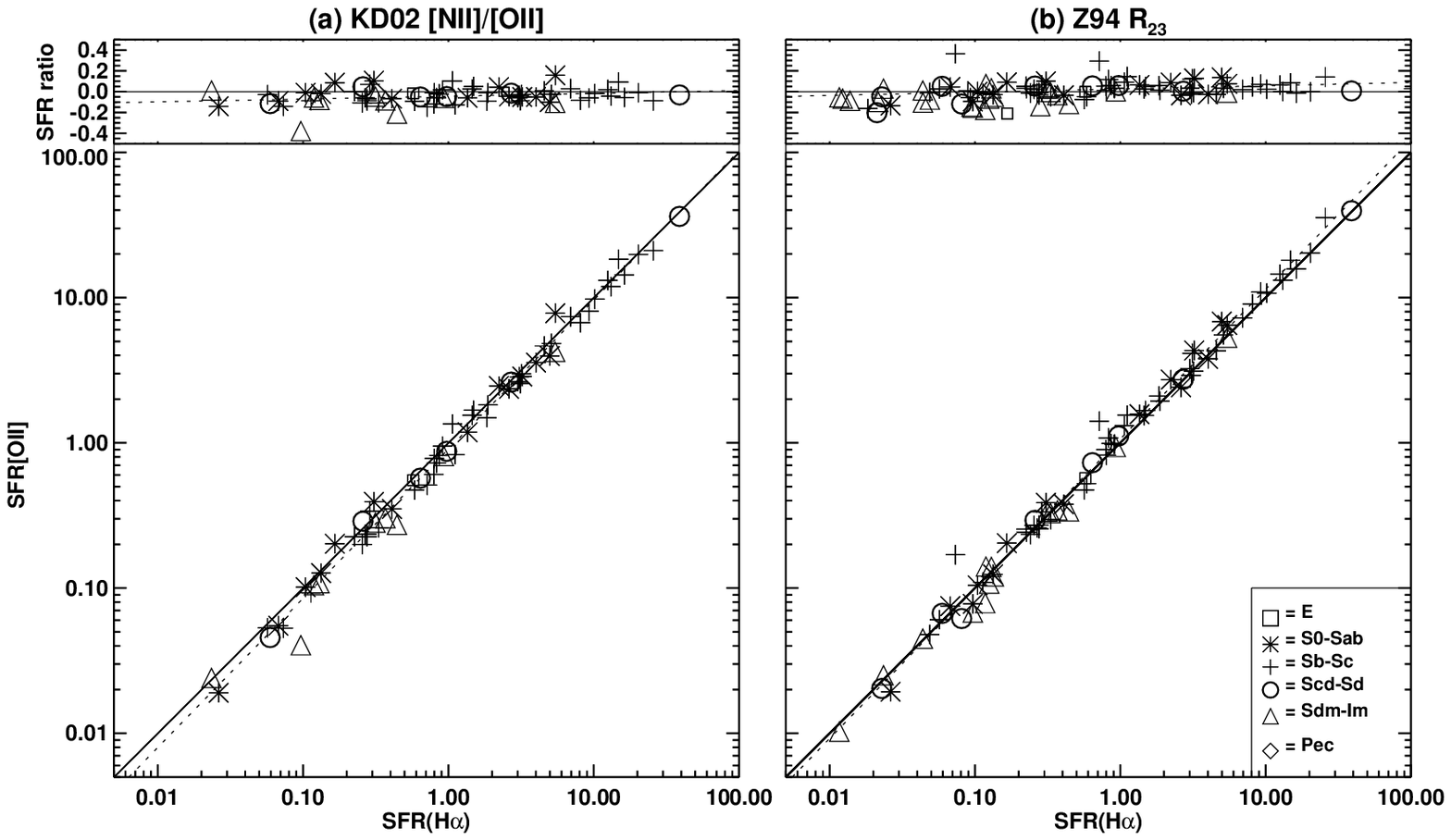}
\caption{Comparison between the SFRs derived from \Ha\ and \OII\ with
SFR(\OII,Z) calculated using our theoretical grids 
(equation~\ref{eq_Grid_SFR_OII}) for an effective 
ionization parameter $q\sim3\times 10^7$cm/s.  
Abundances were calculated using:
(a) the \citet[][; KD02]{Kewley02b} \NIIOII\ method, and 
(b) the \citet[][; Z94]{Zaritsky94} \R23\ method.
The error in SFR(\Ha) is $\sim30$\% and the error in SFR(\OII,Z) is
$\sim35$\%.
\label{Grid_SFRs_KD02_Z94}}
\end{figure}

In Figure~\ref{Grid_SFRs_KD02_Z94}, we compare the \Ha\ and \OII\ SFRs
with SFR(\OII,Z) calculated according to our theoretical models
(equation~\ref{eq_Grid_SFR_OII}) with abundances estimated by
either (a) the KD02 \NIIOII\ method  or (b) the Z94 \R23\ method. 
Table~\ref{SFR_coeffs} contains the slope, intercept, and scatter.
The slope and y-intercept are close to unity and zero respectively,
for both abundance diagnostics.  The scatter is 0.05 dex
compared to 0.08 dex using our SFR(\OII) calibration without correction 
for abundance (equation~\ref{eq_SFR_OII_ratiocorr}; 
Figure~\ref{SFR_Ha_vs_SFR_OII_ratiocorr}).  
Clearly our theoretical SFR(\OII,Z)$_{t}$ calibration is successful in
reducing the scatter observed in the SFR(\OII) estimates. 
This diagnostic will be most useful for deriving a new SFR(\OII) 
calibration for samples which have a different
mean abundance from the NFGS.

\section{The application of SFR(\OII) to high $z$ galaxies \label{high_z}}

\subsection{Reddening Determination}

\begin{figure}
\epsscale{0.5}
\plotone{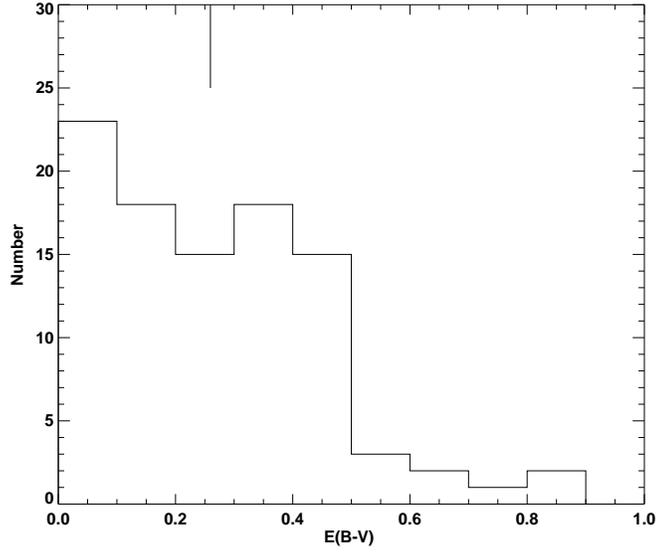}
\caption{The reddening distribution traced by E(B-V) for the NFGS.  
All fluxes have been corrected for Galactic extinction.
The vertical line at
the top shows the mean of the distribution: $0.26 \pm 0.02$.
\label{EB_V_hist}}
\end{figure}

Recently, many investigations have used \OII\ to constrain the
cosmic star formation history for redshifts  $0.4<z<1.6$ 
\citep[e.g.,][]{Hammer97,Hogg98,Rosa-Gonzalez02,Hippelein03}.  
At these redshifts, \Ha\ is usually unavailable and correction for 
reddening using the
methods outlined above is thus impossible.  
Without the Balmer decrement, many investigators 
apply an ``average'' or ``recommended'' mean attenuation of 
${\rm A_{V}}\sim 1$~mag prior to the calculation of either 
SFR(\OII) or abundance.  Assuming 
${\rm R}_{V}={\rm A}_V/{\rm E}(B-V) = 3.1$, ${\rm A_{V}}=1$ 
corresponds to
E($B-V) \sim 0.3$.  Figure~\ref{EB_V_hist} shows the distribution of 
reddening traced by E(B-V) for our sample.  The mean E(B-V) for the
galaxies in our sample (after correction for Galactic extinction) 
is $0.26\pm0.02$, consistent with the common choice of E($B-V) \sim 0.3$.
  
\begin{figure}
\epsscale{0.9}
\plotone{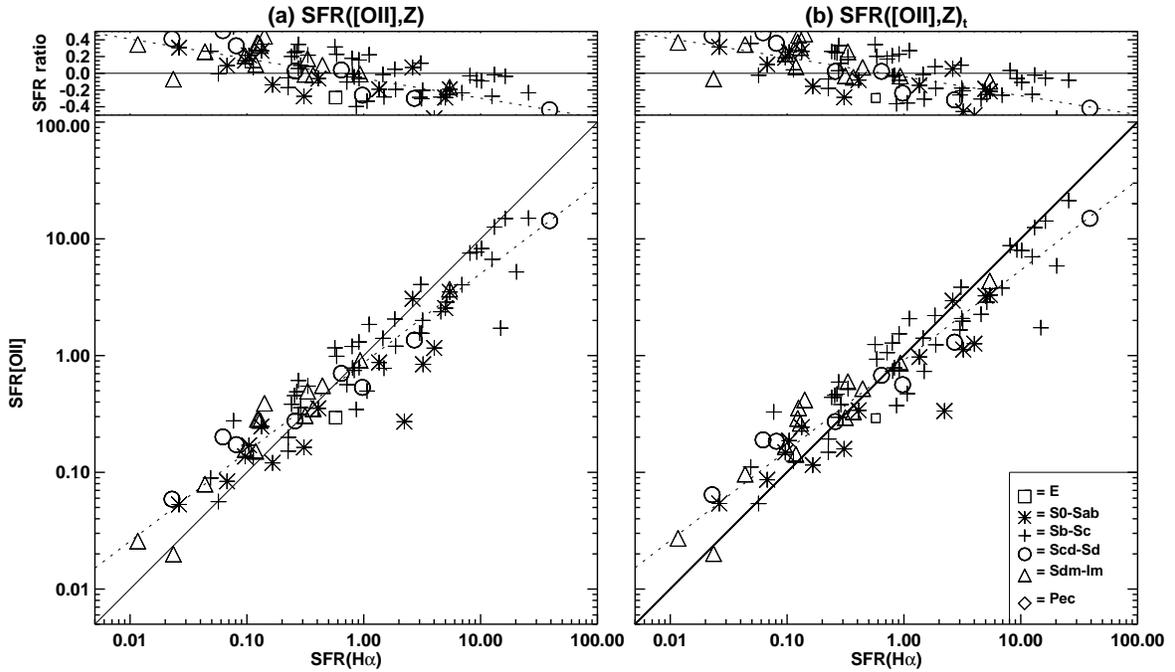}
\caption{Bottom panel: Comparison between the SFR(\Ha) and SFR(\OII) 
with the
average reddening of ${\rm A_{V}}=1$ applied to every galaxy.  The SFR(\OII\ ratio 
is corrected for reddening and abundance according to 
(a) equation~(\ref{eq_SFR_abund_corr_adopt}) and (b) 
equation~(\ref{eq_Grid_SFR_OII})
Abundances are calculated using the
Z94 \R23\ method, which is comparable to the \NIIOII\ method. 
The solid line has a slope of one; the dotted
line is the least squares best fit to the data. The estimated error in 
SFR(\Ha) is $\sim30$\% and the error in SFR(\OII,Z) is
$\sim35$\%.  The SFR(\OII,Z) error does not include the systematic error 
introduced \,by 
using ${\rm A_{V}}=1$ rather than the reddening derived from the Balmer
decrement.
Top panel: The  K98 SFR(\Ha) versus the logarithm of the ratio of 
SFR(\OII) and SFR(\Ha) from the bottom panel.  
\label{SFR_Ha_vs_SFR_OII_Av1}}
\end{figure}

If we apply an E($B-V) = 0.3$ to \R23\ and L(\OII) and use equations~(\ref{eq_SFR_abund_corr_adopt})-(\ref{eq_Z94}) or equation~(\ref{eq_Grid_SFR_OII})
to derive the SFR, the slope is
$a \sim 0.77\pm0.03$ (Figure~\ref{SFR_Ha_vs_SFR_OII_Av1}a,b).
Thus, with a single \EBV\ the SFR(\OII) is a systematic underestimate at 
high SFRs and a 
systematic overestimate at low SFRs.  This effect would be observed if 
the 
galaxies at the highest SFRs are more highly extincted than galaxies with
lower SFRs.  \citet{Wang96} showed that 
the reddening (measured using the \HaHb\ ratio) correlates with FIR luminosity
for a sample of nearby disk galaxies.  A similar effect appears in a 
sample of nuclear starburst and blue compact galaxies by \citet{Calzetti95}.
We also know from \citet{Kewley02a} that the SFR(FIR) agrees to within 10\% on average 
with SFR(\Ha), and we have shown here that there exists a 1:1 relationship between 
SFR(\Ha) and SFR(\OII) after reddening and abundance correction.  It is therefore
not surprising that we observe increasing reddening with SFR(\OII).

\clearpage
\begin{figure}[!t]
\epsscale{0.7}
\plotone{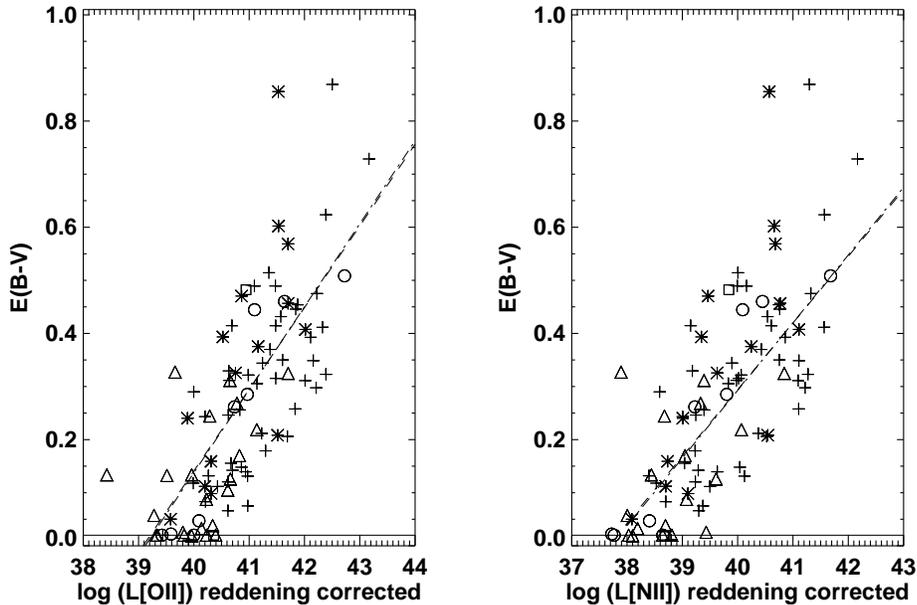}
\caption{Comparison between (a) the L(\OII) and E(B-V), and (b) the
L(\NII) and E(B-V).  L(\OII) and L(\NII) are corrected
for reddening using the Balmer decrement.  The solid horizontal line 
indicates
the detection limit on E(B-V) for our data.  Galaxies lying on or below this line have upper limits for E(B-V).  The dashed line is the least squares
fit to the data if the upper limits are fixed at 0.02.  The dot-dashed line
is the least squares fit if the upper limits are fixed at 0.00.
Estimated errors in the line luminosities are $\sim16$\% while the errors 
in E(B-V) are $\pm 0.04$.
\label{L_OII_vs_EB_V}}
\end{figure}

Figure~\ref{L_OII_vs_EB_V} shows a strong correlation of L(\OII) with E(B-V) 
for our sample with a large scatter (rms=0.13 dex). We corrected L(\OII) 
for reddening using the Balmer decrement. The Spearman Rank 
coefficient is 0.73; the probability of obtaining this coefficient by 
chance is $8\times10^{-16}$.   
We made two fits to the data to estimate the impact of the upper limits: 
one with the upper limits
on E(B-V) set at their maximum value of 0.02 (dashed line in Figure~\ref{L_OII_vs_EB_V}), and the second line with the upper limits set at zero 
(dot-dashed line).  The fits are almost identical: the slope
and y-intercepts agree to within 1\%, well within the errors.  
The best fit (including E(B-V) upper limits as 
either 0.02 or 0.00) is:

\begin{equation}
E(B-V) = (0.174\pm0.035) \log[{\rm L([OII])_{i}}] - (6.84\pm1.44) \label{eq_LOII_EB_V}
\end{equation}

\noindent
where the subscript i indicates that L(\OII) (in ergs/s) has been corrected for
reddening.  The intrinsic \NII\ luminosity is also correlated with
E(B-V) (Figure~\ref{L_OII_vs_EB_V}).  The \NII\ emission-line requires
less correction for reddening than \OII\ and provides an
independent measure of the relationship between the emission-line 
luminosity and reddening for the NFGS sample.

This relationship can be understood physically: the galaxies
with the highest rates of star formation are likely to also produce
larger quantities of dust.  Most of the dust in galaxies is probably
produced by carbon or oxygen-rich stars on the asymptotic giant branch 
\citep[see][for a review]{Mathis90}.  Supernovae may 
also be important because they insert heavy elements into the
surrounding interstellar medium.  If
we assume that the initial mass function is similar for the galaxies in our 
sample, then higher star formation rates enable more carbon and oxygen-rich
stars to reach the asymptotic giant branch, producing larger quantities of 
dust.  This scenario implies that most of the NFGS galaxies must
have been forming stars at least 1.5-2~Gyr ago because it takes 
approximately this long for the low- and intermediate-mass stars in a 
typical stellar population to evolve to the asymptotic giant branch 
\citep[e.g.,][]{Mouhcine02}.

The scatter around the fit in Figure~\ref{L_OII_vs_EB_V} is large 
(0.13 dex).  The amount of dust 
obscuring the observed optical emission from the nebular gas 
varies from galaxy to galaxy.  In addition, because we use global 
spectra, we observe the sum of the emission from the brightest \HII\ 
regions in each galaxy.  Geometry, dust composition, and 
stellar properties are all likely to have an impact on the observed 
optical emission from the brightest \HII\ regions.

Because the reddening is correlated with the intrinsic \OII\ luminosity, 
equation~(\ref{eq_LOII_EB_V}) provides a very crude estimate of the 
reddening for the NFGS.   The relationship between E(B-V) and the 
\OII\ luminosity could be different for other samples and tests
are required to determine whether equation~(\ref{eq_LOII_EB_V}) 
may be applied to non-NFGS galaxies.  In addition, for galaxies 
at high redshift E(B-V) may not be a reliable indicator of
the reddening if the dust does not conform to a foreground screen geometry
\citep{Witt00}.  However, there is some evidence that a
reddening-luminosity 
relationship exists at high redshifts, at least for rapidly star-forming 
galaxies.  \citet{Adelberger00}
found that the sum of the bolometric dust luminosity and
the 1600\AA\ luminosity (${\rm L_{bol, dust} + L_{1600}}$) is correlated with
the ratio of these two quantities (${\rm L_{bol, dust} / L_{1600}}$) for high-$z$ 
and low-$z$ galaxies alike.  The
sum ${\rm L_{bol, dust} + L_{1600}}$ provides a crude estimate of 
the star formation rate; the ratio ${\rm L_{bol, dust} / L_{1600}}$ is
a rough tracer of the dust obscuration.  If such a reddening-luminosity 
relationship holds for non-NFGS samples, then it might be 
feasible to use equation~(\ref{eq_LOII_EB_V}) to derive 
a very rough reddening estimate at distances where \Ha\ is redshifted out 
of the observable wavelength range.

The intrinsic \OII\ luminosity ${\rm L([OII])_{i}}$ is related to the observed 
\OII\ luminosity ${\rm L([OII])_{o}}$ using the 
standard equation \citep[e.g.,][]{Calzetti00}:

\begin{equation}
{\rm L([OII])_{i} (ergs\,s^{-1})} = {\rm L([OII])_{o} (ergs\,s^{-1})}
\times 10^{0.4\,k_{\rm [OII]}\,{\rm E(B-V)}} \label{eq_extinct_correct}
\end{equation}

\noindent
where $k_{\rm [OII]}= 4.771$ using the CCM reddening law.  If we substitute
equation~(\ref{eq_LOII_EB_V}) into equation~(\ref{eq_extinct_correct}) 
for E(B-V), we obtain:

\begin{equation}
{\rm L([OII])_{i}} = 3.11 \times 10^{-20} {\rm L([OII])_{o}^{1.495}} \label{eq_OII_lum}
\end{equation}

\noindent
where the intrinsic and observed luminosities are in units of ergs/s.
The estimated intrinsic luminosity from equation~(\ref{eq_OII_lum}) can 
now be used in equation~(\ref{eq_SFR_abund_corr_adopt}) along with 
\R23\ to calculate the SFR(\OII,Z) for the NFGS.  We emphasize that this
relation may be different for other samples, and should not be applied 
blindly to other galaxies.    

\begin{figure}
\epsscale{0.9}
\plotone{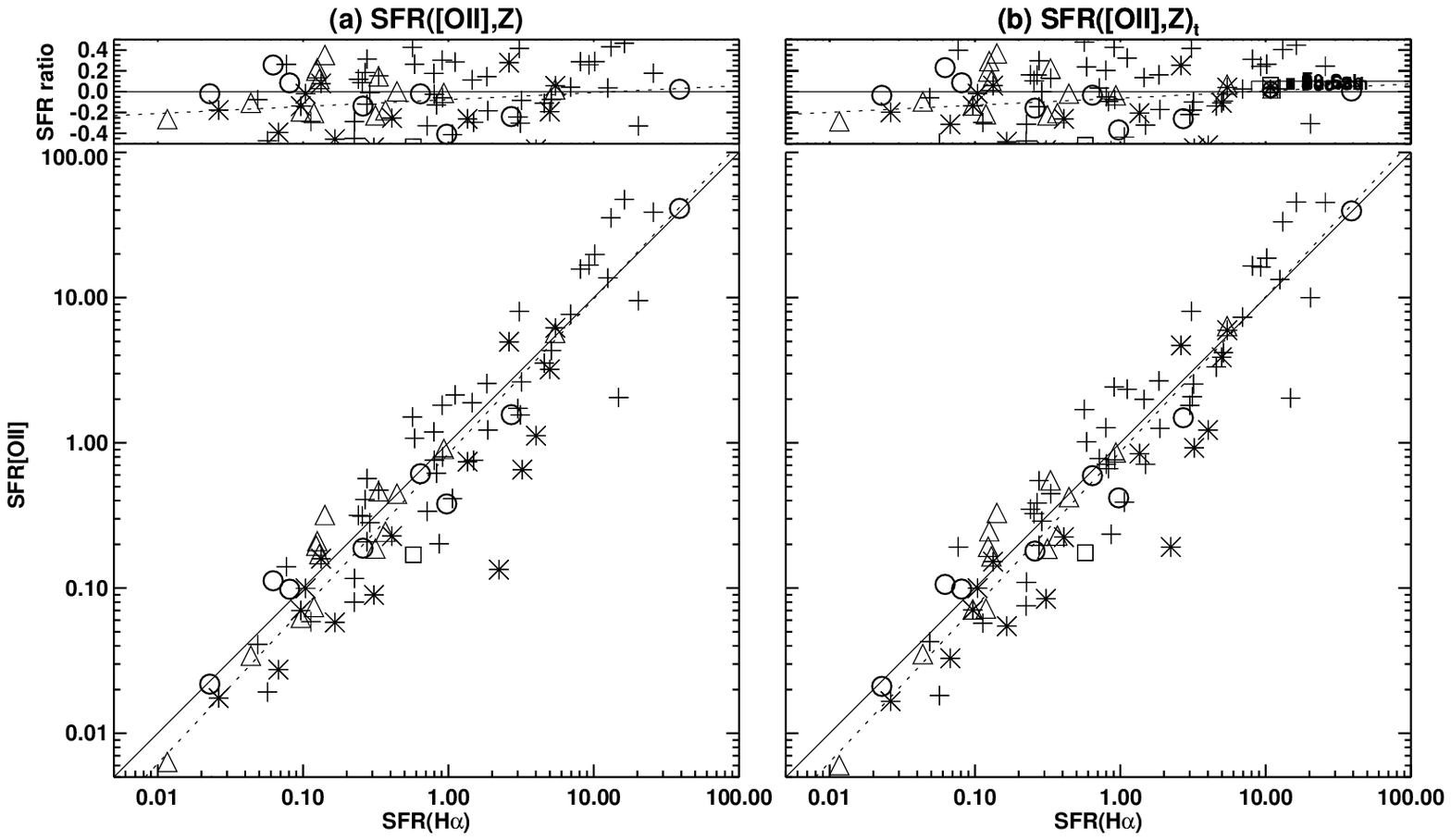}
\caption{Bottom panel: Comparison between the SFR(\Ha) and SFR(\OII,Z).  
SFR(\OII,Z) is corrected for reddening and abundance using the E(B-V) 
derived from the correlation of E(B-V) with \OII\ 
luminosity (equation~\ref{eq_LOII_EB_V}).  In panel (a), 
the abundance was calculated
using equation~(\ref{eq_SFR_abund_corr_adopt}) 
(best-fit to the \OIIHa\ and \OH curve).  In panel (b), the 
abundance was calculated using
equation~(\ref{eq_Grid_SFR_OII}) (derived from our theoretical models).
The dotted line shows the least squares fit to the data.  The error in 
SFR(\Ha) is $\sim30$\% and the error in SFR(\OII,Z) is
$\sim35$\%.  The SFR(\OII,Z) error does not include the error introduced by 
using the \OII\ luminosity rather than the reddening derived from the Balmer
decrement.
Top panel: The SFR(\Ha) versus the logarithm of the ratio of 
SFR(\OII) and SFR(\Ha) from the bottom panel. 
\label{SFR_Ha_vs_SFR_OII_LOIIcorr}}
\end{figure}

Figure~\ref{SFR_Ha_vs_SFR_OII_LOIIcorr} compares SFR(\OII,Z) with
SFR(\Ha) for the NFGS, where SFR(\OII,Z) is effectively corrected for reddening 
and abundance in the following manner:  

\begin{enumerate}
\item We estimate the intrinsic ${\rm L([OII])_{i}}$ 
luminosity from equation~(\ref{eq_OII_lum}).  
\item We estimate E(B-V) from the intrinsic ${\rm L([OII])_{i}}$ luminosity 
with equation~(\ref{eq_LOII_EB_V}).
\item We use the E(B-V) estimate to correct the \R23\ ratio for reddening.
\item We use the reddening-corrected \R23\ ratio in the Z94 diagnostic
(equation~\ref{eq_Z94}) to derive the abundance.
\item We use the abundance and intrinsic ${\rm L([OII])_{i}}$ 
luminosity in equation~(\ref{eq_SFR_abund_corr_adopt}) to derive a
reddening and abundance corrected SFR(\OII,Z) estimate.
\item We use the abundance and intrinsic ${\rm L([OII])_{i}}$ 
luminosity in equation~(\ref{eq_Grid_SFR_OII}) to derive a
reddening and abundance corrected SFR(\OII,Z)$_{t}$ estimate from our
theoretical SFR(\OII,Z)$_{t}$ calibration.
\end{enumerate}

\noindent
Note that we correct the comparison star formation rate
SFR(\Ha) for reddening using the true E(B-V) obtained from the 
Balmer decrement.  
The best-fit lines in Figure~\ref{SFR_Ha_vs_SFR_OII_LOIIcorr} have a slope
$a=1.07 \pm 0.03$ and a y-intercept close to zero 
($b=-0.06 \pm 0.03$). Clearly, if the best-fit slope
is not close to 1 (as is the case when an average A(v)=1 is assumed), 
the difference between the mean estimated SFR and the true 
SFR increases with increasing star formation rate.  Thus the bias introduced
can easily appear as a function of redshift because we observe intrinsically more
luminous galaxies at larger $z$.  The scatter in 
Figure~\ref{SFR_Ha_vs_SFR_OII_LOIIcorr} is 0.21 dex but this scatter is 
less important than the slope for studies of the star formation
history.  

\begin{figure}
\epsscale{0.7}
\plotone{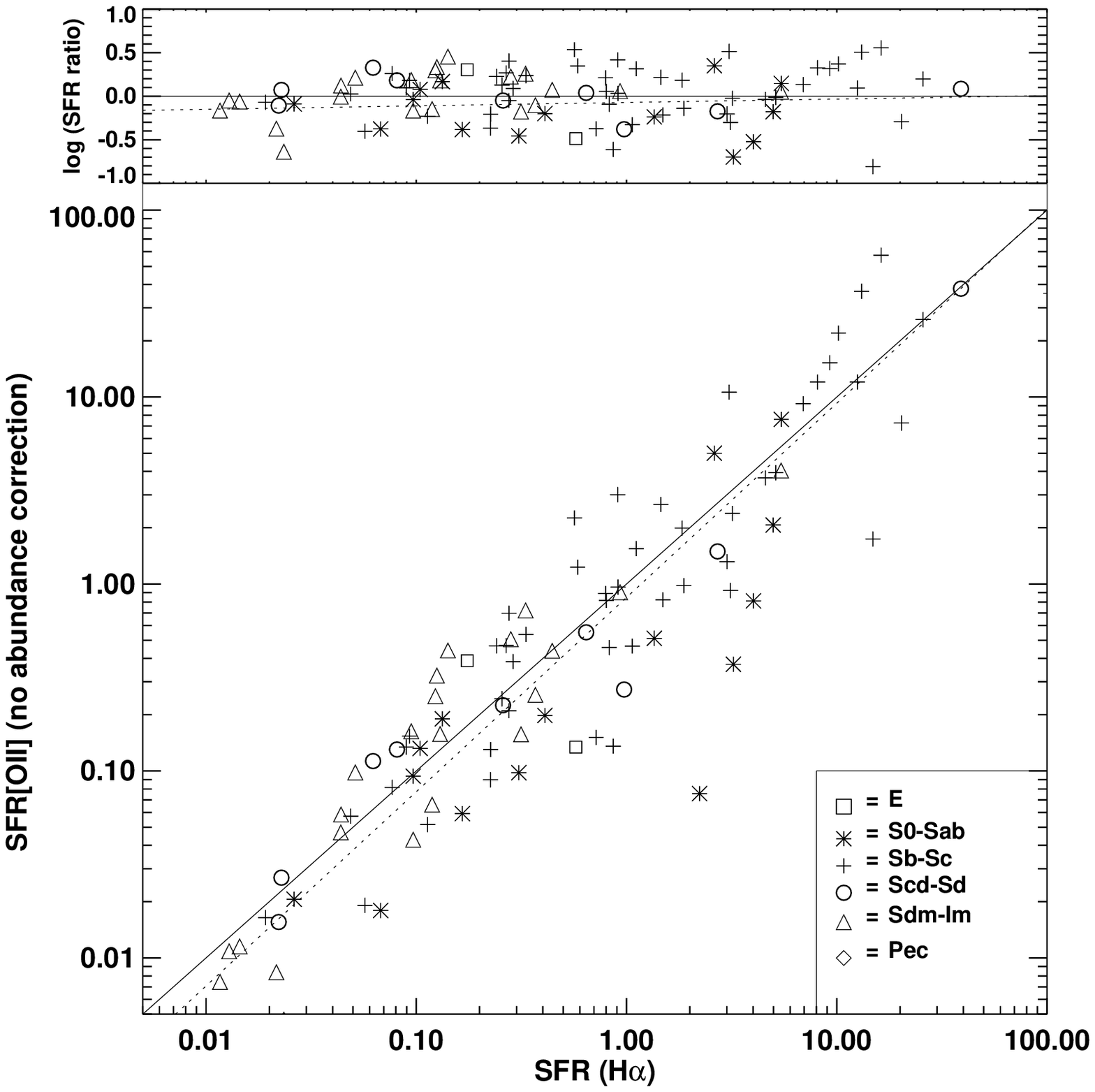}
\caption{Bottom panel: Comparison between the SFR(\Ha) and SFR(\OII).  SFR(\OII) is corrected
for reddening using the E(B-V) derived from the correlation of E(B-V) 
with \OII\ 
luminosity (equation~\ref{eq_LOII_EB_V}).  SFR(\OII) is not explicitly
corrected for abundance.  SFR(\OII) is only corrected for abundance through
the correlation of E(B-V) 
with L(\OII).  The dotted line shows the least squares fit to the data.  
The error in 
SFR(\Ha) is $\sim30$\% and the error in SFR(\OII,Z) is
$\sim35$\%.  The SFR(\OII,Z) error does not include the error introduced by 
using the \OII\ luminosity rather than the reddening derived from the Balmer
decrement.
Top panel: The SFR(\Ha) versus the logarithm of the ratio of 
SFR(\OII) and SFR(\Ha) from the bottom panel.  
\label{SFR_Ha_vs_SFR_OII_LOIIcorr_noabund}}
\end{figure}

Figure~\ref{SFR_Ha_vs_SFR_OII_LOIIcorr_noabund} shows a 
comparison of the SFR(\Ha) with the SFR(\OII) where SFR(\OII) is 
effectively corrected for reddening but only partially for oxygen abundance 
through the correlation of E(B-V) with luminosity.
We calculated the intrinsic ${\rm L([OII])_{i}}$ luminosity  using 
equation~(\ref{eq_OII_lum}), and derived the SFR(\OII) using
equation~(\ref{eq_SFR_OII_ratiocorr}).  The scatter is larger 
than if we apply an abundance correction ($\sim 0.26$ dex vs. 0.21 dex),
but again, with large samples this increased scatter should not be a problem.

As we have stressed, our equations~\ref{eq_SFR_OII_ratiocorr} and 
\ref{eq_SFR_abund_corr_adopt} were derived from an unbiased
local sample.  The excitation and abundance properties of galaxies in
other samples and of galaxies at 
higher redshifts may not be the same as those observed in the local 
galaxy population.  It may be possible to correct ${\rm L([OII])_{o}}$
for reddening using a reddening-luminosity relation such as equation~\ref{eq_OII_lum}, 
but the application of equation~\ref{eq_OII_lum} (or other such relation) 
awaits further testing.  The lack of samples with both integrated spectra 
and Balmer decrement measurements (or other reddening indicator) 
make testing equation~\ref{eq_OII_lum} difficult at present.  
We will test the reddening-luminosity relation for a large, objectively 
selected sample of galaxy pairs in a future paper 
(Kewley et al. {\it in prep}), once aperture effects have been analysed.   
For the reasons outlined above, we do not apply equation~\ref{eq_OII_lum} 
to high-$z$ galaxies.  Instead, we test whether our SFR([OII]) relations 
can be applied to galaxies at higher redshifts with the standard assumption 
of ${\rm A_{V}}=1$.   Evidence 
for a mean ${\rm A_{V}}\sim1$ for galaxies with redshifts $z<0.3$ is 
found by 
\citet{Tresse02}, \citet{Sullivan00}, \citet{Liang03} using the Balmer
decrement.  However \citet{Flores99} concludes that the global opacity 
of the ($z<1$) universe is between $A_{V}=0.5-0.85$ using radio, IR, UV, and 
optical photometry.

\subsection{Testing SFR([OII]) for galaxies $0.5<z<1.6$}

To test our SFR([OII]) relations on galaxies at larger-$z$, we 
use two samples: the sample of \citet{Hicks02} (hereafter H02)
and the sample of \citet{Tresse02} (hereafter T02).  
H02 obtained rest-frame blue spectra for 14 emission-line
galaxies representative
of the population at $0.8<z<1.6$ in the 1999 NICMOS parallel grism 
H$\alpha$ survey.  We use the seven objects in H02 with measured \OII\ and \Ha\ luminosities.  T02 obtained H$\alpha$ measurements for 30 galaxies
in the Canada-France Redshift Survey with redshifts $0.5<z<1.1$.  
Both H02 and T02 report [\ion{O}{2}]/H$\alpha$ ratios that differ 
significantly from those in the NFGS, due either to much larger 
reddening values or intrinsically lower line ratios.  Therefore, these
samples provide a critical test of the applicability of our 
[\ion{O}{2}]--SFR calibration to galaxies at higher redshifts.  

We recalculated the \OII\ and \Ha\ luminosities for the standard
cosmology ($h=0.72$; $\Omega_{m}=0.29$) and computed 
the intrinsic \OII\ and \Ha\ luminosities with the CCM reddening curve, 
assuming ${\rm A_{V}}=1$.   Table~\ref{Hicks_data} lists the relevant 
derived quantities.   The SFR(\Ha) and SFR(\OII) are from  
equations~\ref{eq_SFR_Ha} and \ref{eq_SFR_OII_ratiocorr}. 
SFR(\OII) is significantly lower than SFR(\Ha) 
(Figure~\ref{SFR_Ha_vs_SFR_OII_H02T02}a).  This discrepancy 
may result from one or a combination of (1) poor ($<5\sigma$) S/N at \OII,
(2), sky contamination at \OII, (3) a stronger dependence of \OIIHa\ on 
the ionization parameter than observed in the NFGS, (4) reddening more than 
the assumed ${\rm A_{V}}=1$, (5) reddening may not be a simple forground screen 
for the H02 and T02 galaxies, (6) the H02 and T02 samples may have a 
different mean oxygen abundance than the NFGS.
We discuss these possibilities individually below.

\subsubsection{Poor S/N}
Poor S/N at \OII\ may affect the H02 sample: only two H02 galaxies have 
$>5\sigma$ \OII\ detections.  However, the majority (28/30) of the 
T02 sample have $>5\sigma$ detections, ruling out poor 
S/N as a reason for the SFR discrepancy for the majority of galaxies in 
Figure~(\ref{SFR_Ha_vs_SFR_OII_H02T02}a).  

\subsubsection{Sky Contamination}
Sky contamination at \OII\ affects the majority (5/7 galaxies) 
of the H02 sample, but the T02
sample was specifically selected to avoid sky contamination in the optical
and near-IR.  Inspection of the T02 spectra reveals no evidence for sky
contamination.

\subsubsection{Ionization Parameter}
Ionization parameter estimates can not be made with the T02 and H02 samples 
because \OIII~$\lambda5007$ emission-line fluxes are unavailable.  However,
\OII\ and \OIII\ fluxes are available for the $0.47<z<0.92$ sample by
\citet{Lilly02,Lilly03} (hereafter LCS).  The mean \OIII/\OII\ ratio for
the LCS sample is $0.37\pm0.04$, corresponding to a mean 
ionization parameter of $\sim3.5\times 10^{7}$~cm/s using the 
\OIII/\OII\ ionization 
parameter calibration of \citet{Kewley02b}.  If the mean ionization of 
the H02 and T02 sample is similar to that observed by LCS, then the
ionization parameter cannot account for the SFR discrepancy in 
Figure~(\ref{SFR_Ha_vs_SFR_OII_H02T02}a).

\subsubsection{Reddening}
If the SFR discrepancy is a result of reddening, then the average
extinction required to bring the SFRs into agreement is 
$A_{V}\sim1.6$.   The variation in optical extinction as
a function of redshift is unknown, however the multiwavelength study by 
\citet{Flores99} suggests a mean extinction in $0<z<1$ galaxies of 
$A_{V}=0.5-0.85$, significantly lower than $A_{V}\sim1.6$.   

\subsubsection{Foreground Screen Assumption}

An alternative explanation for the difference between the
\OII\ and \Ha\ SFRs in Figure~(\ref{SFR_Ha_vs_SFR_OII_H02T02}a)
is that the dust geometry may not be a simple
screen.  \citet{Witt00} show that \EBV\ saturates at around 0.2-0.3 mag
for geometries where the dust is mixed with the gas.  In this scenario,
the true reddening may be much larger than predicted by \EBV.  We will
investigate this possibility  in a future study of an unbiased sample of
galaxy pairs and N-tuples \citep{Barton00}.

\subsubsection{Metallicity}
\begin{figure}[!t]
\epsscale{0.9}
\plotone{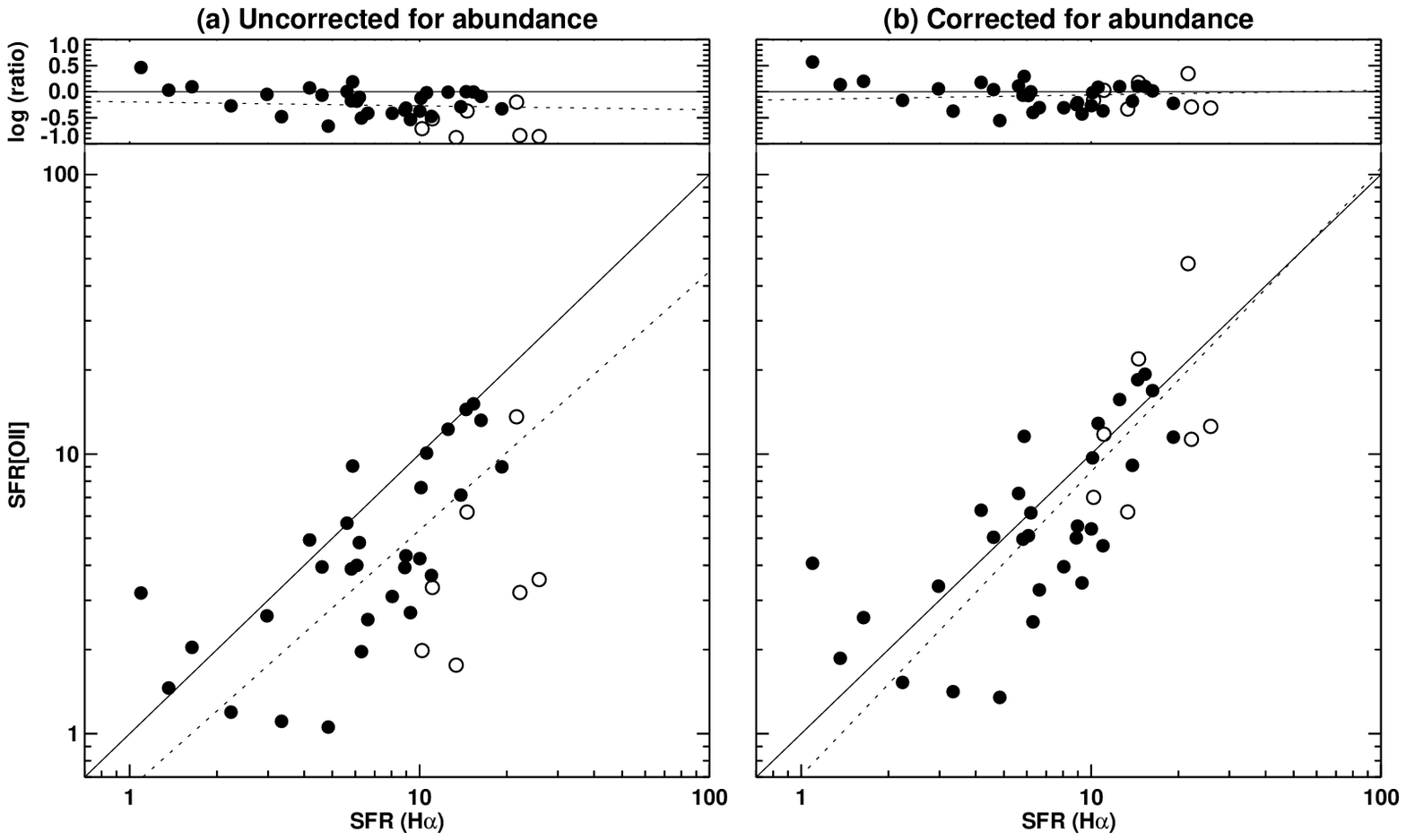}
\caption{Bottom panels: Comparison between the SFR(\Ha) and SFR(\OII) for
the T02 (Tresse et al. 2002) (filled) and the H02 (Hicks et al. 2002) 
(unfilled) samples.  
SFR(\OII) and SFR(\Ha) are corrected for reddening assuming 
${\rm A_{V}}\sim1$.  In figure (a), SFR(\OII) is calculated using
equation~\ref{eq_SFR_OII_ratiocorr} which does not include an explicit
correction for abundance, and assumes \OH$\sim8.6$.  In Figure (b),
SFR(\OII) is calculated according to equation~\ref{eq_SFR_abund_corr},
with mean M91 metallicities of 8.75 (T02) and 8.95 (H02) derived from 
the mean \OIIHa\ ratios of the two samples.
The dotted line shows the least squares fit to the T02 and H02 data.
The solid line shows equality between SFR(\Ha) and SFR(\OII).
Errors in SFR(\Ha) and SFR(\OII) are $\sim30$\% and $\sim35$\% 
respectively, not including errors systematic introduced by using ${\rm A_{V}\sim1}$
rather than the true (unknown) extinction. 
Top panel:  The SFR(\Ha) versus the logarithm of the ratio of 
SFR(\OII) and SFR(\Ha) from the bottom panel.  
\label{SFR_Ha_vs_SFR_OII_H02T02}}
\end{figure}

To investigate the effect of metallicity, we begin with the
mean intrinsic \OIIHa\ ratios: $0.34\pm0.09$
and $0.94 \pm 0.11$ for the H02 and T02 samples, respectively 
(assuming an average  ${\rm A_{V}}\sim 1$~mag).  
Using our theoretical grids 
(equation~\ref{eq_Model_OIIHa_abund}) and 
assuming that the average ionization parameter is $q=3\times10^7$~cm/s, 
the H02 and T02 \OIIHa\ ratios correspond to metallicities of 
${\rm \log(O/H)+12}\sim9.18$ and $9.06$ using our theoretical models, or
${\rm \log(O/H)+12}\sim8.95$ and $8.75$ using the M91 diagnostic
and equation~\ref{eq_abund_vs_OIIHa}.  These values are consistent with
the recent result by LCS that $>75$\% of 65 CFRS 
galaxies between $0.5 < z < 1.0$ have an oxygen abundance 
${\rm \log(O/H)+12} \sim 8.9$ using the M91 method.  The SFR(\OII) 
calibration used in Figure~\ref{SFR_Ha_vs_SFR_OII_H02T02}a assumes that the
mean abundance is the same as that measured for the NFGS.
The absolute value of the abundance is diagnostic dependent 
(compare the x-axes in 
Figures~\ref{OIIHa_vs_abund4}a - ~\ref{OIIHa_vs_abund4}d).  Because 
LCS use the M91 method, we also use this method to
derive the mean NFGS abundance for comparison: 
${\rm \log(O/H)+12} \sim 8.63$.  The difference between the 
NFGS, T02 and H02 metallicities is likely
to result from the different luminosity ranges covered by each sample.
Most high-$z$ samples consist of more intrinsically luminous
galaxies than in local samples. 
Indeed, when LCS compare the mean abundances for the CFRS sample 
with the NFGS selected over the same luminosity range, they find 
that the mean abundances are similar.  

Hydrodynamic cosmological models yield different
predictions about the stellar metallicity distribution as a 
function of redshift.  Models by \citet{Edvardsson93} and \citet{Nagamine01}
predict that the average stellar metallicity is constant up to $z\sim 2$;
the models by \citet{Rocha-Pinto00} predict that earlier stars are
more metal-poor.  Either way, it is likely that most 
samples observed at high-$z$ contain more intrinsically luminous
(and thus higher metallicity) galaxies than local samples.  

To check whether the high metallicities result from a luminosity
selection effect, we compare the mean ${\rm M_{B}}$ and metallicity
with the metallicity-luminosity relations presented in 
\citet{Kobulnicky03}.  The mean ${\rm M_{B}}$ ($\sim -21$) and 
metallicity (${\rm \log(O/H)+12}\sim8.75$) for the T02 sample are consistent
with the metallicity-luminosity relation for the 
DGSS $0.6<Z<0.8$ galaxies in Kobulnicky et al.   On the other hand, the H02  
mean ${\rm M_{B}}$ ($\sim -21$ and metallicity (${\rm \log(O/H)+12}\sim8.95$) are more similar to the upper end of the 
luminosity-metallicity relation in local samples.  This result 
may be caused
by the H02 \Ha\ sample selection.  Such a selection could potentially
bias the sample towards galaxies with intrinsically strong \Ha\ fluxes,
small \OIIHa\ ratios and high metallicities.

We obtain a metallicity-corrected SFR(\OII) with 
either the mean \OIIHa\ ratios and equation~(\ref{eq_SFR_Ha}), 
or equivalently, we use the mean metallicities derived above and equation~(\ref{eq_Grid_SFR_OII})
or (\ref{eq_SFR_abund_corr_adopt}).  Figure~(\ref{SFR_Ha_vs_SFR_OII_H02T02}b) 
compares the metallicity-corrected SFR(\OII)
estimates to SFR(\Ha).
The SFR(\OII) and SFR(\Ha) agree with a mean relative scatter of 0.17 dex.
We conclude that correcting for a mean metallicity consistent
with the mean \OIIHa\ for the T02 and H02 samples produces 
better agreement between SFR(\OII) and SFR(\Ha). The scatter is identical to
that observed for the NFGS with ${\rm A_{V}}\sim 1$~mag (Figure~14).

\newpage
\section{SFR(\OII) and the Cosmic Star Formation History \label{cosmicsfr}}

Analysis of the star formation history of the Universe depends on use of
the \OII\ emission-line as a 
SFR diagnostic in the $z \sim 0.4 - 1.6$ range
\citep[e.g.,][]{Hammer97,Hogg98,Sullivan00,Gallego02,Hippelein03,Teplitz03}.
There are many approaches to calculating the star formation 
history based on the \OII\ luminosity.  One approach
\citep{Hogg98} is to calculate the \OII\ luminosity density as a function
of redshift and then to convert this estimate directly into a star formation 
rate based on one of the SFR(\OII) conversions.  Hogg et al. use the K92 
calibration, and conclude that it (the K92 conversion) may not be directly 
applicable to galaxies at high-$z$. 

\citet{Hippelein03} take another approach.  Hippelein et al. use an
extinction-corrected \OIIHa\ flux ratio (0.9) to convert the K98 \Ha\
SFR calibration into an \OII\ SFR calibration for their sample
which covers the redshift range $0.25<z<1.2$.   
The resulting constant used to convert the intrinsic \OII\ luminosity (in
ergs/s) into a SFR is $8.8\times10^{-42}$.  If the \OIIHa\ ratio that 
Hippelein et al. derive is a good approximation to the average for 
their entire sample, then the constant $8.8\times10^{-42}$ may be used 
to predict a mean abundance.  Equation~(\ref{eq_SFR_abund_corr_adopt}) 
gives a mean abundance of $\sim 8.87$ for the Hippelein et al. 
sample with the M91 abundance diagnostic.  This abundance estimate is 
similar to  the mean abundance of the LCS 
sample ($\sim 8.9$), and is reasonable if the Hippelein et al.
sample contains relatively more luminous galaxies than in local samples.
An alternative explanation of the low \OIIHa\ ratio observed by 
Hippelein et al. is that the ionization state of the gas dominates the
\OIIHa\ ratio and that the average ionization parameter is higher
in the Hippelein et al. sample than in the NFGS.

\citet{Teplitz03} followed a similar method 
for their sample of 71 $0.46<z<1.415$ galaxies 
from the STIS Parallel Survey.  They used the \citet{Jansen01} NFGS relation
between absolute blue magnitude and \OIIHa\ to derive an \OIIHa\ ratio 
(uncorrected for reddening) for
various luminosity ranges in their sample. Teplitz et al. note that 
the \OIIHa\ ratio is highly dependent on the metallicity and reddening of
each individual galaxy.  Teplitz et al. calculate a mean uncorrected 
\OIIHa\ ratio of 0.45, lower than those observed locally 
(Figure~\ref{EB_V_vs_OIIHa}a).   
From these ratios they convert 
the \OII\ luminosity into an \Ha\ luminosity using the K98 SFR(\Ha) 
relation.  Teplitz et al. show that
the comoving star formation density estimated using \Ha\ systematically
exceeds the \OII\ estimate derived for the same redshift range.  However, 
Teplitz et al. use SFR densities which have been 
corrected for reddening in various ways.  They use the uncorrected 
\OII\ luminosity densities with the empirically derived \OIIHa\ to convert
the \OII\ luminosity densities into \Ha\ luminosity densities.  
This process is similar to the one used by K98 to derive his SFR(\OII) 
indicator: the difference in reddening at \OII\ and at \Ha\ is
taken into account, but the reddening of the \Ha\ luminosity is not.  
The derived luminosity densities should therefore be corrected 
for reddening to allow comparison of the resulting SFR densities 
with those calculated from \Ha\ surveys.  Typically, \Ha\ surveys are 
already corrected for reddening based on some assumption about the 
average attenuation.  Many studies of
SFR densities based on \Ha\ make
a correction for reddening based on an assumed ${\rm A_{V}}\sim 1$~mag. 

\begin{figure}
\epsscale{0.6}
\plotone{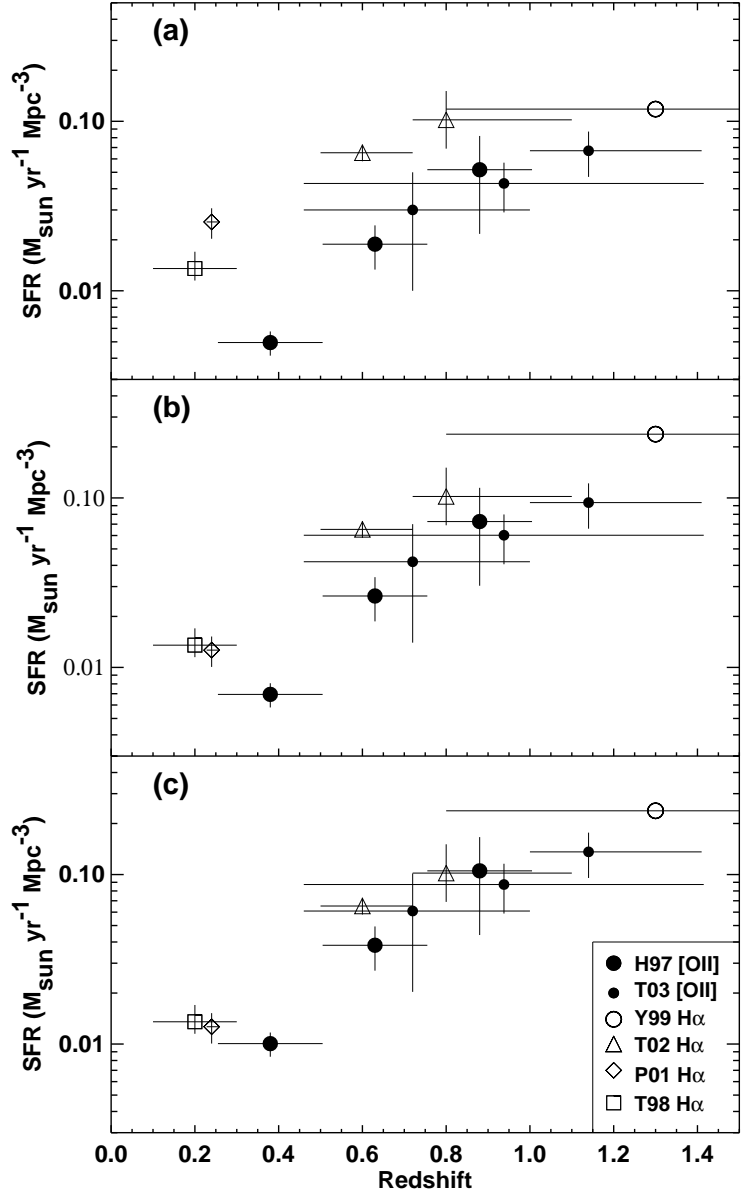}
\caption{Top (a): The cosmic star formation history where the SFR densities
are calculated according to Teplitz et al. (2003):
SFR(H$\alpha$) density is calculated using the K98 SFR(H$\alpha$) 
calibration, SFR(\OII) density is calculated using [OII]/H$\alpha = 0.45$ 
and the K98 SFR(H$\alpha$) calibration.  Reddening is the same as used 
in Teplitz et al.
Middle (b):  The SFR densities calculated using a 
consistent reddening correction.  We used the
K98 SFR(H$\alpha$) calibration, assuming an A(v)=1
(equation~\ref{eq_SFR_OII_ratiocorr};Z=log(O/H)$+12=8.6$).  
Bottom (c): The SFR densities calculated using a consistent
reddening and metallicity correction:
Z=log(O/H)$+12=8.8$ (equation~\ref{eq_SFR_OII_ratiocorr8.8}).   
References correspond to: \citet{Hammer97} (H97), \citet{Teplitz03} (T03),
\citet{Yan99} (Y99), \citet{Tresse02} (T02), \citet{Pascual01} (P01),
\citet{Tresse98} (T98).
\label{Madau_plot}}
\end{figure}

Figure~(\ref{Madau_plot}a) shows the discrepancy observed by 
\citet{Teplitz03} between star formation densities measured with
\Ha\ and those measured with \OII.  
All data on Figure~(\ref{Madau_plot}) have been converted to the
standard cosmology
($h=0.72$, $\Omega_{m}=0.29$; Table~\ref{Madau_table}).
Some of the SFR densities from Figure~(\ref{Madau_plot}a) 
have been corrected or partially corrected for reddening; others 
are uncorrected.

\citet{Gallego02} emphasize the importance of using the same 
assumptions for star formation rate conversion and
reddening for all data points in SFR density comparisons.  
For example, Figure~(5) in \citet{Teplitz03} and 
Figure~(\ref{Madau_plot}a) show that the \Ha\ SFR density point of 
\citet{Pascual01} significantly exceeds that of \citet{Tresse98}.  
The source of this apparent difference is in the assumed reddening.  
\citet{Pascual01}
use an ${\rm A_{H\alpha}}\sim 1$ which corresponds to an E(B-V)$\sim 0.6$
and an attenuation factor of $10^{0.4k(H\alpha)E(B-V)}\sim3.7$; 
Tresse et al assume an average ${\rm A_{V}}\sim 1$ which corresponds
to an E(B-V)$\sim 0.3$, corresponding to an average attenuation of 
$\sim 2.0$.  If we assume the same reddening of ${\rm A_{V}}\sim 1$ 
for both samples, the two \Ha\ SFR density estimates agree to within 10\%
(see Figure~\ref{Madau_plot}b).

Figure~(\ref{Madau_plot}a) also shows \OII\ SFR densities, 
including those by \citet{Hammer97}.  
Because Hammer et al. provide the comoving luminosity density 
(uncorrected for reddening) as a function
of redshift, we convert these into reddening corrected SFR densities 
using our SFR(\OII) 
formula (equation~\ref{eq_SFR_OII_ratiocorr}), assuming an 
average reddening of ${\rm A_{V}}\sim 1$ .  Figure~(\ref{Madau_plot}b) 
shows the new SFR(\OII) 
densities. 
The resulting intrinsic SFR(\OII) densities are $\sim50$\% larger than the 
SFR(\OII) densities in Figure~(\ref{Madau_plot}a), bringing the SFR(\OII) 
values into
closer agreement with the SFR(\Ha) data points.   
This difference is a result of 
reddening correction.  This conclusion can easily be verified by correcting 
the SFR(\Ha) densities (calculated by Teplitz et al. using the Hammer et al. 
data and \OIIHa$=0.45$) for
reddening at \Ha\ using the same extinction curve and assuming 
an ${\rm A_{V}}\sim 1$.   
Even with a 50\% increase, the \OII\ SFR density
estimates are still lower than the \Ha\ SFR estimates, reminiscent of
the T02 and H02 galaxies in Section~\ref{high_z}.

So far, we have used equation~(\ref{eq_SFR_OII_ratiocorr}) to estimate the SFR(\OII)
densities.  The use of this equation assumes that the average abundance 
for the samples at high-$z$ is \OH$\sim8.6$ (M91 diagnostic) as observed 
in the NFGS.  
The LCS sample, and most other high-z samples, contain many intrinsically 
more luminous galaxies than in the NFGS.  Galaxies
with luminosities representative of the local luminosity function are often 
too faint to be included in high redshift samples.
As we have discussed, the star formation rate density for any particular 
redshift is estimated using the \OII\ luminosity density 
and some \OII\ SFR calibration.  The use of any SFR \OII\ calibration 
requires the calculation of or an assumption about the \OIIHa\ ratio.  
In SFR history studies, the luminosity density is, in principle, corrected 
for the missing lower luminosity galaxies but the assumed \OIIHa\ is not
corrected.  The \OIIHa\ ratio for high-$z$ samples will be 
typical of the high luminosity (high-metallicity) galaxies observable, 
despite the fact that the mean metallicity for high redshift galaxies must
actually be lower than is observed locally.  Evidence for an \OIIHa\ 
ratio typical of high luminosity galaxies is easily observed in the
Hippelein et al. and Teplitz et al. samples.  The mean reddening-corrected
\OIIHa\ ratio for the Hippelein sample (0.9) corresponds to a mean abundance
of \OH$\sim8.87$ (M91 method).  The Teplitz sample has a mean uncorrected
\OIIHa\ ratio of $\sim 0.45$.  If we assume an ${\rm A_{V}}\sim 1$ 
\citep[as in][]{Tresse02,Pascual01}, then the 
mean reddening-corrected \OIIHa\ ratio is $\sim0.83$.  This
\OIIHa\ ratio corresponds to a mean metallicity of \OH$\sim8.90$, 
similar to the Hippelein sample. 

Both Teplitz et al. and Hippelein et al. use an \OIIHa\ ratio typical
for high-$z$ galaxies in their SFR density calculation.  
This \OIIHa\ ratio effectively takes the average 
metallicity into account.  We should therefore apply a similar correction.
If we assume that the average oxygen abundance for the Hammer et al. and 
Teplitz et al. galaxies is \OH$\sim8.9$ (using the M91 method) 
then equation~\ref{eq_SFR_OII_ratiocorr} becomes:

\begin{equation}
{\rm SFR (M_{\odot} yr^{-1}) = (9.53 \pm 0.91) \times 10^{-42}\, L([OII])\, (ergs\,s^{-1})}
\label{eq_SFR_OII_ratiocorr8.8}
\end{equation}

\noindent
where we have assumed the NFGS rms dispersion about the mean abundance.
The L(\OII) is corrected for reddening assuming  ${\rm A_{V}}\sim 1$
and a foreground screen geometry.  
The resulting SFR densities are a factor of 1.45 larger
than the estimates with equation~(\ref{eq_SFR_OII_ratiocorr}).   
The \OII\ SFR density 
estimates are still slightly lower than the \Ha\ SFR density estimates,
but the the \Ha\ and \OII\ SFR densities now agree to within 
$\sim 30$\% (Figure~\ref{Madau_plot}c).  

An alternative explanation for the difference between the
\OII\ and \Ha\ SFR densities in Figure~\ref{Madau_plot}b 
is that the dust geometry may not be a simple
screen.  \citet{Witt00} show that \EBV\ saturates at around 0.2-0.3 mag
for geometries where the dust is mixed with the gas.  In this scenario,
the true reddening may be much larger than predicted by \EBV.  

We  conclude that the major differences between \Ha\ and \OII -based 
estimates of the star formation rate density as a function of redshift 
probably result from
inconsistent assumptions about reddening and abundance, and/or from 
failure to correct for the biases they introduce.  Careful correction for 
both the reddening and abundance brings the star formation rate
density estimates into agreement to within $\sim 30$\%.

\section{Conclusions}

We investigate the use of the \OII\ emission-line as a star formation
indicator for a sample of 97 nearby field galaxies.  Our high S/N 
integrated (global) spectra allow correction of the Balmer lines for 
underlying stellar absorption and correction of the emission-line fluxes 
for abundance.  We find:

\begin{itemize}
\item There is a systematic difference between the SFR(\Ha) and SFR(\OII) 
using the \citet{Kennicutt98} calibrations.  The difference results from the use 
of the observed (uncorrected) \OIIHa\ ratio to convert the \OII\ 
luminosity into a SFR indicator.  
\item We derive a new SFR(\OII) indicator 
(equation~\ref{eq_SFR_OII_ratiocorr}) which is independent of the reddening 
between \OII\ and \Ha . This indicator removes the systematic difference 
observed for the Kennicutt calibration.  
\end{itemize}

We estimate the abundances for our sample using four independent 
abundance diagnostics:
\citet{Kewley02b} (\NIIOII), \citet{McGaugh91} (\R23), 
\citet{Zaritsky94} (\R23), and \citet{Charlot01} (\OIIIHb).  
We show that:

\begin{enumerate}
\item There is a strong correlation between abundance and the \OIIHa\ ratio.
\item There is no strong dependence of \OIIHa\ on the ionization 
parameter for \OH$>8.5$  
\item The observed discrepancy between the Kennicutt (1998)
SFR(\Ha) and SFR(\OII)
calibrations result from reddening and abundance differences between the 
K92 sample and other surveys.
\end{enumerate}

Our theoretical stellar population synthesis and photoionization models 
support these conclusions.
We derive a new SFR(\OII,Z) calibration which includes a correction 
for abundance  (equation~\ref{eq_SFR_abund_corr_adopt}).  
This relation reduces the rms residuals between SFR(\Ha) and SFR(\OII) 
significantly from 0.11 to 0.03-0.05 dex. 
Our SFR(\OII,Z) calibration is based on the intrinsic 
(reddening corrected) \OII\ luminosity and the abundance.  This 
calibration can be used to derive a SFR relation 
for samples where the mean abundance differs from the NFGS. 
The SFR(\OII,Z) relation may also be applied to samples where 
abundance estimates are available, but where \Hb\ has too low a S/N to
be used to calculate a star formation rate based on \Ha. 

We observe a strong correlation between the intrinsic \OII\ luminosity 
and E(B-V) for the NFGS.  Using the observed \OII\ luminosity to 
estimate the reddening produces no systematic shift between the 
SFR(\Ha) and SFR(\OII), but there is a large dispersion (0.21 dex).
Further investigation is required to verify whether a similar 
reddening-luminosity relation holds for other samples.  

We discuss the application of our SFR(\OII) calibrations to studies 
of the cosmic star formation history.  We show that we can remove
the previously observed discrepancy 
between the star formation rates based on \OII\ and those based on 
\Ha\ by correcting for reddening and abundance.  
  The agreement  between
SFR(\Ha) and SFR(\OII,Z) emphasizes the importance of aquiring data which
yield internally consistent reddening and abundance estimates over the entire
redshift range of the sample.

\acknowledgments
The authors wish to thank the anonymous referee for many insightful comments
which have led to improvements in this paper. We also thank R. Kennicutt, 
S. Kenyon, and M. Dopita for useful comments and suggestions.
L. J. Kewley is supported by a Harvard-Smithsonian 
CfA Fellowship.   M. J. Geller is supported by the Smithsonian Institution.

\clearpage 

\begin{deluxetable}{llrrrrrrrrrrrrrr}
\tablecolumns{16}
\tablewidth{25cm}
\tabletypesize{\scriptsize}
\rotate
\tablecaption{The NFGS Sample with \OII, \OIII, \NII, \Ha, and \Hb\ fluxes \label{sample_table}}
\tablehead{
 ID &  Name
&  $cz$
&  \multicolumn{2}{c}{E(B-V)}
&  \multicolumn{4}{c}{$\log(\frac{\rm O}{\rm H})+12$}
&  ${\rm L(H\alpha)}$\tablenotemark{c}
&  ${\rm L([OII]))}$\tablenotemark{d}
&  SFR(\Ha)
&  SFR([OII])
&  SFR([OII])\tablenotemark{e}
&  SFR([OII],Z)\tablenotemark{f}
&  SFR([OII])$_{L}$\tablenotemark{g}
\\
&    
&  (km/s)
&  real\tablenotemark{a}
&  est\tablenotemark{b}
&  KD02 
&  Z94 
&  M91 
&  C01
&  $(log(\frac{L}{\rm L_{\odot}}))$
&  $(log(\frac{L}{\rm L_{\odot}})$ 
&  (K98) 
&  (K98) 
&  (eq~\ref{eq_SFR_OII_ratiocorr}) 
&  (eq~\ref{eq_Grid_SFR_OII})
&  (eq~\ref{eq_OII_lum},~\ref{eq_SFR_abund_corr_adopt})\\
}
\startdata
  2 &  A00289+0556 &  2055 & 0.30 & 0.20 & 8.71 & 8.73 & 8.57 & 8.85 &  6.93 &  7.15 &   0.26 &   0.41 &   0.36 &   0.29 &   0.19 \\ 
  4 &  A00389-0159 &  5302 & 0.47 & 0.37 & 8.94 & 9.09 & 8.90 & 9.25 &  8.22 &  8.12 &   4.99 &   2.68 &   3.32 &   6.84 &   3.21 \\ 
  5 &  A00442+3224 &  4859 & 0.40 & 0.38 & 8.92 & 8.94 & 8.78 & 8.82 &  8.02 &  8.02 &   3.19 &   2.48 &   2.64 &   3.12 &   2.63 \\ 
  9 &  A01047+1625 &   158 & 0.16 & 0.00 & 7.6-8.2 & \nodata & 8.15 & 8.61 &  4.69 &  4.84 &   0.001 & 0.003 &  0.002 &  0.001 &   0.0002 \\ 
 12 &  A01123-0046 & 10184 & 0.64 & 0.46 & 8.96 & 8.96 & 8.80 & 8.84 &  8.83 &  8.80 &  20.31 &   9.18 &  16.03 &  20.26 &   9.51 \\ 
 15 &  A01300+1804 &   686 & 0.16 & 0.00 & 7.6-8.2 & \nodata & 8.04 & 8.63 &  5.85 &  5.93 &   0.02 &   0.03 &   0.02 &   0.02 &   0.01 \\ 
 16 &  A01344+2838 &  7756 & 0.46 & 0.48 & 8.78 & 8.76 & 8.62 & 8.80 &  8.36 &  8.53 &   6.93 &   7.06 &   8.49 &   7.26 &   7.67 \\ 
 17 &  A01346+0438 &  3158 & 0.32 & 0.30 & 8.77 & 8.80 & 8.66 & 8.81 &  7.42 &  7.56 &   0.80 &   1.01 &   0.91 &   0.82 &   0.76 \\ 
 19 &      NGC 695 &  9705 & 0.80 & 0.58 & 8.89 & 8.93 & 8.76 & 8.90 &  9.54 &  9.58 & 104.19 &  39.26 &  96.36 & 111.49 &  47.36 \\ 
 21 &  A02008+2350 &  2669 & 0.30 & 0.30 & 8.80 & 8.88 & 8.73 & 8.83 &  7.49 &  7.55 &   0.93 &   1.04 &   0.90 &   0.94 &   0.91 \\ 
 23 &       IC 197 &  6332 & 0.45 & 0.41 & 8.87 & 8.83 & 8.69 & 8.76 &  8.18 &  8.26 &   4.57 &   3.81 &   4.57 &   4.31 &   3.54 \\ 
 24 &      IC 1776 &  3405 & 0.20 & 0.39 & 8.54 & 8.58 & 8.49 & 8.68 &  7.48 &  7.71 &   0.91 &   1.83 &   1.30 &   0.98 &   1.82 \\ 
 25 &  A02056+1444 &  4405 & 0.47 & 0.33 & 8.90 & 8.98 & 8.81 & 8.91 &  8.00 &  7.99 &   3.01 &   1.98 &   2.46 &   3.24 &   1.72 \\ 
 28 &  A02257-0134 &  1762 & 0.25 & 0.11 & 8.72 & 8.91 & 8.72 & 9.06 &  6.60 &  6.70 &   0.12 &   0.16 &   0.13 &   0.14 &   0.07 \\ 
 32 &  A02493-0122 &  1508 & 0.06 & 0.07 & 8.85 & 9.04 & 8.89 & 8.73 &  6.50 &  6.22 &   0.10 &   0.08 &   0.04 &   0.07 &   0.06 \\ 
 34 &  A03202-0205 &  8227 & 0.44 & 0.46 & 8.93 & 8.83 & 8.64 & 9.04 &  8.26 &  8.43 &   5.44 &   5.92 &   6.83 &   6.47 &   6.20 \\ 
 41 &     NGC 2799 &  1882 & 0.31 & 0.17 & 8.84 & 8.92 & 8.76 & 8.89 &  7.02 &  7.07 &   0.32 &   0.33 &   0.30 &   0.34 &   0.19 \\ 
 43 &     NGC 2844 &  1486 & 0.39 & 0.10 & 8.81 & 8.82 & 8.60 & 9.17 &  6.74 &  6.94 &   0.17 &   0.21 &   0.22 &   0.20 &   0.06 \\ 
 45 &     NGC 3009 &  4666 & 0.32 & 0.25 & 8.93 & 9.05 & 8.86 & 9.23 &  7.44 &  7.40 &   0.83 &   0.69 &   0.63 &   1.08 &   0.62 \\ 
 46 &      IC 2520 &  1226 & 0.49 & 0.16 & 8.84 & 8.86 & 8.71 & 8.81 &  7.28 &  7.36 &   0.57 &   0.45 &   0.58 &   0.57 &   0.17 \\ 
 47 &  A09557+4758 &  1172 & 0.04 & 0.23 & 8.44 & 8.52 & 8.50 & 8.56 &  6.62 &  6.76 &   0.13 &   0.28 &   0.14 &   0.11 &   0.21 \\ 
 48 &     NGC 3075 &  3566 & 0.16 & 0.30 & 8.97 & 9.08 & 8.90 & 8.94 &  7.42 &  7.28 &   0.79 &   0.73 &   0.48 &   0.90 &   1.19 \\ 
 49 &  A09579+0439 &  4185 & 0.49 & 0.30 & 8.76 & 8.76 & 8.60 & 8.87 &  7.69 &  7.89 &   1.49 &   1.52 &   1.96 &   1.67 &   0.76 \\ 
 50 &     NGC 3104 &   604 & 0.03 & 0.10 & 8.19 & \nodata & 8.06 & 8.61 &  6.16 &  6.24 &   0.04 &   0.09 &   0.04 &   0.03 &   0.05 \\ 
 51 &  A10042+4716 &   571 & 0.06 & 0.00 & 8.54 & 8.72 & 8.62 & 8.67 &  5.59 &  5.70 &   0.01 &   0.02 &   0.01 &   0.01 &   0.01 \\ 
 53 &  A10114+0716 &  1228 & 0.08 & 0.17 & 7.6-8.2 & \nodata & 8.13 & 8.63 &  6.49 &  6.63 &   0.09 &   0.19 &   0.11 &   0.08 &   0.11 \\ 
 55 &  A10171+3853 &  2008 & 0.09 & 0.17 & 8.85 & 8.96 & 8.79 & 8.91 &  6.63 &  6.64 &   0.13 &   0.20 &   0.11 &   0.14 &   0.17 \\ 
 57 &     NGC 3264 &   929 & $\le$0.02 & 0.21 & 8.50 & 8.64 & 8.60 & 8.55 &  6.61 &  6.65 &   0.12 &   0.22 &   0.11 &   0.08 &   0.20 \\ 
 59 &  A10321+4649 &  3338 & 0.31 & 0.17 & 9.06 & 9.21 & 9.02 & 9.31 &  7.38 &  7.06 &   0.72 &   0.32 &   0.29 &   1.41 &   0.34 \\ 
 60 &  A10337+1358 &  2997 & 0.46 & 0.21 & 8.89 & 8.99 & 8.81 & 8.98 &  7.51 &  7.51 &   0.97 &   0.68 &   0.82 &   1.12 &   0.38 \\ 
 61 &      IC 2591 &  6755 & 0.21 & 0.49 & 8.79 & 8.80 & 8.67 & 8.75 &  8.01 &  8.11 &   3.08 &   4.41 &   3.23 &   2.91 &   8.02 \\ 
 62 &  A10365+4812 &   854 & $\le$0.02 & 0.00 & 8.25 & \nodata & 8.05 & 8.56 &  5.80 &  5.86 &   0.02 &   0.04 &   0.02 &   0.01 &   0.01 \\ 
 63 &  A10368+4811 &  1534 & 0.13 & 0.16 & 8.14 & \nodata & 8.26 & 8.57 &  6.47 &  6.68 &   0.09 &   0.20 &   0.12 &   0.09 &   0.09 \\ 
 69 &  A10504+0454 &  5793 & 0.24 & 0.43 & 8.91 & 8.92 & 8.77 & 8.78 &  7.94 &  7.93 &   2.62 &   2.81 &   2.15 &   2.41 &   4.96 \\ 
 70 &     NGC 3454 &  1153 & 0.24 & 0.09 & 8.79 & 8.89 & 8.74 & 8.80 &  6.57 &  6.62 &   0.11 &   0.13 &   0.10 &   0.11 &   0.06 \\ 
 71 &  A10592+1652 &  2936 & 0.16 & 0.25 & 8.52 & 8.62 & 8.53 & 8.65 &  6.90 &  7.09 &   0.24 &   0.48 &   0.31 &   0.23 &   0.32 \\ 
 73 &     NGC 3510 &   704 & 0.05 & 0.16 & 8.43 & 8.62 & 8.56 & 8.58 &  6.43 &  6.51 &   0.08 &   0.16 &   0.08 &   0.06 &   0.10 \\ 
 76 &  A11040+5130 &  2204 & 0.07 & 0.28 & 8.74 & 8.83 & 8.69 & 8.75 &  6.96 &  7.03 &   0.28 &   0.50 &   0.27 &   0.26 &   0.57 \\ 
 77 &       IC 673 &  3851 & 0.40 & 0.26 & 8.96 & 9.04 & 8.87 & 8.98 &  7.65 &  7.58 &   1.36 &   0.89 &   0.96 &   1.57 &   0.74 \\ 
 79 &  A11072+1302 & 12743 & 0.32 & 0.62 & 8.79 & 8.85 & 8.71 & 8.79 &  8.73 &  8.80 &  16.24 &  17.57 &  16.07 &  15.74 &  47.31 \\ 
 82 &     NGC 3633 &  2553 & 0.88 & 0.12 & 8.91 & 8.95 & 8.75 & 9.17 &  7.87 &  7.94 &   2.23 &   0.77 &   2.22 &   2.72 &   0.13 \\ 
 88 &  A11336+5829 &  1225 & 0.12 & 0.10 & 8.46 & 8.57 & 8.50 & 8.63 &  6.21 &  6.41 &   0.05 &   0.11 &   0.06 &   0.05 &   0.04 \\ 
 89 &    NGC 3795A &  1154 & $\le$0.02 & 0.15 & 8.62 & 8.91 & 8.77 & 8.79 &  6.31 &  6.42 &   0.06 &   0.13 &   0.07 &   0.07 &   0.11 \\ 
 90 &  A11372+2012 & 10964 & 0.30 & 0.58 & 8.92 & 8.96 & 8.80 & 8.85 &  8.64 &  8.63 &  13.13 &  12.29 &  10.67 &  13.19 &  35.39 \\ 
 91 &     NGC 3795 &  1091 & 0.29 & 0.01 & 8.72 & 8.82 & 8.66 & 8.84 &  6.28 &  6.42 &   0.06 &   0.08 &   0.07 &   0.06 &   0.02 \\ 
 92 &  A11378+2840 &  1821 & 0.16 & 0.16 & 8.61 & 8.66 & 8.55 & 8.70 &  6.54 &  6.73 &   0.10 &   0.21 &   0.13 &   0.10 &   0.10 \\ 
 93 &  A11392+1615 &   786 & 0.06 & 0.02 & 8.62 & 8.65 & 8.59 & 8.59 &  5.94 &  6.00 &   0.03 &   0.05 &   0.02 &   0.02 &   0.02 \\ 
 94 &     NGC 3846 &  1396 & $\le$0.02 & 0.25 & 8.59 & 8.71 & 8.61 & 8.66 &  6.67 &  6.81 &   0.14 &   0.32 &   0.16 &   0.12 &   0.32 \\ 
 96 &  A11476+4220 &  1033 & 0.24 & 0.01 & 8.91 & 9.02 & 8.84 & 8.94 &  6.35 &  6.30 &   0.07 &   0.07 &   0.05 &   0.08 &   0.03 \\ 
 98 &       IC 746 &  5027 & 0.32 & 0.39 & 8.66 & 8.66 & 8.54 & 8.73 &  7.68 &  7.90 &   1.46 &   2.21 &   1.99 &   1.54 &   1.89 \\ 
100 &     NGC 3978 &  9978 & 0.41 & 0.56 & 9.02 & 9.13 & 8.95 & 9.11 &  8.93 &  8.74 &  25.74 &  12.71 &  13.95 &  35.60 &  38.74 \\ 
103 &     NGC 4034 &  2384 & $\le$0.02 & 0.12 & 8.91 & 9.17 & 8.99 & 9.22 &  6.41 &  6.32 &   0.08 &   0.10 &   0.05 &   0.17 &   0.14 \\ 
104 &  A11592+6237 &  1120 & $\le$0.02 & 0.14 & 8.22 & \nodata & 8.18 & 8.54 &  6.23 &  6.38 &   0.05 &   0.12 &   0.06 &   0.04 &   0.07 \\ 
105 &  A12001+6439 &  1447 & 0.33 & 0.19 & 8.83 & 8.87 & 8.73 & 8.77 &  7.13 &  7.17 &   0.41 &   0.40 &   0.37 &   0.38 &   0.23 \\ 
107 &     NGC 4120 &  2251 & 0.15 & 0.26 & 8.69 & 8.80 & 8.68 & 8.72 &  7.04 &  7.12 &   0.33 &   0.51 &   0.33 &   0.30 &   0.47 \\ 
109 &     NGC 4141 &  1980 & 0.08 & 0.37 & 8.53 & 8.63 & 8.56 & 8.61 &  7.27 &  7.39 &   0.57 &   1.13 &   0.62 &   0.47 &   1.51 \\ 
110 &     NGC 4159 &  1761 & 0.30 & 0.21 & 8.64 & 8.77 & 8.65 & 8.72 &  7.09 &  7.20 &   0.37 &   0.45 &   0.40 &   0.34 &   0.24 \\ 
112 &     NGC 4238 &  2771 & 0.15 & 0.33 & 8.75 & 8.82 & 8.69 & 8.73 &  7.29 &  7.35 &   0.59 &   0.89 &   0.56 &   0.52 &   1.07 \\ 
113 &     NGC 4248 &   484 & 0.33 & 0.00 & 8.70 & 8.74 & 8.60 & 8.80 &  5.89 &  6.08 &   0.02 &   0.03 &   0.03 &   0.03 &   0.00 \\ 
114 &  A12167+4938 &  3639 & 0.34 & 0.31 & 8.74 & 8.76 & 8.61 & 8.82 &  7.48 &  7.66 &   0.91 &   1.20 &   1.14 &   0.98 &   0.80 \\ 
116 &     NGC 4288 &   532 & 0.13 & 0.08 & 8.58 & 8.55 & 8.48 & 8.63 &  6.16 &  6.38 &   0.04 &   0.10 &   0.06 &   0.04 &   0.03 \\ 
121 &  A12295+4007 &   685 & $\le$0.02 & 0.00 & 7.94 & \nodata & 7.90 & 8.57 &  5.87 &  5.84 &   0.02 &   0.03 &   0.02 &   0.01 &   0.01 \\ 
123 &  A12304+3754 &   503 & 0.02 & 0.04 & 8.53 & 8.70 & 8.60 & 8.67 &  5.88 &  6.00 &   0.02 &   0.05 &   0.03 &   0.02 &   0.02 \\ 
124 &     NGC 4509 &   907 & 0.04 & 0.17 & 8.22 & \nodata & 8.15 & 8.51 &  6.50 &  6.55 &   0.09 &   0.18 &   0.09 &   0.07 &   0.11 \\ 
125 &  A12331+7230 &  6959 & 0.46 & 0.41 & 8.85 & 8.92 & 8.75 & 8.90 &  8.23 &  8.29 &   5.13 &   4.06 &   4.95 &   5.54 &   4.32 \\ 
126 &  A12446+5155 &   502 & 0.02 & 0.00 & 7.6-8.2 & \nodata & 8.02 & 8.66 &  5.63 &  5.74 &   0.01 &   0.03 &   0.01 &   0.01 &   0.01 \\ 
127 &     NGC 4758 &  1244 & 0.42 & 0.13 & 8.57 & 8.69 & 8.54 & 8.81 &  6.87 &  7.11 &   0.23 &   0.29 &   0.32 &   0.25 &   0.08 \\ 
135 &  A13194+4232 &  3396 & 0.29 & 0.27 & 8.83 & 8.95 & 8.77 & 8.98 &  7.33 &  7.38 &   0.64 &   0.72 &   0.61 &   0.73 &   0.61 \\ 
140 &     NGC 5230 &  6855 & 0.35 & 0.54 & 8.88 & 8.92 & 8.76 & 8.88 &  8.53 &  8.57 &  10.19 &   9.82 &   9.47 &  10.74 &  19.85 \\ 
141 &  A13361+3323 &  2364 & 0.17 & 0.29 & 8.41 & 8.51 & 8.47 & 8.59 &  7.04 &  7.24 &   0.33 &   0.66 &   0.44 &   0.33 &   0.46 \\ 
143 &  A13422+3526 &  2502 & 0.33 & 0.16 & 8.70 & 8.77 & 8.62 & 8.82 &  6.87 &  7.05 &   0.23 &   0.30 &   0.28 &   0.24 &   0.12 \\ 
148 &     NGC 5425 &  2062 & 0.25 & 0.19 & 8.76 & 8.85 & 8.71 & 8.80 &  6.96 &  7.04 &   0.28 &   0.35 &   0.28 &   0.27 &   0.21 \\ 
153 &     NGC 5541 &  7698 & 0.48 & 0.50 & 8.94 & 8.98 & 8.80 & 9.01 &  8.62 &  8.64 &  12.53 &   8.82 &  11.03 &  14.53 &  13.64 \\ 
155 &     NGC 5608 &   662 & $\le$0.02 & 0.00 & 8.33 & \nodata & 7.99 & 8.67 &  5.68 &  5.75 &   0.01 &   0.03 &   0.01 &   0.01 &   0.01 \\ 
156 &  A14305+1149 &  2234 & 0.13 & 0.25 & 8.77 & 8.84 & 8.69 & 8.78 &  6.95 &  7.03 &   0.27 &   0.45 &   0.27 &   0.26 &   0.41 \\ 
158 &     NGC 5762 &  1788 & 0.10 & 0.19 & 8.83 & 8.82 & 8.68 & 8.75 &  6.64 &  6.73 &   0.13 &   0.23 &   0.13 &   0.12 &   0.16 \\ 
159 &  A14489+3547 &  1215 & $\le$0.02 & 0.24 & 8.31 & \nodata & 8.09 & 8.42 &  6.76 &  6.78 &   0.17 &   0.29 &   0.15 &   0.10 &   0.28 \\ 
161 &      IC 1066 &  1613 & 0.49 & 0.14 & 8.72 & 8.71 & 8.50 & 9.04 &  7.01 &  7.28 &   0.31 &   0.37 &   0.48 &   0.39 &   0.09 \\ 
165 &     NGC 5874 &  3128 & 0.52 & 0.25 & 8.80 & 8.78 & 8.57 & 9.11 &  7.55 &  7.77 &   1.07 &   1.10 &   1.50 &   1.31 &   0.41 \\ 
166 &    NGC 5875A &  2470 & 0.13 & 0.20 & 8.92 & 9.02 & 8.86 & 8.89 &  6.93 &  6.85 &   0.26 &   0.29 &   0.18 &   0.27 &   0.31 \\ 
168 &      IC 1124 &  5242 & 0.64 & 0.24 & 8.97 & 9.08 & 8.89 & 9.22 &  8.03 &  7.95 &   3.22 &   1.27 &   2.23 &   4.33 &   0.65 \\ 
170 &  A15314+6744 &  6461 & 0.23 & 0.36 & 8.99 & 9.08 & 8.90 & 8.95 &  7.78 &  7.64 &   1.84 &   1.48 &   1.10 &   2.10 &   2.56 \\ 
171 &     NGC 5993 &  9578 & 0.27 & 0.50 & 9.02 & 9.09 & 8.92 & 8.92 &  8.43 &  8.25 &   8.11 &   5.45 &   4.46 &   9.05 &  15.73 \\ 
174 &     NGC 6007 & 10548 & 0.33 & 0.52 & 8.94 & 9.04 & 8.86 & 9.00 &  8.49 &  8.43 &   9.27 &   7.32 &   6.75 &  10.93 &  16.80 \\ 
175 &  A15523+1645 &  2191 & 0.27 & 0.24 & 8.62 & 8.59 & 8.47 & 8.74 &  6.98 &  7.25 &   0.29 &   0.55 &   0.45 &   0.34 &   0.28 \\ 
179 &     NGC 6131 &  5054 & 0.13 & 0.34 & 8.99 & 9.13 & 8.95 & 9.13 &  7.57 &  7.39 &   1.11 &   1.00 &   0.62 &   1.55 &   2.14 \\ 
181 &     NGC 7077 &  1142 & 0.15 & 0.13 & 8.55 & 8.56 & 8.52 & 8.56 &  6.50 &  6.62 &   0.10 &   0.16 &   0.10 &   0.08 &   0.07 \\ 
183 &  A22306+0750 &  1995 & 0.57 & 0.16 & 8.87 & 8.95 & 8.77 & 8.99 &  7.46 &  7.51 &   0.86 &   0.54 &   0.82 &   0.99 &   0.20 \\ 
186 &  A22426+0610 &  1925 & 0.16 & 0.25 & 8.81 & 8.92 & 8.78 & 8.71 &  7.17 &  7.08 &   0.44 &   0.46 &   0.30 &   0.34 &   0.45 \\ 
187 & A22551+1931N &  5682 & 0.61 & 0.30 & 8.90 & 8.93 & 8.77 & 8.80 &  8.12 &  8.12 &   4.01 &   2.00 &   3.32 &   3.79 &   1.12 \\ 
189 &     NGC 7460 &  3296 & 0.48 & 0.31 & 8.97 & 9.10 & 8.91 & 9.13 &  8.01 &  7.90 &   3.12 &   1.58 &   1.98 &   4.14 &   1.56 \\ 
190 &     NGC 7537 &  2648 & 0.41 & 0.31 & 8.93 & 8.96 & 8.79 & 8.87 &  7.79 &  7.80 &   1.87 &   1.43 &   1.57 &   1.94 &   1.23 \\ 
192 &  A23176+1541 &  4380 & 0.49 & 0.34 & 8.81 & 8.84 & 8.69 & 8.81 &  7.95 &  8.06 &   2.71 &   2.24 &   2.87 &   2.74 &   1.56 \\ 
193 &     NGC 7620 &  9565 & 0.56 & 0.59 & 8.88 & 8.92 & 8.76 & 8.85 &  9.11 &  9.14 &  38.98 &  23.30 &  34.87 &  39.65 &  41.12 \\ 
195 &      IC 1504 &  6306 & 0.92 & 0.35 & 8.78 & 8.77 & 8.57 & 9.04 &  8.69 &  8.92 &  14.86 &   6.76 &  20.94 &  18.08 &   2.05 \\ 
196 &     NGC 7752 &  4902 & 0.37 & 0.42 & 8.97 & 9.04 & 8.87 & 8.84 &  8.25 &  8.12 &   5.43 &   3.31 &   3.32 &   5.30 &   5.71 \\ 
198 &  A23542+1633 &  1788 & 0.11 & 0.26 & 8.11 & \nodata & 8.08 & 8.56 &  6.97 &  7.03 &   0.28 &   0.46 &   0.27 &   0.20 &   0.38 \\ 
\enddata
\tablenotetext{a}{E(B-V) is the observed reddening calculated using the 
Balmer Decrement as described in Section~\ref{sample}. E(B-V) includes
Galactic extinction which is $\sim0.015$ on average for the NFGS}
\tablenotetext{b}{E(B-V) is calculated using the uncorrected
(observed) [OII] luminosity from equations~(\ref{eq_LOII_EB_V}) and (\ref{eq_OII_lum}), 
described in Section~\ref{high_z}.}
\tablenotetext{c}{The \Ha\ luminosity is corrected for underlying
stellar absorption and for reddening
using the E(B-V) derived from the Balmer decrement. Units are in the logarithmic scale in term of \Lsun}
\tablenotetext{d}{The [OII] luminosity is corrected for reddening
using the E(B-V) derived from the Balmer decrement.  Units are in the logarithmic scale in terms of \Lsun}
\tablenotetext{e}{SFR([OII]) is derived using equation~(\ref{eq_SFR_OII_ratiocorr})
and L([OII]) corrected for reddening using the E(B-V) derived from the 
Balmer decrement.}
\tablenotetext{f}{SFR([OII]) is derived using 
equation~(\ref{eq_Grid_SFR_OII})  with the Z94 abundance.
\R23 and L([OII]) have been corrected for reddening using the E(B-V) derived 
from the Balmer decrement.}
\tablenotetext{g}{SFR([OII]) is derived using the intrinsic luminosity
estimated from the observed luminosity (equation~\ref{eq_OII_lum}).
Equation~(\ref{eq_SFR_abund_corr_adopt}) was used to derive the SFR([OII]).}
\end{deluxetable}

\clearpage

\begin{deluxetable}{lllllll}
\tabletypesize{\small}
\tablecaption{Slope, y-intercept, and scatter for the line of best fit in equation~\ref{eq_abund_vs_OIIHa} (Figures~\ref{OIIHa_vs_abund4}a-d).
\label{coeffs}}
\tablehead{Figure
& Abundance
& slope
& y-intercept
& rms
& \multicolumn{2}{c}{Spearman Rank} \\
Number 
& Diagnostic 
& $a$& $b$ & $\sigma$
& Coefficient 
& Probability (\%)\\
}
\startdata
\ref{OIIHa_vs_abund4}a & KD02 \NIIOII\ & $-1.99 \pm 0.32$ & $18.67 \pm 2.83$ & 0.10 & -0.79 & $1.75\times 10^{-16}$\\
\ref{OIIHa_vs_abund4}b & Z94 \R23\ & $-1.75 \pm 0.25$ & $16.73 \pm 2.23$ & 0.07 & -0.88 & $3.80\times10^{-26}$\\
\ref{OIIHa_vs_abund4}c & M91 \R23\ &  $-2.29 \pm 0.41$ & $21.21 \pm 3.66$ & 0.04 & -0.93 & $\lesssim 10^{-30}$\\
\ref{OIIHa_vs_abund4}d & C01 \OIIIHb\ & $-0.88 \pm 0.17$ & $8.78 \pm 1.53$ & 0.24 & -0.36 & 
0.05 \\
\enddata
\end{deluxetable}

\begin{deluxetable}{llrrr}
\tabletypesize{\small}
\tablecaption{Slope, y-intercept, and scatter for the line of best fit to
the SFR(\Ha) versus SFR(\OII) plots.
\label{SFR_coeffs}}
\tablehead{Figure
& Abundance
& slope
& y-intercept 
& rms\\
Number 
& Diagnostic 
& $a$& $b$ &$\sigma$\\
}
\startdata
\ref{SFR_Ha_vs_SFR_OII_K98} (K98) & none & $0.83\pm0.02$ & $0.01\pm0.02$ & 0.11\\
\ref{SFR_Ha_vs_SFR_OII_ratiocorr} & none & $0.97 \pm 0.02$ & $-0.03 \pm 0.02$ & 0.08\\ \hline
\ref{SFR_Ha_vs_SFR_OII_4abund}a & KD02 \NIIOII\ & $ 1.04 \pm 0.03$ & $-0.007 \pm 0.03$ & 0.05 \\
\ref{SFR_Ha_vs_SFR_OII_4abund}b & Z94 \R23\ & $1.02 \pm 0.03$ & $0.01 \pm 0.03$ & 0.04 \\
\ref{SFR_Ha_vs_SFR_OII_4abund}c & M91 \R23\ &  $1.01 \pm 0.03$ & $-0.012 \pm 0.03$ & 0.03 \\
\ref{SFR_Ha_vs_SFR_OII_4abund}d & C01 \OIIIHb\ &$1.01 \pm 0.03$ & $ 0.04\pm 0.03$ & 0.08\\ \hline
\ref{Grid_SFRs_KD02_Z94}a & KD02 \NIIOII\  & $1.03 \pm 0.04$ & $-0.04\pm0.03$
& 0.05 \\
\ref{Grid_SFRs_KD02_Z94}b & Z94 \R23\ & $1.03 \pm 0.02$ & $+0.03\pm0.01$ 
& 0.05 \\
\ref{SFR_Ha_vs_SFR_OII_Av1}a & Z94 (${\rm A_{V}}=1$) & $0.76 \pm 0.03$ &
$-0.06 \pm 0.02$ & 0.17 \\
\ref{SFR_Ha_vs_SFR_OII_Av1}b & Grid (${\rm A_{V}}=1$) & $0.77 \pm 0.03$ &
$-0.04 \pm 0.02$ & 0.17 \\
\ref{SFR_Ha_vs_SFR_OII_LOIIcorr}a,b & Z94; E(B-V) from L(\OII) & 
$1.07 \pm 0.03$ & $-0.06 \pm 0.03$ & 0.21 \\
\ref{SFR_Ha_vs_SFR_OII_H02T02}a & none (${\rm A_{V}}=1$) &  $0.93\pm0.12$  & $-0.20 \pm 0.11$ & 0.21 \\
\ref{SFR_Ha_vs_SFR_OII_H02T02}b & M91 mean (${\rm A_{V}}=1$) &  $1.09\pm0.11$  & $-0.15\pm0.10$ & 0.17 \\

\enddata
\end{deluxetable}

\clearpage

\begin{deluxetable}{lllll}
\tabletypesize{\small}
\tablecaption{Coefficients for the  best fit to
the theoretical curves for abundance versus \OIIHa.
\label{Model_coeffs}}
\tablehead{$q$ (cm/s) 
& $a$
& $b$
& $c$
& $d$\\
}
\startdata
5$\times10^6$ & -2564.67 & 847.554 & -92.9404 & 3.38261 \\
1$\times10^7$ & -2877.94 & 955.234 & -105.245 & 3.85016 \\
2$\times10^7$ & -2281.80 & 754.840 & -82.8388 & 3.01667 \\
4$\times10^7$ & -1432.67 & 470.545 & -51.2139 & 1.84750 \\
8$\times10^7$ & -786.096 & 256.059 & -27.6048 & 0.984957 \\
\enddata

\end{deluxetable}

\begin{deluxetable}{llrcrrrccccrrr}
\tabletypesize{\scriptsize}
\rotate
\tablecaption{Luminosity and SFRs for the seven Hicks et al. (2002; H02) galaxies and 30 Tresse et al. (2002; T02) galaxies 
with \OII\ and \Ha\ luminosities (${\rm A_{V}}=1$ assumed).
\label{Hicks_data}}
\tablehead{ID & Ref  
&  $z$
&  log
&  F[OII]
&  F(\Ha)
&  log
&  log
&  log
&  log
&  SFR[OII]
&  SFR[OII]
&  SFR\\
& & & $(\frac{[{\rm OII}]}{\rm H\alpha})_{o}$  &  
 \multicolumn{2}{c}{($10^{-18} {\rm erg\,s^{-1}\,cm^{-}2}$)}  
& L([OII])$_{o}$  
& L([OII])$_{i}$ 
& L(\Ha)$_{o}$ 
& L(\Ha)$_{i}$
& (8.6)\tablenotemark{a}
& (8.75,8.95)\tablenotemark{b}
& \Ha \\
}
\startdata
 2 & H02 & 0.76 &  -0.56 &  98.61 & 358.03 &  41.40 &  41.97 &  41.96 &  42.27 &   6.20 &  21.91 &  14.58 \\ 
 4 & H02 & 1.14 &  -0.39 &  79.21 & 194.44 &  41.74 &  42.32 &  42.13 &  42.44 &  13.60 &  48.02 &  21.61 \\ 
 5 & H02 & 1.45 &  -1.07 &   5.64 &  66.26 &  40.85 &  41.43 &  41.92 &  42.23 &   1.76 &   6.21 &  13.38 \\ 
 7 & H02 & 1.06 &  -0.71 &  23.27 & 119.34 &  41.13 &  41.70 &  41.84 &  42.15 &   3.33 &  11.78 &  11.07 \\ 
 9 & H02 & 1.49 &  -1.05 &  10.67 & 119.72 &  41.16 &  41.73 &  42.21 &  42.51 &   3.56 &  12.57 &  25.85 \\ 
10 & H02 & 1.37 &  -1.03 &  11.81 & 126.55 &  41.11 &  41.69 &  42.14 &  42.45 &   3.20 &  11.30 &  22.19 \\ 
11 & H02 & 1.13 &  -0.90 &  11.81 &  93.81 &  40.91 &  41.48 &  41.81 &  42.11 &   1.98 &   7.01 &  10.20 \\ 
 0.0338 & T02 & 1.03 &  -0.20 & 113.90 & 178.50 &  41.79 &  42.36 &  41.98 &  42.29 &  15.12 &  19.32 &  15.34 \\ 
 0.0564 & T02 & 0.61 &  -0.67 &  30.10 & 140.30 &  40.65 &  41.23 &  41.32 &  41.63 &   1.11 &   1.41 &   3.34 \\ 
 0.0874 & T02 & 0.71 &  -0.46 &  22.30 &  64.50 &  40.69 &  41.26 &  41.15 &  41.45 &   1.19 &   1.53 &   2.24 \\ 
 3.0167 & T02 & 0.60 &  -0.09 &  57.30 &  71.30 &  40.92 &  41.49 &  41.01 &  41.32 &   2.04 &   2.60 &   1.64 \\ 
 3.0422 & T02 & 0.71 &  -0.72 &  50.10 & 265.60 &  41.04 &  41.61 &  41.77 &  42.07 &   2.71 &   3.46 &   9.30 \\ 
 3.0480 & T02 & 0.61 &  -0.19 & 155.60 & 238.50 &  41.36 &  41.93 &  41.55 &  41.85 &   5.66 &   7.24 &   5.62 \\ 
 3.0485 & T02 & 0.61 &  -0.52 & 249.70 & 823.20 &  41.56 &  42.14 &  42.08 &  42.39 &   9.01 &  11.50 &  19.22 \\ 
 3.0488 & T02 & 0.61 &  -0.21 & 278.20 & 450.50 &  41.61 &  42.19 &  41.82 &  42.13 &  10.09 &  12.88 &  10.57 \\ 
 3.0570 & T02 & 0.65 &  -0.37 &  94.00 & 220.70 &  41.21 &  41.78 &  41.58 &  41.89 &   4.00 &   5.11 &   6.08 \\ 
 3.0589 & T02 & 0.72 &  -0.60 &  57.00 & 228.90 &  41.10 &  41.67 &  41.70 &  42.01 &   3.10 &   3.96 &   8.05 \\ 
 3.0595 & T02 & 0.61 &  -0.69 &  54.40 & 269.40 &  40.90 &  41.48 &  41.60 &  41.90 &   1.97 &   2.51 &   6.30 \\ 
 3.0615 & T02 & 1.05 &  -0.20 &  88.10 & 139.10 &  41.70 &  42.27 &  41.90 &  42.20 &  12.27 &  15.68 &  12.54 \\ 
 3.0879 & T02 & 0.60 &  -0.16 &  41.10 &  59.40 &  40.77 &  41.35 &  40.93 &  41.24 &   1.46 &   1.86 &   1.36 \\ 
 3.0952 & T02 & 0.86 &  -0.12 &  58.10 &  76.00 &  41.30 &  41.87 &  41.42 &  41.72 &   4.93 &   6.30 &   4.18 \\ 
 3.0984 & T02 & 0.70 &  -0.85 &  20.40 & 144.50 &  40.63 &  41.21 &  41.48 &  41.79 &   1.06 &   1.35 &   4.84 \\ 
 3.1027 & T02 & 1.04 &  -0.19 & 106.20 & 164.40 &  41.77 &  42.34 &  41.96 &  42.26 &  14.45 &  18.45 &  14.48 \\ 
 3.1032 & T02 & 0.62 &   0.28 &  84.10 &  44.60 &  41.11 &  41.69 &  40.84 &  41.14 &   3.19 &   4.07 &   1.09 \\ 
 3.1242 & T02 & 0.77 &  -0.51 &  66.90 & 214.30 &  41.25 &  41.82 &  41.75 &  42.06 &   4.33 &   5.53 &   8.98 \\ 
 3.1309 & T02 & 0.62 &  -0.54 & 104.10 & 364.10 &  41.20 &  41.78 &  41.75 &  42.05 &   3.93 &   5.02 &   8.90 \\ 
 3.1345 & T02 & 0.62 &  -0.24 &  70.00 & 121.90 &  41.03 &  41.60 &  41.27 &  41.58 &   2.64 &   3.37 &   2.97 \\ 
 3.1349 & T02 & 0.62 &  -0.30 & 128.60 & 255.20 &  41.29 &  41.87 &  41.59 &  41.89 &   4.83 &   6.16 &   6.20 \\ 
 3.1534 & T02 & 0.80 &  -0.00 & 127.90 & 128.10 &  41.57 &  42.14 &  41.57 &  41.87 &   9.07 &  11.58 &   5.88 \\ 
 3.9003 & T02 & 0.62 &  -0.48 & 187.70 & 564.10 &  41.46 &  42.04 &  41.94 &  42.24 &   7.14 &   9.12 &  13.89 \\ 
22.0429 & T02 & 0.63 &  -0.60 &  65.30 & 261.20 &  41.02 &  41.59 &  41.62 &  41.92 &   2.56 &   3.27 &   6.63 \\ 
22.0637 & T02 & 0.54 &  -0.56 & 151.60 & 554.80 &  41.24 &  41.81 &  41.80 &  42.10 &   4.23 &   5.40 &  10.02 \\ 
22.0764 & T02 & 0.82 &  -0.36 &  51.30 & 118.70 &  41.20 &  41.77 &  41.56 &  41.87 &   3.89 &   4.96 &   5.82 \\ 
22.0770 & T02 & 0.82 &  -0.31 & 100.40 & 206.70 &  41.49 &  42.06 &  41.80 &  42.11 &   7.59 &   9.70 &  10.12 \\ 
22.0779 & T02 & 0.93 &  -0.26 &  38.60 &  69.60 &  41.21 &  41.78 &  41.46 &  41.77 &   3.95 &   5.04 &   4.61 \\ 
22.0843 & T02 & 0.92 &  -0.66 &  37.00 & 170.60 &  41.18 &  41.75 &  41.84 &  42.14 &   3.68 &   4.70 &  10.99 \\ 
22.1406 & T02 & 0.82 &  -0.28 & 175.10 & 333.50 &  41.73 &  42.30 &  42.01 &  42.31 &  13.21 &  16.88 &  16.29 \\ 
\enddata
\tablenotetext{a}{SFR([OII]) is calculated using equation~(\ref{eq_SFR_OII_ratiocorr}).  This equation assumes 
a metallicity of ${\rm log(O/H)+12} = 8.6$ (the mean metallicity of our NFGS sample using the M91 method).}
\tablenotetext{b}{SFR([OII]) is calculated using equation~(\ref{eq_SFR_abund_corr}) with 
${\rm log(O/H)+12} = 8.75$ (M91 method) for the T02 sample and ${\rm log(O/H)+12} = 8.95$ (M91 method) 
for the H02 sample}
\end{deluxetable}

\begin{deluxetable}{llllll}
\tabletypesize{\small}
\tablecaption{Star formation rate densities in Figure~\ref{Madau_plot}.
\label{Madau_table}}
\tablehead{
 $z$
& Dataset
& line
& \multicolumn{3}{c}{SFR density (${\rm M_{\odot}\,yr^{-1}\,Mpc^{-3}}$)}
\\
& & & 
orig\tablenotemark{a} & new\tablenotemark{b} 
& new\tablenotemark{c} \\
& & & & (Z=8.6) & (Z=8.9) 
}
\startdata
0.20 & Tresse \& Maddox 1998 (M98) & H$\alpha$ & 0.01 & 0.01 & 0.01 \\
0.24 & Pascual et al. 2001 (P01) & H$\alpha$ & 0.03 & 0.01 & 0.01 \\
0.38 & Hammer et al. 1997 (H97) & [OII] & 0.005 & 0.007 &  0.01 \\
0.60 & Tresse et al. 2002 (T02) & H$\alpha$ & 0.07 & 0.07 & 0.07 \\
0.63 & Hammer et al. 1997 (H97) & [OII] & 0.02 & 0.03 & 0.04 \\
0.72 & Teplitz et al. 2003 (T03) & [OII] & 0.03 & 0.04 & 0.06 \\
0.80 & Tresse et al. 2002 (T02) & H$\alpha$ & 0.10 & 0.10 & 0.10 \\
0.88 & Hammer et al. 1997 (H97) & [OII] & 0.05 & 0.07 & 0.11 \\
0.94 & Teplitz et al. 2003 (T03) & [OII] & 0.04 & 0.06 & 0.09 \\
1.14 & Teplitz et al. 2003 (T03) & [OII] & 0.07 & 0.09 & 0.14 \\
1.30 & Yan et al. 1999 (Y99) & H$\alpha$ & 0.12 & 0.24 & 0.24 \\
\enddata
\tablenotetext{a}{SFR densities are calculated according to 
Teplitz et al. (2003):
SFR(H$\alpha$) density is calculated using the K98 SFR(H$\alpha$) calibration,
SFR([OII]) density is calculated using [OII]/H$\alpha = 0.45$ and the 
K98 SFR(H$\alpha$) calibration.  Reddening is kept as 
in Teplitz et al. and is not the same for each data set.}
\tablenotetext{b}{SFR(H$\alpha$) densities calculated using the 
K98 SFR(H$\alpha$) calibration, assuming an A(v)=1. 
SFR([OII]) densities calculated using our SFR([OII]) calibration 
(equation~\ref{eq_SFR_OII_ratiocorr}), assuming A(v)=1 and
Z=log(O/H)$+12=8.6$.}
\tablenotetext{c}{SFR(H$\alpha$) densities calculated using the 
K98 SFR(H$\alpha$) calibration, assuming an A(v)=1. 
SFR([OII]) densities calculated using our SFR([OII]) calibration 
(equation~\ref{eq_SFR_OII_ratiocorr8.8}), assuming A(v)=1 and
Z=log(O/H)$+12=8.9$.}
\end{deluxetable}

\end{document}